# Reconstructing Biological Pathways by Applying Selective Incremental Learning to (Very) Small Language Models

Pranta Saha
Vaccine and Infectious Disease Organization
University of Saskatchewan
Saskatoon, SK, Canada
prs527@mail.usask.ca

Joyce Reimer
Vaccine and Infectious Disease Organization
University of Saskatchewan
Saskatoon, SK, Canada
joyce.reimer@usask.ca

Brook Byrns
Advanced Research Computing
University of Saskatchewan
Saskatoon, SK, Canada
brook.byrns@usask.ca

Connor Burbridge
Advanced Research Computing
University of Saskatchewan
Saskatoon, SK, Canada
connor.burbridge@usask.ca

Neeraj Dhar
Vaccine and Infectious Disease Organization
University of Saskatchewan
Saskatoon, SK, Canada
neeraj.dhar@usask.ca

Jeffrey Chen
Vaccine and Infectious Disease Organization
University of Saskatchewan
Saskatoon, SK, Canada
jeffrey.chen@usask.ca

Steven Rayan
Centre for Quantum Topology and Its Applications (quanTA)
University of Saskatchewan
Saskatoon, SK, Canada
steven.rayan@usask.ca

Gordon Broderick
Vaccine and Infectious Disease Organization
University of Saskatchewan
Saskatoon, SK, Canada
awr794@mail.usask.ca

## ABSTRACT

The use of generative artificial intelligence (AI) models is becoming ubiquitous in many fields. Though progress continues to be made, general purpose large language AI models (LLM) show a tendency to deliver creative answers, often called "hallucinations", which have slowed their application in the medical and biomedical fields where accuracy is paramount. We propose that the design and use of much smaller, domain and even task-specific LM may be a more rational and appropriate use of this technology in biomedical research. In this work we apply a very small LM by today's standards to the specialized task of predicting regulatory interactions between molecular components to fill gaps in our current understanding of intracellular pathways. Toward this we attempt to correctly posit known pathway-informed interactions recovered from manually curated pathway databases by selecting and using only the most informative examples as part of an active learning scheme. With this example we show that a small (~110 million parameters) LM based on a Bidirectional Encoder Representations from Transformers (BERT) architecture can propose molecular interactions relevant to tuberculosis persistence and transmission with over 80% accuracy using less than 25% of the ~520 regulatory relationships in question. Using information entropy as a metric for the iterative selection of new tuning examples, we also find that increased accuracy is driven by favoring the use of the incorrectly assigned statements with the highest certainty (lowest entropy). In contrast, the concurrent use of correct but least certain examples contributed little and may have even been detrimental to the learning rate.

## 1 Introduction

The presence and use of generative artificial intelligence (AI), in particular pretrained Large Language Models (LLM), is growing rapidly in an ever-expanding range of fields, becoming even ubiquitous in some [1]. Still, there remain significant challenges to their use including high energy costs, data privacy concerns, and the lack of trustworthiness, rigorous logic and transparency [1]. The latter are of special concern in high-risk fields such as medicine where accuracy supported by established facts is paramount and the consequences of a "flight of fancy" can be severe and even life-threatening [2]. Even in the fields of pharmaceutical and biomedical research, pursuing an unsubstantiated lead can be very costly in both time and resources. Despite advances in explainable AI (XAI) directed at improving transparency, the detection and mitigation of these hallucinations often fail to generalize well across datasets, suggesting again that truthfulness may not be universal but rather contextual and multifaceted [3, 4]. In fairness, very large general-purpose models are expected to cover an incredibly broad range of topics undoubtedly leading to immense hunger for data [5] and an almost unavoidably superficial understanding that lends itself to fanciful best guesses or hallucinations. Attempting to escape the adage "Jack of all trades, master of none", the development and use of domain-specific LLM has emerged as a potentially promising strategy for improving accuracy and believability.

However, so awash with superficial common knowledge are LLM, that attempts at redirecting the latter to more focused domains have been shown in certain circumstances to further increase their appetite for hallucination [4]. The sheer size of such models imparts a significant inertia to altering their foundational beliefs, or calibrated parameter sets, to refocus and tune to new data that is too specialized or too far outside topics represented in their pre-training data sets [6].

In this work, we pursue a highly focused task in a very narrow domain, that involves the reconstruction of biological molecular pathways. As we do not require our model to generate examples outside the immediate domain of molecular cell biology, we can readily take advantage of the known improvements in accuracy associated with domain-specific models. Moreover, this narrow focus also allows for the use of smaller models pre-trained on more limited and potentially higher quality training sets. In contrast to current models approaching 10 T parameters in size [7], we explore herein how very small LM, with parameter counts of 3 B, 1 B and 100 M, might respond to iterative tuning with limited amounts of high quality manually curated biological pathway data describing accepted relationships governing the regulation of proteins in living cells. In addition to exploring model size, we show the benefit of sequential incremental tuning on small segments of data using active learning [8], a static variant of the reinforcement learning strategy recently used to great advantage by DeepSeek (Hangzhou, China) [9]. In this specific use case, we find that policies for the selection of new tuning examples that focus on incorrect prediction made with high certainty favor a broader sampling of the statement space, more robustly avoiding over-represented data elements and leading to better generalization. This strategy applied to a very small 100M parameter BERT LM was able to produce molecular relationships with over 80% accuracy using less than 25% of the overall reference set.

## 2   Methods

### 2.1   Language Models

We evaluated two pretrained-language model architectures, namely an autoregressive language model architecture Llama (Meta, Menlo Park, CA), of 2 different sizes, and a masked-language model 'bert-base-uncased' (Google, Mountain View, CA). The autoregressive (AR) Llama is a language model based on the Bidirectional Encoder Representations from Transformers (BERT) [10] architecture, but with the variation of using a casual decoder only stack. In other words, it predicts the next token in a sequence given previous tokens using the activation function SwiGLU (Swish-Gated Linear Unit) instead of the standard GeLU or ReLU [11]. Available in a multitude of sizes, we used the Llama 3.2 (text only) variant with 1B parameters as a comparator. The architecture and the number of parameters is summarized in Table 1. Llama 3.2 was pretrained using a context length of 128k on up to 9 trillion tokens of text from publicly available sources across 9 languages.

The second architecture consisted of a base uncased implementation of the BERT language model architecture. We also considered a transformer-based model, with the main difference versus Llama being its use of an encoder stack, where each layer uses *self-attention* to look at the entirety of the sentence at once. By including sentence segments positioned both before and after each word, this model attempts to better capture the complete context of each word. For our study, we utilized the BERT-base uncased model consisting of 12-layer transformer blocks and 12 heads as well as 768 hidden units for 110 M parameters in total (Table 1). The model was pretrained on BookCorpus, a dataset consisting of 11,038 unpublished books [12] and English Wikipedia [13]. Pretraining was conducted using a Masked Language Model (MLM) [14] protocol whereby a subset of words in a sentence were randomly masked or hidden and the LM asked to predict those missing words.

### 2.2 A biological use case: data collection and processing

In this preliminary example, we explore the use of LM in proposing as yet undocumented molecular interactions to address gaps in our current understanding of biological pathways important in the response of immune cells to infection by the Mycobacterium tuberculosis (Mtb). Specifically, we applied named entity recognition (NER) to two peer-reviewed publications, extracting a list of 22 host proteins involved in persistence of Mtb infection within macrophages [15] and 9 host proteins mediating its transmission by cough [16]. Here, a domain-informed NER was carried out with a custom implementation that used the PyTesseract (Google, Mountain View, CA) [17] for converting PDF files to text and performing Optical Character Recognition (OCR), then recursively applying the LM pretrained in the biomedical domain, such as 'en_ner_jnlpba_md', 'en_ner_bc5cdr_md', 'en_ner_bionlp13cg_md' available under the ScispaCy Python library [18], to extract text elements and label them where appropriate to putative protein names. These text elements were then verified against standard protein names the HGNC (HUGO Gene Nomenclature Committee)[19] and UniProt identifiers [20] and reconciled where possible.

A reference dataset consisting of trusted manually curated relationships linking these 31 proteins to each other and to their first neighbors was constructed by recovering the latter from the Pathway Commons database [21], itself an amalgamation of 23 pathway databases, and the BEL Selventa Large Corpus (https://github.com/cthoyt/selventa-knowledge) [22] using the suite of tools available within Integrated Network and Dynamical Reasoning Assembler (INDRA v1.23.0 [23]). Extracted relationships were reconciled into INDRA statements describing a regulatory action of an upstream molecular mediator on a downstream target and labeled as activation, Inhibition, IncreaseAmount, and DecreaseAmount relationship classes. Following the removal of contradictory and redundant INDRA statements, the result was an expanded reference data set consisting now of 399 proteins linked by 517 unique regulatory relationships, with 401 of these classified as positive or upregulation (Upregulates label) and 116 classified as negative or downregulation (Downregulates label) of a downstream target by an upstream mediator (Figure 1; Supplemental Table S1). These statements taken collectively form a partial regulatory network with the interconnectivity of these proteins described by standard graph theoretical measures in Supplemental Table S2.

| Model Type | Model Name | Year Release | Architecture | Number of Paramters | Context Window | Accessed Through |
|---|---|---|---|---|---|---|
| Autoregressive Language Model | Llama 3.2 | 2024 | Transformer based, Decoder Only Stack | 1.3 billion | 128000 tokens | ollama/llama3.2:1b |
| Masked Language Model | bert-base-uncased | 2018 | Encoder architecture of transformer with 12 layers and 768 hidden layers | 110 million | 512 tokens | HuggingFace(google-bert/bert-base-uncased) |

**Table 1. Language model features.**

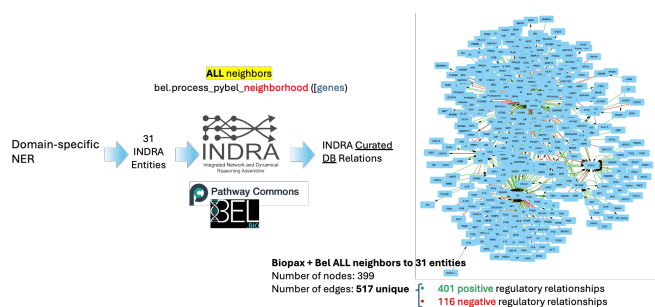

**Figure 1. Assembly of a reference data set.** Outline of assembly process used for recovery of entities of interest from text and relationships from manually curated pathway databases.

### 2.3 Standardization of data and query formulation

From the post-processed INDRA statements, we created an initial dataset in tabular format with columns: Source, Target and Relationship (upregulates, downregulates). Modern Language Models, such as BERT, are highly robust machine learning models and these models at their core operate on numerical data. They perform mathematical operations such as matrix multiplications, additions, non-linear transformations and cannot directly understand or process text string like "Upregulates" and "Downregulates" as labels. Therefore, to represent categorical information for model training we converted classes to binary labels upregulates to 0 and downregulates to 1. We then use these entries to populate the following LM query (Eq.1).

$$\text{"Does protein [SOURCE\_PROTEIN] regulate protein [TARGET\_PROTEIN] positively or negatively?"} \quad (1)$$

To explore a minimal but informative supervision scenario, we created a baseline data subset consisting of only 10 upregulates and 10 downregulates relations under the constraint that each protein appears only once per class as source and target. The baseline model ($M(0)$) is tuned and a separate random subset (sampling without replacement) of 10 examples from each class are selected to validate this first model. At each iteration, 2 incorrectly assigned statements are selected from each class based on criteria described below, added to the base tuning set and the model retuned. As such, the tuning set grows as 2x 10 + *iterations* x (4) examples (Figure 2).

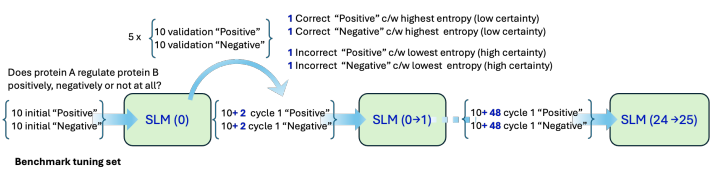

**Figure 2. Sequential tuning and data selection.** Repeated sequential cycles of active learning with 100% memory using AdamW optimization algorithm using new examples selected as highly confident (low information entropy) but incorrect assignments.

### 2.4 Entropy-based sample selection

To assess the level of certainty with which each validation querie is assigned to weak or strong affinity class, we first use the Softmax activation function to transform and normalize the LM raw output scores, also known as "logits", into a probability distribution (Eq. 2) [24]:

$$p_i = \frac{\exp(z_i)}{\Sigma_j \exp(z_j)} \quad (2)$$

where $z_i$ is the logit corresponding to class $i$ normalized to the interval [0,1], and where all output values sum up to 1, thereby providing an estimate of the model's confidence in assigning a query response to that class. The predicted class was determined by selecting the class with highest probability. We then also used the Shannon entropy over the Softmax output distribution to quantify our model's uncertainty:

$$H(p) = -\sum_{i=1}^{n} p_i \, log_2(p_i) \quad (3)$$

where $p_i$ is the probability of assignment to the $i$-th class. If the Softmax output for a given prediction is highly concentrated on a single class, the entropy will be low, which means the model is very confident about its prediction. If the output is spread more evenly across multiple classes, the entropy will be high, which will mean the model is uncertain about which class the input belongs to. In this work, each query was submitted to the LM in replicate a total of 10 times to reduce stochastic variation and compute stable average entropy and confidence values.

### 2.5 LM fine tuning

At each iteration, 5 sets of 10 queries (5 per class) were randomly drawn from the unlabeled pool and evaluated by the model in replicate 10 times each. From these predictions, we selected 1 high-entropy correct but uncertain prediction and 1 low-entropy incorrect but certain prediction from each class, resulting in a maximum of 4 new samples per iteration. As high entropy but correct predictions reflect model uncertainty, this initial criteria was intended to provide positive reinforcement and encourage boundary refinement. Conversely, low entropy but incorrect predictions reflect model overconfidence and their selection for tuning is intended to provide negative reinforcement and the correction of misadjusted beliefs. Selected samples were appended to the training dataset and marked as "used" to avoid re-selection in future iterations. In this cumulative tuning set, both pre-existing and new additions are assigned the same weight supporting a 100% persistent memory framework.

After each selection round, the model was fine-tuned on the updated dataset for 3 epochs using the AdamW optimizer (learning rate: 2e-5, $\varepsilon$ = 1e-8) [25]. We used Ray Tune, a widely recognized Python library for distributed hyperparameter optimization [26], to explore search spaces for learning rate, epochs, batch size and weight decay. Models were fine-tuned on both unique and random datasets under identical conditions to assess performance across different data configurations. The hyperparameter selection is summarized in Table 2. Tuning was conducted with a batch size of 8, and validation accuracy was recorded using a consistent held-out validation set. This iterative cycle was repeated for twenty-five iterations.

| Hyperparameter | Search Space | Best Value |
|---|---|---|
| Optimizer | AdamW | AdamW |
| Weight Decay | 0.1, 0.01, 0.001 | 0.01 |
| Learning Rate | 1e-5 - 5e-5 | 3.00E-05 |
| Batch Size | 4, 8, 16 | 8 |
| Epochs | 2, 4, 8 | 4 |
| Device | CUDA | CUDA |

**Table 2. Hyperparameter selection for BERT model tuning using AdamW search algorithm.**

## 3 Results

### 3.1 Zero-shot inference with small and very small LM

To assess the out-of-the-box performance of both small and very small general-purpose LM on a specialized task with only one round of fine-tuning, we conducted a zero-shot inference task tailored to their respective architectures. Llama (version 3.2 1B) is an autoregressive generative language model, capable of processing open-ended prompts without requiring explicit formatting or fine-tuning. Therefore, we formulated the question as: "Does protein [SOURCE_PROTEIN] regulate protein [TARGET_PROTEIN] positively or negatively?". The model was expected to generate a direct answer (e.g. "positively" or "negatively"), reflecting its understanding of directed regulatory action. Since BERT (bert-base-uncased) is a masked language model (MLM), it requires contextualized input with a token to predict. To guide the model's understanding of protein-protein regulatory interactions, we first provided 10 example sentences, such as: "STAT3 positively regulates ERK", "TLR4 negatively regulates TNF". These question-and-answer examples served as a contextual priming. The model was then queried using the format: "Does protein [SOURCE_PROTEIN] regulate protein [TARGET_PROTEIN] positively or negatively? It regulates it [MASK]." Here, the model predicted the masked token, producing an output of either "positively" or "negatively" based on the learned context.

In these limited evaluations, we fine-tuned each LM on 80% of the examples in the smallest class or 93 positive regulation and 93 negative regulation examples. We then submitted the remaining 20% or 23 randomly selected source and target protein pairs as validation queries, replicating this 3 times (Figure 3). The 1B parameter Llama 3.2 LM demonstrated an overwhelming bias to one class, delivering an overall accuracy of 50% (0.43 std. error) with a low entropy of 0.43 for each class underscoring a high certainty of these class assignments (Supplemental Table S3). The much smaller BERT uncased base model delivered an equivalent albeit slightly higher accuracy (54%, p=0.27) but with a much more balanced performance and with a significantly higher uncertainty (entropy of 0.95, 0.91 vs 0.43, p=0.03, 0.04).

### 3.2 Sequential active learning in BERT base LM

Given the equivalent or better zero-shot performance offered by the much smaller BERT LM, we next then assessed how performance might improve further with additional rounds of tuning using minimal amounts of new data carefully selected. As described in Section 2.5 we randomly selected a baseline initial fine-tuning subset of data consisting of 10 example queries from each class. These were used in a first round of fine-tuning of the out-of-the-box BERT base uncased LM, with performance evaluated against 10 validation queries never before seen by the LM randomly selected from each class. Uncertainty in the form of information entropy was computed for each predicted validation set assignment. In a first experiment we select for each class the correctly predicted validation set example with the highest entropy and the incorrectly predicted validation set example with the lowest entropy. In other words, the most certain incorrect predictions and the least certain incorrect predictions, when available. This was repeated over incremental fine-tuning 25 cycles in two replicate numerical experiments based on the same initial baseline tuning set with the only difference being the random selection of subsequent validation sets. Initial results suggest that the results can be quite variable with repeat experiments achieving an overall accuracy of 69% to 83% (Figure 4; Table 3), using roughly 21% (109 and 110 of 517 relationship statements) of the overall data available.

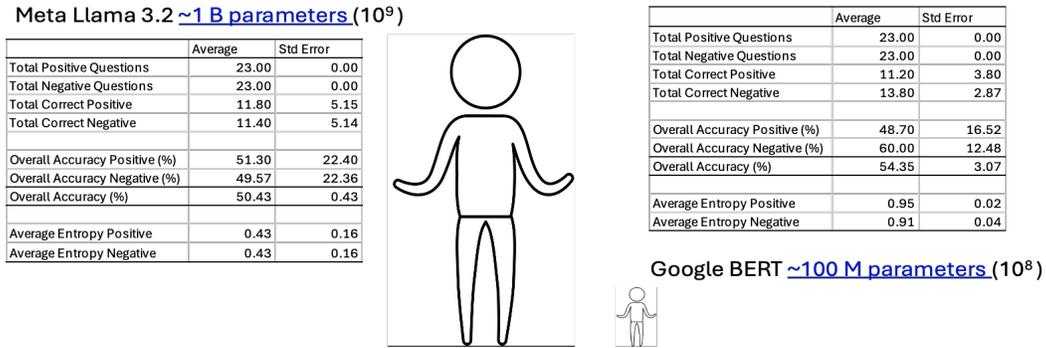

**Figure 3. David and Goliath.** A zero-shot inference using 80% of the overall reference set for tuning and 20% for predictive validation sampled in 5 repeat queries consisting of 23 questions for each class.

To explore the contribution of criteria used in the selection of fine-tuning examples, we examined the trajectory in entropy values of these validation sets across iterations and compared them across experiments. Results shown in Figure 5 show that the trajectory across iterations for entropy values corresponding to incorrect assignments made with high certainty or low entropy in experiment #1 are quite different from those in experiment #2 for both classes. This is especially visible mid-course in iterations 10 through 20 where entropy values for experiment #2 assignments are markedly higher (less certain). In comparison, the trajectories for entropy values assigned to correct assignments made with low certainty or high entropy essentially overlap in experiments #1 and #2. As the distinguishing feature between experiments #1 and #2 appear driven by fluctuations in the entropy of incorrect and overconfident assignments we attempted to leverage this effect further by focusing on these overconfident assignments only and increasing their number. In a new round of experiments, we now select the 2 incorrect assignments made in each class with the lowest and second lowest entropy values only. Focusing only on the addition of 4 incorrect and overconfident examples at each iteration we now obtain an overall accuracy exceeding 90% using only 23% (117 out of 517 relationship statements) of the overall data available over the same 25 tuning cycles (Figure 6; Table 3).

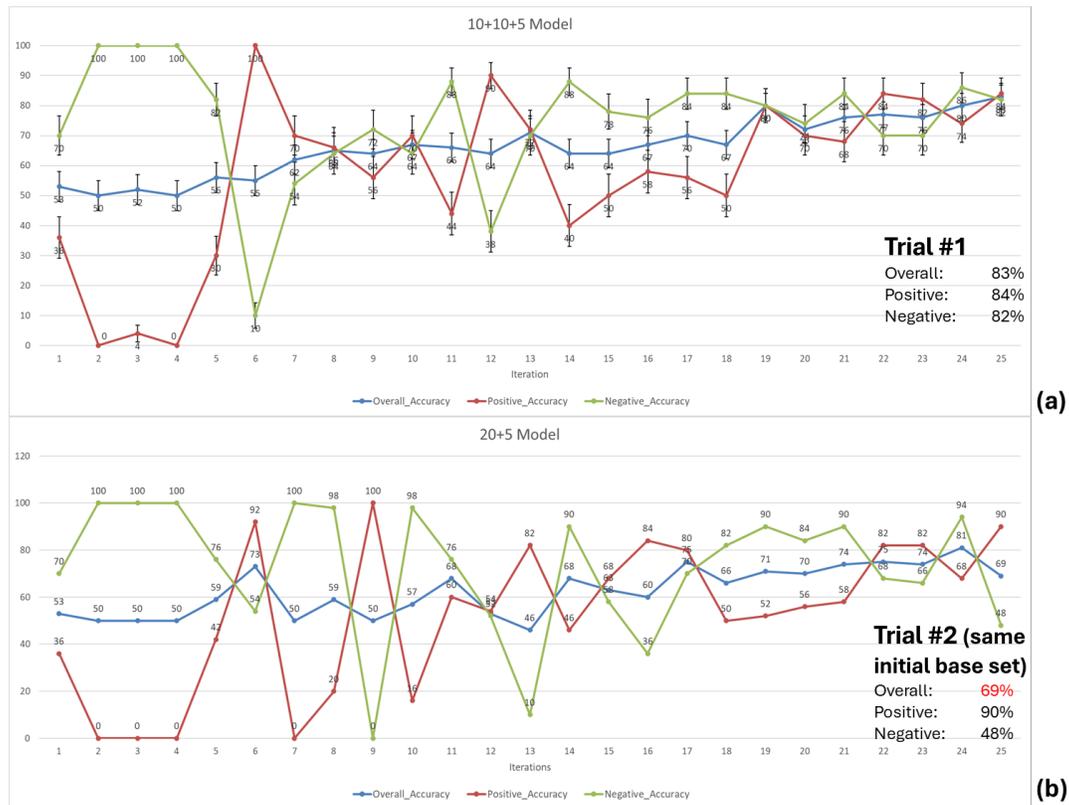

**Figure 4. A joint reward and penalty strategy.** Predictive accuracy achieved with incremental active learning using both underconfident and overconfident examples.

|  | Correct high E + incorrect low E | | | | | Incorrect low E only | | | | | |
|---|---|---|---|---|---|---|---|---|---|---|---|
|  | Trial #1 | Trial #2 | Trial #3 | Average | Std Error | Trial #1 | Trial #2 | Trial #3 | Average | Std Error | p value |
| Positive Accuracy | 84 | 90 | 74 | 82.67 | 4.67 | 88 | 80 | 100 | 89.33 | 5.81 | 0.424 |
| Negative Accuracy | 82 | 48 | 76 | 68.67 | 10.48 | 94 | 100 | 96 | 96.67 | 1.76 | 0.112 |
| Overall Accuracy | 83 | 69 | 75 | 75.67 | 4.06 | 91 | 90 | 98 | 93.00 | 2.52 | **0.030** |
| Unique Sources | | | | | | | | | | | |
| Edge Count > 1 | 27 | 26 | 32 | 28.33 | 1.86 | 26 | 32 | 29 | 29.00 | 1.73 | 0.806 |
| Edge Count =1 | 16 | 21 | 26 | 21.00 | 2.89 | 23 | 19 | 15 | 19.00 | 2.31 | 0.619 |
| Unique Targets | | | | | | | | | | | |
| Edge Count > 1 | 30 | 26 | 21 | 25.67 | 2.60 | 28 | 20 | 20 | 22.67 | 2.67 | 0.466 |
| Edge Count =1 | _44_ | _40_ | 40 | 41.33 | 1.33 | _47_ | 37 | 38 | 40.67 | 3.18 | 0.860 |
| Total Unique | 117 | 113 | 119 | 116.33 | 1.76 | 124 | 108 | 102 | 111.33 | 6.57 | 0.530 |
| Positive Samples Added | 47 | 45 | 48 | 46.67 | 0.88 | 50 | 43 | 37 | 43.33 | 3.76 | 0.471 |
| Negative Samples Added | 47 | 45 | 47 | 46.33 | 0.67 | 47 | 39 | 36 | 40.67 | 3.28 | 0.223 |
| Total samples | 94 | 90 | 95 | 93.00 | 1.53 | 97 | 82 | 73 | 84.00 | 7.00 | 0.326 |

**Table 3. Summary of active learning results.**

### 3.3 Selecting informative learning subsets

To better understand how differences in example selection criteria or policy affect learning rate and accuracy, we examined the properties of the examples selected as informative across the 25 fine-tuning iterations. The representation of each molecular species in the role of upstream mediator or downstream target is a product of its location and connectivity as a node in the greater regulatory interaction network. The node level statistics describing the overall connectivity (total edge count), the number of upstream mediators (total indegree), the number of downstream targets (total outdegree), and other connectivity metrics of each node are described in Table S2. Unsurprisingly, inflammatory immune master regulators like TNF and IL6 have the highest connectivity in this network, regulating 177 and 67 downstream targets respectively, which in turn makes them high frequency query subject proteins in the reference set. Likewise, they also rank highly as downstream targets as do STAT1, CRP and MYD88 to a lesser extent, making these well represented as query object proteins in the reference set. The relevance of this over-representation to selection during active learning may be assessed by examining the specific subset of statements extracted over the 25 iterations. Results summarized in Table 3 show that overall predictive accuracy was significantly higher for the policy using only incorrect predictions made with low entropy or high certainty (93.0% vs 75.7%, p=0.03). In examining each regulatory class, we find that while assignment to a positive class was similar in accuracy (82.7% vs 89.3%, p=0.424) across policies, the difference in assignment to a negative class was much more noticeable (68.7% vs 96.7%, p=0.112), though not statistically significant. In looking more closely at the samples selected by each policy, we found that on average the focus on incorrect low entropy examples resulted in the recruitment of somewhat fewer not more learning examples (84 vs 93, p=0.326) though again the variability was such that this difference did not achieve statistical significance. The composition of these learning sets was quite similar in terms of the overall number of unique molecular species selected (116 and 111 species). Likewise, differences in the frequency with which high versus low connectivity source and target species were selected across policies was unremarkable (Supplemental Table S4) even when examining those species uniquely selected by one policy or the other (Supplemental Table S5). However, when examining the neighborhood connectivity of the species selected differentially in one policy over the other, we find that the use of only incorrect low entropy examples appears to favor a markedly higher sampling of target species connected specifically to network hub mediators (Supplemental Table S5). Indeed, in those few reference set relationships that these species do maintain, they appear as targets of very well-connected master regulators (e.g. TNF with 177 downstream targets). In this example, the policy of selecting only incorrect predictions made with low entropy or high confidence uniquely recruited 24 hub-mediated species compared to 16 such species for the rival policy of also considering low confidence or high entropy but correct predictions.

### 4 Discussion

In this preliminary exploration we attempt to assess the feasibility of using progressively smaller generative AI Language Models (LM) to propose regulatory relationships between molecules currently missing from our understanding of biological pathways. Using a set of regulatory relationships extracted from manually curated databases as a reference truth, we show that even when tuned on 80% of the available data, the 100M parameter BERT model performs at least as well in terms of predictive accuracy as the much larger Llama 3.2 1B and 3 B LMs. Interestingly, the latter display a significantly higher level of certainty (lower entropy) despite delivering essentially random class assignments (~50% accuracy). With added size and complexity delivering little value in this use case, we then show that even the parsimonious 100M parameter basic BERT model can achieve classification accuracies in excess of 80% using little more than 20% of the overall data if we incrementally introduce small numbers of well-chosen examples instead of indiscriminate tuning on much larger amounts of data. This in and of itself is not all that surprising and the advantages of incremental elective tuning in delivering high accuracies at low resource usage have drawn significant industry attention recently as demonstrated by DeepSeek's (Hangzhou, China) [9] use of reinforcement learning. In the current analysis, we use active learning (AL) as a framework for the incremental selection of informative tuning examples which might be described as being akin to a static form of reinforcement learning.

**Figure 5. Contributing policy elements.** Selection of overconfident examples is distinguishing feature in better performing active learning trial.

**Figure 6. A reward focused approach.** Predictive accuracy using a penalty-based approach selecting only high confidence but incorrect learning examples.

Given the significant and often dramatic reduction in data requirement, there is a rapidly evolving focus on comparing various policies for the selection of the most informative examples with the most fundamental approaches using metrics for prediction uncertainty, as well as example diversity and similarity [27]. Here we focused on prediction uncertainty [28] as a selection metric where the standard selection strategy consists of recruiting examples with the most uncertain prediction or highest entropy. However, as labeled data was available, we compared the selection of high entropy but correctly predicted examples with the recruitment of low entropy but incorrectly classified examples. While labels have been used as a measure of consistency to infer uncertainty [29] and estimates of correctness based on greedy decoding have been used to adjust uncertainty measures [30], there is little work exploring the direct use of true labels in conjunction with an information theoretic measure such as entropy. Indeed, our exploratory analysis suggests that low certainty, but correct predictions may be less informative than incorrect predictions made with high certainty. Focusing exclusively on the latter appears to favor a more active sampling of species that though poorly connected to other species overall, are selectively mediated by well-connected master regulators. Such well-connected mediators would appear frequently as source terms in the reference query set, which may explain the low entropy values or high certainty of validation statements including these mediators as sources. As a result, predictions might inherit a false confidence (low entropy) from such a frequently occurring source term while struggling to predict a rarely occurring target term or poorly connected target species. This early observation may serve to further modify, inform and streamline the label-aware strategy proposed by Lin et al. [30], offering an alternative to the greedy decoding model that may deliver an even more efficient use of data. Indeed, the concurrent prediction of correctness used by Lin [30] might be modified to focus on infrequent target terms leading to generally incorrect answers by leveraging our knowledge of the small world or hub-centric network architectures common in biology [31]. Such a modified label-aware strategy could be used as a means of proposing the next best query to verify, serving as a means of designing new maximally informative experiments.

Certainly, these observations are derived from a very limited and specific use case and these concepts should be more broadly tested to assess how well they might generalize. In further exploiting the advantages of domain-focused models, a migration from the basic BERT pre-trained model used here to the BioBERT model pre-trained on biomedical content in PubMed would be worth assessing. Regardless of base model, as our group continues to explore definitions and metrics of reliability and trustworthiness, we expect to also enlist concepts from Confident Learning [32]. In addition, while a small reference data set was used here this may not always be the case and efforts will inevitably become focused on improving the compactness of the overall reference set (currently 100% retention) by leveraging parameter-efficient fine-tuning approaches (PEFT) [33] such as LoRa/ QLoRa [34]. Ultimately, these results will help craft more robust approaches for predicting the most compact sets of maximally informative new queries. Specifically, new queries for which no labels exist yet and that, if answered experimentally, would optimally enrich the refence data set.

## ACKNOWLEDGEMENTS

This work was supported by the University of Saskatchewan's Centre for Quantum Topology and Its Applications (quanTA) and by the Vaccine and Infectious Disease Organization (VIDO). VIDO receives operational funding from the Canada Foundation for Innovation (CFI) through the Major Science Initiatives Fund and from the Government of Saskatchewan through Innovation Saskatchewan and the Ministry of Agriculture. The quanTA Centre's high-performance computational work has been advanced through a CFI John R. Evans Leaders Fund grant. We thank the University of Saskatchewan Advanced Research Computing (ARC) team for their efforts and close collaboration in creating an excellent local environment for this computational work. This article is submitted with the permission of the Director of VIDO.

**Table S1**. Reference truth statemenst extracted from Pathway Commons and BEL Large Corpus

| Source | Target | Relationship Action |
|--------|--------|---------------------|
| IL6 | A2M | Activation |
| IL6 | ABCB1 | Activation |
| TNF | ABCB1 | Activation |
| IL6 | ABCC1 | Activation |
| IL6 | ABCG2 | Activation |
| IL6 | ADAMTS5 | Activation |
| TNF | AGT | Activation |
| AGT | AIM2 | Activation |
| IL6 | AKT1 | Activation |
| TNF | AKT2 | Activation |
| TNF | ALPL | Activation |
| IL6 | ANXA1 | Activation |
| TNF | AREG | Activation |
| TNF | ATF2 | Activation |
| TNF | ATF3 | Activation |
| TNF | BMP2 | Activation |
| TNF | C1R | Activation |
| TNF | C3 | Activation |
| TNF | C5AR1 | Activation |
| IL6 | C8A | Activation |
| IL6 | C8G | Activation |
| AP1 | CA2 | Activation |
| NKX2-1 | CA2 | Activation |
| PKC | CA2 | Activation |
| TNF | CASP1 | Activation |
| TNF | CASP2 | Activation |
| TNF | CASP6 | Activation |
| TNF | CASP7 | Activation |
| MYD88 | CASP8 | Activation |
| TNF | CASP8 | Activation |
| IL6 | CBL | Activation |
| TNF | CCL11 | Activation |
| CRP | CCL2 | Activation |
| TNF | CCL2 | Activation |
| TNF | CCL3 | Activation |
| CRP | CCR2 | Activation |
| CEBP | CD14 | Activation |
| IFNA1 | CD14 | Activation |
| KITLG | CD14 | Activation |
| PI3K | CD14 | Activation |
| PRKAC | CD14 | Activation |
| TNF | CD14 | Activation |
| IL6 | CD163 | Activation |
| IL6 | CD1A | Activation |
| TNF | CD1A | Activation |
| TNF | CD38 | Activation |
| TNF | CD40 | Activation |
| TNF | CD40LG | Activation |
| TNF | CD55 | Activation |
| TNF | CD59 | Activation |
| TNF | CD80 | Activation |
| TNF | CD86 | Activation |
| STAT1 | CDKN1A | Activation |
| TNF | CDKN1A | Activation |
| IL6 | CDKN1B | Activation |
| IL6 | CEBPB | Activation |
| TNF | CEBPB | Activation |
| TNF | CFB | Activation |
| TNF | CFL2 | Activation |
| TNF | CHUK | Activation |
| TNF | CK2 | Activation |
| TNF | CLEC3B | Activation |
| TNF | CLIC4 | Activation |
| IL6 | CMA1 | Activation |
| CD32 | CRP | Activation |
| CXCL8 | CRP | Activation |
| ERK | CRP | Activation |
| HNF1A | CRP | Activation |
| IL1B | CRP | Activation |
| IL6 | CRP | Activation |
| LEP | CRP | Activation |
| NFkappaB | CRP | Activation |
| p38 | CRP | Activation |
| PTGS2 | CRP | Activation |
| TNF | CRP | Activation |
| TNF | CSF1 | Activation |
| TNF | CSF2 | Activation |
| TNF | CSF3 | Activation |
| TNF | CTSB | Activation |
| IL6 | CTSL | Activation |
| IL6 | CTSS | Activation |
| TNF | CX3CL1 | Activation |
| TNF | CXCL10 | Activation |
| TNF | CXCL12 | Activation |
| TNF | CXCL5 | Activation |
| TLR9 | CXCL8 | Activation |
| TNF | CXCL9 | Activation |
| TNF | CXCR3 | Activation |
| TNF | DAPK1 | Activation |
| IL6 | EGFR | Activation |
| TNF | EHF | Activation |
| IRF3 | EP300 | Activation |
| PTPA | ERK | Activation |
| IL6 | ESR1 | Activation |
| TNF | F2RL1 | Activation |
| TNF | F2RL3 | Activation |
| TNF | FAS | Activation |
| TNF | FCER2 | Activation |
| CRP | FCGR1A | Activation |
| TNF | FCGR3B | Activation |
| IL6 | FN1 | Activation |
| TNF | FOXO4 | Activation |
| TNF | FUT4 | Activation |
| IL6 | FYN | Activation |
| IL6 | GBF1 | Activation |
| CAMP | GSK3B | Activation |
| IL6 | HBEGF | Activation |
| IL6 | HCK | Activation |
| TNF | HIF1A | Activation |
| TNF | HLA-A | Activation |
| TNF | HLA-B | Activation |
| TNF | HLA-C | Activation |
| TNF | HLA-F | Activation |
| IL6 | HP | Activation |
| TNF | HSD11B1 | Activation |
| IL6 | HSF1 | Activation |
| TNF | HSF1 | Activation |
| IL6 | HSP90B1 | Activation |
| HOXD3 | IFI16 | Activation |
| IFNG | IFI16 | Activation |
| LIF | IFI16 | Activation |
| RAF1 | IFI16 | Activation |
| STAT3 | IFI16 | Activation |
| IRF3 | IFNA1 | Activation |
| IRF3 | IFNB1 | Activation |
| STAT1 | IFNG | Activation |
| TNF | IFNG | Activation |
| TNF | IKBKB | Activation |
| TNF | IKBKG | Activation |
| MYD88 | IKK_family | Activation |
| TNF | IL11 | Activation |
| TNF | IL12A | Activation |
| TNF | IL12B | Activation |
| IL6 | IL17A | Activation |
| TNF | IL18 | Activation |
| TNF | IL4R | Activation |
| ADRB2 | IL6 | Activation |
| AGT | IL6 | Activation |
| AHR | IL6 | Activation |
| AP1 | IL6 | Activation |
| AR | IL6 | Activation |
| BSG | IL6 | Activation |
| CCL2 | IL6 | Activation |
| CD40 | IL6 | Activation |
| CD40LG | IL6 | Activation |
| CXCL12 | IL6 | Activation |
| EGFR | IL6 | Activation |
| ERBB2 | IL6 | Activation |
| F10 | IL6 | Activation |
| F2 | IL6 | Activation |
| GAB1 | IL6 | Activation |
| HCK | IL6 | Activation |
| IFNG | IL6 | Activation |
| IL13 | IL6 | Activation |
| IL17A | IL6 | Activation |
| IL17F | IL6 | Activation |
| IL18 | IL6 | Activation |
| IL1B | IL6 | Activation |
| IL1R1 | IL6 | Activation |
| IL4 | IL6 | Activation |
| IL6 | IL6 | Activation |
| IL6R | IL6 | Activation |
| IL6ST | IL6 | Activation |
| INS | IL6 | Activation |
| ITGA5 | IL6 | Activation |
| KLF7 | IL6 | Activation |
| LEPR | IL6 | Activation |
| LMX1B | IL6 | Activation |
| MAPK1 | IL6 | Activation |
| NFkappaB | IL6 | Activation |
| NGF | IL6 | Activation |
| OSM | IL6 | Activation |
| p38 | IL6 | Activation |
| PARP1 | IL6 | Activation |
| PIK3C3 | IL6 | Activation |
| PKC | IL6 | Activation |
| PNPT1 | IL6 | Activation |
| PTGER4 | IL6 | Activation |
| PTK2 | IL6 | Activation |
| PTPN11 | IL6 | Activation |
| TGFB1 | IL6 | Activation |
| TIRAP | IL6 | Activation |
| TLR4 | IL6 | Activation |
| TNFRSF1A | IL6 | Activation |
| VEGF | IL6 | Activation |
| XDH | IL6 | Activation |
| IL6 | IL6ST | Activation |
| MYD88 | IRAK1 | Activation |
| MYD88 | IRAK4 | Activation |
| BEGAIN | IRF3 | Activation |
| CXCL12 | IRF3 | Activation |
| IKBKE | IRF3 | Activation |
| MAVS | IRF3 | Activation |
| TICAM1 | IRF3 | Activation |
| TLR3 | IRF3 | Activation |
| TLR4 | IRF3 | Activation |
| MYD88 | IRF5 | Activation |
| TLR9 | IRF7 | Activation |
| TNF | ITGA6 | Activation |
| TNF | ITGAV | Activation |
| TNF | JAK | Activation |
| TNF | JUN | Activation |
| IL6 | KRAS | Activation |
| IL6 | LYN | Activation |
| TNF | LYZ | Activation |
| TNF | MAP2K1 | Activation |
| TNF | MAP2K2 | Activation |
| TNF | MAP2K6 | Activation |
| TNF | MAP3K1 | Activation |
| TNF | MAP3K14 | Activation |
| TNF | MAP3K5 | Activation |
| TNF | MAP4K3 | Activation |
| TNF | MAP4K4 | Activation |
| TNF | MAPK1 | Activation |
| IL6 | MAPK14 | Activation |
| IL6 | MAPK3 | Activation |
| IL6 | MAPK7 | Activation |
| TNF | MAPK9 | Activation |
| TNF | MAPKAPK3 | Activation |
| IL6 | MAZ | Activation |
| IL6 | MEF2 | Activation |
| IL6 | MGAT4B | Activation |
| IL6 | MGAT5 | Activation |
| TNF | MIF | Activation |
| TNF | MMP1 | Activation |
| TNF | MMP14 | Activation |
| TNF | MMP3 | Activation |
| IL6 | MMP9 | Activation |
| TNF | MUC5AC | Activation |
| IFNG | MYD88 | Activation |
| IRF5 | MYD88 | Activation |
| IRF7 | MYD88 | Activation |
| TLR1 | MYD88 | Activation |
| **TLR2** | **MYD88** | **Activation** |
| TLR5 | MYD88 | Activation |
| TLR6 | MYD88 | Activation |
| TLR7 | MYD88 | Activation |
| TLR8 | MYD88 | Activation |
| TLR9 | MYD88 | Activation |
| TNF | MYLK | Activation |
| TNF | NCAM1 | Activation |
| CD14 | NFkappaB | Activation |
| IL6 | NFkappaB | Activation |
| **MYD88** | **NFkappaB** | **Activation** |
| TBK1 | NFkappaB | Activation |
| TLR9 | NFkappaB | Activation |
| TNF | NFKB1 | Activation |
| TNF | NFKB2 | Activation |
| TNF | NOS1 | Activation |
| IL6 | ORM1 | Activation |
| TNF | PAWR | Activation |
| TNF | PFN1 | Activation |
| IL6 | PHB1 | Activation |
| p38 | PI3 | Activation |
| IL6 | PI3K | Activation |
| TNF | PLA2G2A | Activation |
| TNF | PLAU | Activation |
| IL6 | POMC | Activation |
| IL6 | POU2F1 | Activation |
| TNF | PPP1R15A | Activation |
| CAMP | PRKAC | Activation |
| TNF | PRKCI | Activation |
| TNF | PRKCZ | Activation |
| IL6 | PRKD1 | Activation |
| TNF | Proteasome | Activation |
| TNF | PSMB9 | Activation |
| TNF | PSME1 | Activation |
| TNF | PTEN | Activation |
| TNF | PTGES | Activation |
| TNF | PTGS2 | Activation |
| TNF | PTHLH | Activation |
| TNF | PTX3 | Activation |
| TNF | RELA | Activation |
| TNF | RIPK1 | Activation |
| TNF | RIPK2 | Activation |
| TNF | RIPK3 | Activation |
| IL6 | RORA | Activation |
| TNF | SELE | Activation |
| IL6 | SERPINA3 | Activation |
| TNF | SERPINE1 | Activation |
| TNF | SERPING1 | Activation |
| TNF | SERPINH1 | Activation |
| TNF | SLC6A4 | Activation |
| TNF | SLC6A6 | Activation |
| TNF | SMPD1 | Activation |
| STAT1 | SOCS1 | Activation |
| TNF | SP1 | Activation |
| TNF | SPP1 | Activation |
| TNF | ST3GAL6 | Activation |
| BMX | STAT1 | Activation |
| EGFR | STAT1 | Activation |
| ETS1 | STAT1 | Activation |
| FGFR1 | STAT1 | Activation |
| FGFR3 | STAT1 | Activation |
| HDAC | STAT1 | Activation |
| HDAC1 | STAT1 | Activation |
| HDAC2 | STAT1 | Activation |
| HDAC3 | STAT1 | Activation |
| IFNA1 | STAT1 | Activation |
| IFNG | STAT1 | Activation |
| IFNGR1 | STAT1 | Activation |
| IFNL1 | STAT1 | Activation |
| IFNL2 | STAT1 | Activation |
| IL10 | STAT1 | Activation |
| IL6ST | STAT1 | Activation |
| IRAK1 | STAT1 | Activation |
| IRF5 | STAT1 | Activation |
| JAK2 | STAT1 | Activation |
| LPL | STAT1 | Activation |
| MAPK1 | STAT1 | Activation |
| MAPK3 | STAT1 | Activation |
| NFkappaB | STAT1 | Activation |
| OSM | STAT1 | Activation |
| PDGFRB | STAT1 | Activation |
| PRKCD | STAT1 | Activation |
| PRKCE | STAT1 | Activation |
| PRMT1 | STAT1 | Activation |
| RET | STAT1 | Activation |
| TNF | STAT1 | Activation |
| TYK2 | STAT1 | Activation |
| IL6 | STAT2 | Activation |
| IL6 | STAT3 | Activation |
| TNF | STAT3 | Activation |
| TNF | TAP1 | Activation |
| TNF | TAP2 | Activation |
| TLR3 | TBK1 | Activation |
| TNF | TGFA | Activation |
| TNF | THBS1 | Activation |
| TNF | TIMP1 | Activation |
| TNF | TIMP3 | Activation |
| CD14 | TLR2 | Activation |
| TNF | TLR2 | Activation |
| TBK1 | TLR3 | Activation |
| TGFA | TLR9 | Activation |
| ADAM17 | TNF | Activation |
| ADRB2 | TNF | Activation |
| AGT | TNF | Activation |
| AGTR1 | TNF | Activation |
| AKR1B1 | TNF | Activation |
| ALOX5 | TNF | Activation |
| AP1 | TNF | Activation |
| BSG | TNF | Activation |
| Caspase | TNF | Activation |
| CD14 | TNF | Activation |
| CD40LG | TNF | Activation |
| CENPJ | TNF | Activation |
| CHUK | TNF | Activation |
| CRP | TNF | Activation |
| DCN | TNF | Activation |
| DCTN1 | TNF | Activation |
| DHX9 | TNF | Activation |
| EGR4 | TNF | Activation |
| EP300 | TNF | Activation |
| HMGB1 | TNF | Activation |
| HNRNPU | TNF | Activation |
| HSPA4 | TNF | Activation |
| HSPB1 | TNF | Activation |
| ICMT | TNF | Activation |
| IFI16 | TNF | Activation |
| IKBKB | TNF | Activation |
| IKBKG | TNF | Activation |
| IL17A | TNF | Activation |
| IL18 | TNF | Activation |
| IL1B | TNF | Activation |
| IRF9 | TNF | Activation |
| MAP3K1 | TNF | Activation |
| MAP3K7 | TNF | Activation |
| MAPK14 | TNF | Activation |
| MAPK3 | TNF | Activation |
| MMP12 | TNF | Activation |
| NAMPT | TNF | Activation |
| NFAT | TNF | Activation |
| NFkappaB | TNF | Activation |
| NFKBIA | TNF | Activation |
| NGF | TNF | Activation |
| NOS3 | TNF | Activation |
| NOX4 | TNF | Activation |
| NRAS | TNF | Activation |
| OLR1 | TNF | Activation |
| PARP1 | TNF | Activation |
| PDE3B | TNF | Activation |
| PLA2G4A | TNF | Activation |
| PTK2 | TNF | Activation |
| RALBP1 | TNF | Activation |
| RIPK2 | TNF | Activation |
| TARDBP | TNF | Activation |
| TIMP3 | TNF | Activation |
| TLR1 | TNF | Activation |
| TLR2 | TNF | Activation |
| TLR4 | TNF | Activation |
| TONSL | TNF | Activation |
| TRAF2 | TNF | Activation |
| VEGF | TNF | Activation |
| TNF | TNFRSF1A | Activation |
| TNF | TNFRSF1B | Activation |
| TNF | TP53 | Activation |
| TNF | TRAF1 | Activation |
| TNF | TRAF2 | Activation |
| TNF | TRAF3 | Activation |
| MYD88 | TRAF6 | Activation |
| NGF | TRPV1 | Activation |
| TNF | TYMP | Activation |
| CRP | VCAM1 | Activation |
| TNF | VCAM1 | Activation |
| IL6 | VEGFA | Activation |
| TNF | VEGFA | Activation |
| TNF | VIM | Activation |
| TNF | XDH | Activation |
| TNF | ABCC2 | Inhibition |
| IL6 | ADIPOQ | Inhibition |
| TNF | ADIPOQ | Inhibition |
| IL6 | AHSG | Inhibition |
| IL6 | AKT | Inhibition |
| PTPA | AKT | Inhibition |
| IL6 | ALB | Inhibition |
| IL6 | APOE | Inhibition |
| TNF | C3AR1 | Inhibition |
| IL6 | CASP9 | Inhibition |
| IFI16 | CCND1 | Inhibition |
| TGFB1 | CD14 | Inhibition |
| ZNF175 | CD14 | Inhibition |
| IL6 | CD33 | Inhibition |
| TNF | CP | Inhibition |
| IL4 | CRP | Inhibition |
| PPARA | CRP | Inhibition |
| TNF | CST3 | Inhibition |
| TNF | CTNNB1 | Inhibition |
| TNF | EGFR | Inhibition |
| TNF | ERBB2 | Inhibition |
| TNF | ERG | Inhibition |
| STAT1 | ESR1 | Inhibition |
| **TNF** | **ESR1** | **Inhibition** |
| IL6 | F12 | Inhibition |
| IL6 | FOXO1 | Inhibition |
| TNF | FSCN1 | Inhibition |
| IL6 | GADD45A | Inhibition |
| PTPA | GLI3 | Inhibition |
| STAT1 | IFNA1 | Inhibition |
| TNF | IGFBP3 | Inhibition |
| IL6 | IL1B | Inhibition |
| BMP6 | IL6 | Inhibition |
| ERK | IL6 | Inhibition |
| GH1 | IL6 | Inhibition |
| MEOX2 | IL6 | Inhibition |
| POMC | IL6 | Inhibition |
| PPARA | IL6 | Inhibition |
| PPARG | IL6 | Inhibition |
| SOCS3 | IL6 | Inhibition |
| SOD1 | IL6 | Inhibition |
| IL6 | INS | Inhibition |
| TNF | INSR | Inhibition |
| TNF | JUP | Inhibition |
| PTPA | MAPK1 | Inhibition |
| TNF | MAPK3 | Inhibition |
| IL6 | NCOA1 | Inhibition |
| STAT1 | NFkappaB | Inhibition |
| CRP | NFKBIA | Inhibition |
| TNF | NFKBIB | Inhibition |
| TNF | NOS2 | Inhibition |
| TNF | NOS3 | Inhibition |
| TNF | OCLN | Inhibition |
| PTPA | PDE4 | Inhibition |
| TNF | POU2F1 | Inhibition |
| TNF | PPM1B | Inhibition |
| TNF | PROCR | Inhibition |
| IL6 | PROS1 | Inhibition |
| TNF | RB1 | Inhibition |
| TNF | RETN | Inhibition |
| TNF | SCARB1 | Inhibition |
| TNF | SCD | Inhibition |
| TNF | SLC10A1 | Inhibition |
| TNF | SLC2A4 | Inhibition |
| TNF | SLCO1A2 | Inhibition |
| ADRB | STAT1 | Inhibition |
| PIAS1 | STAT1 | Inhibition |
| PTPN11 | STAT1 | Inhibition |
| SOCS3 | STAT1 | Inhibition |
| CRP | TERT | Inhibition |
| TNF | TFAP2C | Inhibition |
| TNF | THBD | Inhibition |
| ADIPOQ | TNF | Inhibition |
| ADRB | TNF | Inhibition |
| AMPK | TNF | Inhibition |
| EGF | TNF | Inhibition |
| GSK3B | TNF | Inhibition |
| IFNB1 | TNF | Inhibition |
| IL10 | TNF | Inhibition |
| IL10RA | TNF | Inhibition |
| IL13 | TNF | Inhibition |
| INS | TNF | Inhibition |
| P2RY6 | TNF | Inhibition |
| PEBP1 | TNF | Inhibition |
| PIAS3 | TNF | Inhibition |
| PNPLA3 | TNF | Inhibition |
| POU2F1 | TNF | Inhibition |
| PRKCH | TNF | Inhibition |
| PTEN | TNF | Inhibition |
| PTGS2 | TNF | Inhibition |
| RHOB | TNF | Inhibition |
| RNF216 | TNF | Inhibition |
| RORA | TNF | Inhibition |
| SALL4 | TNF | Inhibition |
| SERPINA1 | TNF | Inhibition |
| SERPINA3 | TNF | Inhibition |
| SFTPA1 | TNF | Inhibition |
| SH3BGRL3 | TNF | Inhibition |
| SOD2 | TNF | Inhibition |
| STAT1 | TNF | Inhibition |
| STAT3 | TNF | Inhibition |
| TNFAIP3 | TNF | Inhibition |
| TNFRSF1B | TNF | Inhibition |
| TXN | TNF | Inhibition |
| USP31 | TNF | Inhibition |
| VIPR2 | TNF | Inhibition |
| XIAP | TNF | Inhibition |
| ZFAND5 | TNF | Inhibition |
| ZFP36 | TNF | Inhibition |

| | |
|---|---|
| Number of Inhibition Statements | 109 |
| Number of Activation Statements | 408 |
| Total Number of Statements | 517 |

\*\* *Indicates statement reconciled across duplicates with majority rule*

**Table S2.** Connectivity properties or relationship statement density of individual sources and targets

| Name | Indegree | Outdegree | EdgeCount | Betweenness Centrality | Closeness Centrality | Clustering Coefficient | Eccentricity | Average Shortest Path Length | Neighborhood Connectivity |
|---|---|---|---|---|---|---|---|---|---|
| TNF | 96 | 177 | 273 | 56.46 | 0.65 | 0.00 | 3.00 | 1.55 | 1.57 |
| IL6 | 59 | 67 | 126 | 23.56 | 0.40 | 0.00 | 5.00 | 2.53 | 1.79 |
| STAT1 | 35 | 7 | 42 | 14.99 | 0.44 | 0.01 | 4.00 | 2.26 | 8.39 |
| CRP | 13 | 7 | 20 | 4.73 | 0.44 | 0.05 | 4.00 | 2.29 | 21.32 |
| MYD88 | 10 | 7 | 17 | 2.45 | 0.34 | 0.01 | 5.00 | 2.98 | 2.38 |
| CD14 | 8 | 3 | 11 | 1.19 | 0.44 | 0.03 | 4.00 | 2.30 | 27.40 |
| NFkappaB | 6 | 4 | 10 | 4.96 | 0.48 | 0.14 | 4.00 | 2.08 | 57.13 |
| IRF3 | 7 | 3 | 10 | 2.10 | 0.29 | 0.00 | 5.00 | 3.48 | 1.90 |
| IFI16 | 5 | 2 | 7 | 1.92 | 0.40 | 0.07 | 4.00 | 2.53 | 37.57 |
| IFNG | 2 | 4 | 6 | 3.04 | 0.37 | 0.15 | 5.00 | 2.71 | 85.60 |
| TLR9 | 1 | 4 | 5 | 1.08 | 0.34 | 0.10 | 5.00 | 2.96 | 6.00 |
| PTPA | 0 | 5 | 5 | 0.00 | 0.27 | 0.00 | 6.00 | 3.69 | 2.20 |
| AGT | 1 | 3 | 4 | 0.77 | 0.47 | 0.00 | 4.00 | 2.11 | 122.67 |
| IFNA1 | 2 | 2 | 4 | 0.14 | 0.31 | 0.00 | 5.00 | 3.25 | 19.33 |
| IL1B | 1 | 3 | 4 | 0.63 | 0.48 | 0.50 | 4.00 | 2.09 | 128.67 |
| EGFR | 2 | 2 | 4 | 0.55 | 0.36 | 0.33 | 5.00 | 2.77 | 135.00 |
| STAT3 | 2 | 2 | 4 | 0.90 | 0.40 | 0.17 | 4.00 | 2.53 | 124.67 |
| MAPK1 | 2 | 2 | 4 | 0.79 | 0.36 | 0.17 | 5.00 | 2.77 | 102.50 |
| MAPK3 | 2 | 2 | 4 | 0.68 | 0.40 | 0.33 | 4.00 | 2.53 | 135.00 |
| TLR2 | 2 | 2 | 4 | 0.60 | 0.40 | 0.33 | 4.00 | 2.49 | 92.33 |
| AP1 | 0 | 3 | 3 | 0.00 | 0.47 | 0.00 | 4.00 | 2.11 | 123.33 |
| CA2 | 3 | 0 | 3 | 0.00 | 0.00 | 0.00 | 0.00 | 0.00 | 2.00 |
| CCL2 | 2 | 1 | 3 | 0.81 | 0.28 | 0.50 | 6.00 | 3.52 | 128.67 |
| CD40LG | 1 | 2 | 3 | 0.50 | 0.47 | 0.00 | 4.00 | 2.11 | 183.50 |
| ERK | 1 | 2 | 3 | 0.28 | 0.36 | 0.17 | 5.00 | 2.80 | 46.67 |
| p38 | 0 | 3 | 3 | 0.00 | 0.36 | 0.17 | 5.00 | 2.79 | 45.33 |
| PTGS2 | 1 | 2 | 3 | 0.00 | 0.40 | 1.00 | 4.00 | 2.53 | 135.00 |
| CXCL12 | 1 | 2 | 3 | 1.27 | 0.29 | 0.00 | 6.00 | 3.46 | 125.67 |
| ESR1 | 3 | 0 | 3 | 0.00 | 0.00 | 0.33 | 0.00 | 0.00 | 135.00 |
| IL17A | 1 | 2 | 3 | 0.63 | 0.47 | 0.00 | 4.00 | 2.11 | 183.50 |
| IL18 | 1 | 2 | 3 | 0.50 | 0.47 | 0.00 | 4.00 | 2.11 | 183.50 |
| IL6ST | 1 | 2 | 3 | 0.05 | 0.36 | 0.00 | 5.00 | 2.77 | 77.00 |
| INS | 1 | 2 | 3 | 0.63 | 0.47 | 0.00 | 4.00 | 2.11 | 183.50 |
| NGF | 0 | 3 | 3 | 0.00 | 0.47 | 0.00 | 4.00 | 2.11 | 122.67 |
| TLR4 | 0 | 3 | 3 | 0.00 | 0.48 | 0.00 | 4.00 | 2.09 | 125.67 |
| TLR3 | 1 | 2 | 3 | 0.00 | 0.25 | 0.00 | 6.00 | 4.00 | 6.00 |
| IRF5 | 1 | 2 | 3 | 0.01 | 0.31 | 0.00 | 5.00 | 3.21 | 27.00 |
| TBK1 | 1 | 2 | 3 | 0.16 | 0.33 | 0.00 | 5.00 | 3.04 | 5.00 |
| POU2F1 | 2 | 1 | 3 | 0.63 | 0.39 | 0.00 | 4.00 | 2.55 | 183.50 |
| ADIPOQ | 2 | 1 | 3 | 0.63 | 0.39 | 0.00 | 4.00 | 2.55 | 183.50 |
| ABCB1 | 2 | 0 | 2 | 0.00 | 0.00 | 0.00 | 0.00 | 0.00 | 183.50 |
| PKC | 0 | 2 | 2 | 0.00 | 0.29 | 0.00 | 6.00 | 3.51 | 59.50 |
| CASP8 | 2 | 0 | 2 | 0.00 | 0.00 | 0.00 | 0.00 | 0.00 | 133.50 |
| PI3K | 1 | 1 | 2 | 0.06 | 0.30 | 0.00 | 5.00 | 3.28 | 63.00 |
| PRKAC | 1 | 1 | 2 | 0.02 | 0.30 | 0.00 | 5.00 | 3.29 | 6.00 |
| CD1A | 2 | 0 | 2 | 0.00 | 0.00 | 0.00 | 0.00 | 0.00 | 183.50 |
| CD40 | 1 | 1 | 2 | 0.50 | 0.28 | 0.00 | 6.00 | 3.51 | 183.50 |
| CDKN1A | 2 | 0 | 2 | 0.00 | 0.00 | 1.00 | 0.00 | 0.00 | 144.50 |
| CEBPB | 2 | 0 | 2 | 0.00 | 0.00 | 0.00 | 0.00 | 0.00 | 183.50 |
| CHUK | 1 | 1 | 2 | 0.00 | 0.39 | 0.00 | 4.00 | 2.55 | 251.00 |
| CXCL8 | 1 | 1 | 2 | 0.01 | 0.31 | 0.00 | 5.00 | 3.28 | 12.00 |
| EP300 | 1 | 1 | 2 | 1.01 | 0.39 | 0.00 | 4.00 | 2.54 | 130.50 |
| CAMP | 0 | 2 | 2 | 0.00 | 0.28 | 0.00 | 5.00 | 3.52 | 2.00 |
| GSK3B | 1 | 1 | 2 | 0.52 | 0.39 | 0.00 | 4.00 | 2.54 | 126.50 |
| HCK | 1 | 1 | 2 | 0.00 | 0.28 | 0.00 | 6.00 | 3.52 | 116.00 |
| HSF1 | 2 | 0 | 2 | 0.00 | 0.00 | 0.00 | 0.00 | 0.00 | 183.50 |
| IFNB1 | 1 | 1 | 2 | 1.01 | 0.39 | 0.00 | 4.00 | 2.54 | 130.50 |
| IKBKB | 1 | 1 | 2 | 0.00 | 0.39 | 0.00 | 4.00 | 2.55 | 251.00 |
| IKBKG | 1 | 1 | 2 | 0.00 | 0.39 | 0.00 | 4.00 | 2.55 | 251.00 |
| ADRB2 | 0 | 2 | 2 | 0.00 | 0.47 | 0.00 | 4.00 | 2.11 | 183.50 |
| BSG | 0 | 2 | 2 | 0.00 | 0.47 | 0.00 | 4.00 | 2.11 | 183.50 |
| ERBB2 | 1 | 1 | 2 | 0.50 | 0.28 | 0.00 | 6.00 | 3.51 | 183.50 |
| IL13 | 0 | 2 | 2 | 0.00 | 0.47 | 0.00 | 4.00 | 2.11 | 183.50 |
| IL4 | 0 | 2 | 2 | 0.00 | 0.36 | 0.50 | 5.00 | 2.80 | 67.50 |
| OSM | 0 | 2 | 2 | 0.00 | 0.36 | 0.00 | 5.00 | 2.77 | 77.00 |
| PARP1 | 0 | 2 | 2 | 0.00 | 0.47 | 0.00 | 4.00 | 2.11 | 183.50 |
| PTK2 | 0 | 2 | 2 | 0.00 | 0.47 | 0.00 | 4.00 | 2.11 | 183.50 |
| PTPN11 | 0 | 2 | 2 | 0.00 | 0.36 | 0.00 | 5.00 | 2.77 | 77.00 |
| TGFB1 | 0 | 2 | 2 | 0.00 | 0.36 | 0.00 | 5.00 | 2.79 | 63.00 |
| TNFRSF1A | 1 | 1 | 2 | 0.50 | 0.28 | 0.00 | 6.00 | 3.51 | 183.50 |
| VEGF | 0 | 2 | 2 | 0.00 | 0.47 | 0.00 | 4.00 | 2.11 | 183.50 |
| XDH | 1 | 1 | 2 | 0.50 | 0.28 | 0.00 | 6.00 | 3.51 | 183.50 |
| IRAK1 | 1 | 1 | 2 | 0.01 | 0.31 | 0.00 | 5.00 | 3.25 | 27.00 |
| IRF7 | 1 | 1 | 2 | 0.00 | 0.25 | 0.50 | 6.00 | 3.96 | 10.50 |
| MAP3K1 | 1 | 1 | 2 | 0.00 | 0.39 | 0.00 | 4.00 | 2.55 | 251.00 |
| MAPK14 | 1 | 1 | 2 | 0.63 | 0.39 | 0.00 | 4.00 | 2.54 | 183.50 |
| TLR1 | 0 | 2 | 2 | 0.00 | 0.40 | 0.00 | 4.00 | 2.49 | 133.50 |
| POMC | 1 | 1 | 2 | 0.00 | 0.28 | 0.00 | 6.00 | 3.52 | 116.00 |
| PTEN | 1 | 1 | 2 | 0.00 | 0.39 | 0.00 | 4.00 | 2.55 | 251.00 |
| RIPK2 | 1 | 1 | 2 | 0.00 | 0.39 | 0.00 | 4.00 | 2.55 | 251.00 |
| RORA | 1 | 1 | 2 | 0.63 | 0.39 | 0.00 | 4.00 | 2.54 | 183.50 |
| SERPINA3 | 1 | 1 | 2 | 0.63 | 0.39 | 0.00 | 4.00 | 2.54 | 183.50 |
| IL10 | 0 | 2 | 2 | 0.00 | 0.40 | 1.00 | 4.00 | 2.53 | 144.50 |
| TGFA | 1 | 1 | 2 | 0.81 | 0.25 | 0.00 | 6.00 | 3.95 | 128.00 |
| TIMP3 | 1 | 1 | 2 | 0.00 | 0.39 | 0.00 | 4.00 | 2.55 | 251.00 |
| NFKBIA | 1 | 1 | 2 | 0.00 | 0.39 | 1.00 | 4.00 | 2.54 | 135.00 |
| NOS3 | 1 | 1 | 2 | 0.00 | 0.39 | 0.00 | 4.00 | 2.55 | 251.00 |
| TRAF2 | 1 | 1 | 2 | 0.00 | 0.39 | 0.00 | 4.00 | 2.55 | 251.00 |
| TNFRSF1B | 1 | 1 | 2 | 0.00 | 0.39 | 0.00 | 4.00 | 2.55 | 251.00 |
| VCAM1 | 2 | 0 | 2 | 0.00 | 0.00 | 1.00 | 0.00 | 0.00 | 135.00 |
| VEGFA | 2 | 0 | 2 | 0.00 | 0.00 | 0.00 | 0.00 | 0.00 | 183.50 |
| AKT | 2 | 0 | 2 | 0.00 | 0.00 | 0.00 | 0.00 | 0.00 | 60.50 |
| PPARA | 0 | 2 | 2 | 0.00 | 0.36 | 0.50 | 5.00 | 2.80 | 67.50 |
| SOCS3 | 0 | 2 | 2 | 0.00 | 0.36 | 0.00 | 5.00 | 2.77 | 77.00 |
| ADRB | 0 | 2 | 2 | 0.00 | 0.40 | 1.00 | 4.00 | 2.53 | 144.50 |
| A2M | 1 | 0 | 1 | 0.00 | 0.00 | 0.00 | 0.00 | 0.00 | 116.00 |
| ABCC1 | 1 | 0 | 1 | 0.00 | 0.00 | 0.00 | 0.00 | 0.00 | 116.00 |
| ABCG2 | 1 | 0 | 1 | 0.00 | 0.00 | 0.00 | 0.00 | 0.00 | 116.00 |
| ADAMTS5 | 1 | 0 | 1 | 0.00 | 0.00 | 0.00 | 0.00 | 0.00 | 116.00 |
| AIM2 | 1 | 0 | 1 | 0.00 | 0.00 | 0.00 | 0.00 | 0.00 | 3.00 |
| AKT1 | 1 | 0 | 1 | 0.00 | 0.00 | 0.00 | 0.00 | 0.00 | 116.00 |
| AKT2 | 1 | 0 | 1 | 0.00 | 0.00 | 0.00 | 0.00 | 0.00 | 251.00 |
| ALPL | 1 | 0 | 1 | 0.00 | 0.00 | 0.00 | 0.00 | 0.00 | 251.00 |
| ANXA1 | 1 | 0 | 1 | 0.00 | 0.00 | 0.00 | 0.00 | 0.00 | 116.00 |
| AREG | 1 | 0 | 1 | 0.00 | 0.00 | 0.00 | 0.00 | 0.00 | 251.00 |
| ATF2 | 1 | 0 | 1 | 0.00 | 0.00 | 0.00 | 0.00 | 0.00 | 251.00 |
| ATF3 | 1 | 0 | 1 | 0.00 | 0.00 | 0.00 | 0.00 | 0.00 | 251.00 |
| BMP2 | 1 | 0 | 1 | 0.00 | 0.00 | 0.00 | 0.00 | 0.00 | 251.00 |
| C1R | 1 | 0 | 1 | 0.00 | 0.00 | 0.00 | 0.00 | 0.00 | 251.00 |
| C3 | 1 | 0 | 1 | 0.00 | 0.00 | 0.00 | 0.00 | 0.00 | 251.00 |
| C5AR1 | 1 | 0 | 1 | 0.00 | 0.00 | 0.00 | 0.00 | 0.00 | 251.00 |
| C8A | 1 | 0 | 1 | 0.00 | 0.00 | 0.00 | 0.00 | 0.00 | 116.00 |
| C8G | 1 | 0 | 1 | 0.00 | 0.00 | 0.00 | 0.00 | 0.00 | 116.00 |
| NKX2-1 | 0 | 1 | 1 | 0.00 | 1.00 | 0.00 | 1.00 | 1.00 | 3.00 |
| CASP1 | 1 | 0 | 1 | 0.00 | 0.00 | 0.00 | 0.00 | 0.00 | 251.00 |
| CASP2 | 1 | 0 | 1 | 0.00 | 0.00 | 0.00 | 0.00 | 0.00 | 251.00 |
| CASP6 | 1 | 0 | 1 | 0.00 | 0.00 | 0.00 | 0.00 | 0.00 | 251.00 |
| CASP7 | 1 | 0 | 1 | 0.00 | 0.00 | 0.00 | 0.00 | 0.00 | 251.00 |
| CBL | 1 | 0 | 1 | 0.00 | 0.00 | 0.00 | 0.00 | 0.00 | 116.00 |
| CCL11 | 1 | 0 | 1 | 0.00 | 0.00 | 0.00 | 0.00 | 0.00 | 251.00 |
| CCL3 | 1 | 0 | 1 | 0.00 | 0.00 | 0.00 | 0.00 | 0.00 | 251.00 |
| CCR2 | 1 | 0 | 1 | 0.00 | 0.00 | 0.00 | 0.00 | 0.00 | 19.00 |
| CEBP | 0 | 1 | 1 | 0.00 | 0.30 | 0.00 | 5.00 | 3.29 | 10.00 |
| KITLG | 0 | 1 | 1 | 0.00 | 0.30 | 0.00 | 5.00 | 3.29 | 10.00 |
| CD163 | 1 | 0 | 1 | 0.00 | 0.00 | 0.00 | 0.00 | 0.00 | 116.00 |
| CD38 | 1 | 0 | 1 | 0.00 | 0.00 | 0.00 | 0.00 | 0.00 | 251.00 |
| CD55 | 1 | 0 | 1 | 0.00 | 0.00 | 0.00 | 0.00 | 0.00 | 251.00 |
| CD59 | 1 | 0 | 1 | 0.00 | 0.00 | 0.00 | 0.00 | 0.00 | 251.00 |
| CD80 | 1 | 0 | 1 | 0.00 | 0.00 | 0.00 | 0.00 | 0.00 | 251.00 |
| CD86 | 1 | 0 | 1 | 0.00 | 0.00 | 0.00 | 0.00 | 0.00 | 251.00 |
| CDKN1B | 1 | 0 | 1 | 0.00 | 0.00 | 0.00 | 0.00 | 0.00 | 116.00 |
| CFB | 1 | 0 | 1 | 0.00 | 0.00 | 0.00 | 0.00 | 0.00 | 251.00 |
| CFL2 | 1 | 0 | 1 | 0.00 | 0.00 | 0.00 | 0.00 | 0.00 | 251.00 |
| CK2 | 1 | 0 | 1 | 0.00 | 0.00 | 0.00 | 0.00 | 0.00 | 251.00 |
| CLEC3B | 1 | 0 | 1 | 0.00 | 0.00 | 0.00 | 0.00 | 0.00 | 251.00 |
| CLIC4 | 1 | 0 | 1 | 0.00 | 0.00 | 0.00 | 0.00 | 0.00 | 251.00 |
| CMA1 | 1 | 0 | 1 | 0.00 | 0.00 | 0.00 | 0.00 | 0.00 | 116.00 |
| CD32 | 0 | 1 | 1 | 0.00 | 0.30 | 0.00 | 5.00 | 3.28 | 19.00 |
| HNF1A | 0 | 1 | 1 | 0.00 | 0.30 | 0.00 | 5.00 | 3.28 | 19.00 |
| LEP | 0 | 1 | 1 | 0.00 | 0.30 | 0.00 | 5.00 | 3.28 | 19.00 |
| CSF1 | 1 | 0 | 1 | 0.00 | 0.00 | 0.00 | 0.00 | 0.00 | 251.00 |
| CSF2 | 1 | 0 | 1 | 0.00 | 0.00 | 0.00 | 0.00 | 0.00 | 251.00 |
| CSF3 | 1 | 0 | 1 | 0.00 | 0.00 | 0.00 | 0.00 | 0.00 | 251.00 |
| CTSB | 1 | 0 | 1 | 0.00 | 0.00 | 0.00 | 0.00 | 0.00 | 251.00 |
| CTSL | 1 | 0 | 1 | 0.00 | 0.00 | 0.00 | 0.00 | 0.00 | 116.00 |
| CTSS | 1 | 0 | 1 | 0.00 | 0.00 | 0.00 | 0.00 | 0.00 | 116.00 |
| CX3CL1 | 1 | 0 | 1 | 0.00 | 0.00 | 0.00 | 0.00 | 0.00 | 251.00 |
| CXCL10 | 1 | 0 | 1 | 0.00 | 0.00 | 0.00 | 0.00 | 0.00 | 251.00 |
| CXCL5 | 1 | 0 | 1 | 0.00 | 0.00 | 0.00 | 0.00 | 0.00 | 251.00 |
| CXCL9 | 1 | 0 | 1 | 0.00 | 0.00 | 0.00 | 0.00 | 0.00 | 251.00 |
| CXCR3 | 1 | 0 | 1 | 0.00 | 0.00 | 0.00 | 0.00 | 0.00 | 251.00 |
| DAPK1 | 1 | 0 | 1 | 0.00 | 0.00 | 0.00 | 0.00 | 0.00 | 251.00 |
| EHF | 1 | 0 | 1 | 0.00 | 0.00 | 0.00 | 0.00 | 0.00 | 251.00 |
| F2RL1 | 1 | 0 | 1 | 0.00 | 0.00 | 0.00 | 0.00 | 0.00 | 251.00 |
| F2RL3 | 1 | 0 | 1 | 0.00 | 0.00 | 0.00 | 0.00 | 0.00 | 251.00 |
| FAS | 1 | 0 | 1 | 0.00 | 0.00 | 0.00 | 0.00 | 0.00 | 251.00 |
| FCER2 | 1 | 0 | 1 | 0.00 | 0.00 | 0.00 | 0.00 | 0.00 | 251.00 |
| FCGR1A | 1 | 0 | 1 | 0.00 | 0.00 | 0.00 | 0.00 | 0.00 | 19.00 |
| FCGR3B | 1 | 0 | 1 | 0.00 | 0.00 | 0.00 | 0.00 | 0.00 | 251.00 |
| FN1 | 1 | 0 | 1 | 0.00 | 0.00 | 0.00 | 0.00 | 0.00 | 116.00 |
| FOXO4 | 1 | 0 | 1 | 0.00 | 0.00 | 0.00 | 0.00 | 0.00 | 251.00 |
| FUT4 | 1 | 0 | 1 | 0.00 | 0.00 | 0.00 | 0.00 | 0.00 | 251.00 |
| FYN | 1 | 0 | 1 | 0.00 | 0.00 | 0.00 | 0.00 | 0.00 | 116.00 |
| GBF1 | 1 | 0 | 1 | 0.00 | 0.00 | 0.00 | 0.00 | 0.00 | 116.00 |
| HBEGF | 1 | 0 | 1 | 0.00 | 0.00 | 0.00 | 0.00 | 0.00 | 116.00 |
| HIF1A | 1 | 0 | 1 | 0.00 | 0.00 | 0.00 | 0.00 | 0.00 | 251.00 |
| HLA-A | 1 | 0 | 1 | 0.00 | 0.00 | 0.00 | 0.00 | 0.00 | 251.00 |
| HLA-B | 1 | 0 | 1 | 0.00 | 0.00 | 0.00 | 0.00 | 0.00 | 251.00 |
| HLA-C | 1 | 0 | 1 | 0.00 | 0.00 | 0.00 | 0.00 | 0.00 | 251.00 |
| HLA-F | 1 | 0 | 1 | 0.00 | 0.00 | 0.00 | 0.00 | 0.00 | 251.00 |
| HP | 1 | 0 | 1 | 0.00 | 0.00 | 0.00 | 0.00 | 0.00 | 116.00 |
| HSD11B1 | 1 | 0 | 1 | 0.00 | 0.00 | 0.00 | 0.00 | 0.00 | 251.00 |
| HSP90B1 | 1 | 0 | 1 | 0.00 | 0.00 | 0.00 | 0.00 | 0.00 | 116.00 |
| HOXD3 | 0 | 1 | 1 | 0.00 | 0.28 | 0.00 | 5.00 | 3.52 | 7.00 |
| LIF | 0 | 1 | 1 | 0.00 | 0.28 | 0.00 | 5.00 | 3.52 | 7.00 |
| RAF1 | 0 | 1 | 1 | 0.00 | 0.28 | 0.00 | 5.00 | 3.52 | 7.00 |
| IKK_family | 1 | 0 | 1 | 0.00 | 0.00 | 0.00 | 0.00 | 0.00 | 16.00 |
| IL11 | 1 | 0 | 1 | 0.00 | 0.00 | 0.00 | 0.00 | 0.00 | 251.00 |
| IL12A | 1 | 0 | 1 | 0.00 | 0.00 | 0.00 | 0.00 | 0.00 | 251.00 |
| IL12B | 1 | 0 | 1 | 0.00 | 0.00 | 0.00 | 0.00 | 0.00 | 251.00 |
| IL4R | 1 | 0 | 1 | 0.00 | 0.00 | 0.00 | 0.00 | 0.00 | 251.00 |
| AHR | 0 | 1 | 1 | 0.00 | 0.28 | 0.00 | 6.00 | 3.52 | 116.00 |
| AR | 0 | 1 | 1 | 0.00 | 0.28 | 0.00 | 6.00 | 3.52 | 116.00 |
| F10 | 0 | 1 | 1 | 0.00 | 0.28 | 0.00 | 6.00 | 3.52 | 116.00 |
| F2 | 0 | 1 | 1 | 0.00 | 0.28 | 0.00 | 6.00 | 3.52 | 116.00 |
| GAB1 | 0 | 1 | 1 | 0.00 | 0.28 | 0.00 | 6.00 | 3.52 | 116.00 |
| IL17F | 0 | 1 | 1 | 0.00 | 0.28 | 0.00 | 6.00 | 3.52 | 116.00 |
| IL1R1 | 0 | 1 | 1 | 0.00 | 0.28 | 0.00 | 6.00 | 3.52 | 116.00 |
| IL6R | 0 | 1 | 1 | 0.00 | 0.28 | 0.00 | 6.00 | 3.52 | 116.00 |
| ITGA5 | 0 | 1 | 1 | 0.00 | 0.28 | 0.00 | 6.00 | 3.52 | 116.00 |
| KLF7 | 0 | 1 | 1 | 0.00 | 0.28 | 0.00 | 6.00 | 3.52 | 116.00 |
| LEPR | 0 | 1 | 1 | 0.00 | 0.28 | 0.00 | 6.00 | 3.52 | 116.00 |
| LMX1B | 0 | 1 | 1 | 0.00 | 0.28 | 0.00 | 6.00 | 3.52 | 116.00 |
| PIK3C3 | 0 | 1 | 1 | 0.00 | 0.28 | 0.00 | 6.00 | 3.52 | 116.00 |
| PNPT1 | 0 | 1 | 1 | 0.00 | 0.28 | 0.00 | 6.00 | 3.52 | 116.00 |
| PTGER4 | 0 | 1 | 1 | 0.00 | 0.28 | 0.00 | 6.00 | 3.52 | 116.00 |
| TIRAP | 0 | 1 | 1 | 0.00 | 0.28 | 0.00 | 6.00 | 3.52 | 116.00 |
| IRAK4 | 1 | 0 | 1 | 0.00 | 0.00 | 0.00 | 0.00 | 0.00 | 16.00 |
| BEGAIN | 0 | 1 | 1 | 0.00 | 0.22 | 0.00 | 6.00 | 4.47 | 10.00 |
| IKBKE | 0 | 1 | 1 | 0.00 | 0.22 | 0.00 | 6.00 | 4.47 | 10.00 |
| MAVS | 0 | 1 | 1 | 0.00 | 0.22 | 0.00 | 6.00 | 4.47 | 10.00 |
| TICAM1 | 0 | 1 | 1 | 0.00 | 0.22 | 0.00 | 6.00 | 4.47 | 10.00 |
| ITGA6 | 1 | 0 | 1 | 0.00 | 0.00 | 0.00 | 0.00 | 0.00 | 251.00 |
| ITGAV | 1 | 0 | 1 | 0.00 | 0.00 | 0.00 | 0.00 | 0.00 | 251.00 |
| JAK | 1 | 0 | 1 | 0.00 | 0.00 | 0.00 | 0.00 | 0.00 | 251.00 |
| JUN | 1 | 0 | 1 | 0.00 | 0.00 | 0.00 | 0.00 | 0.00 | 251.00 |
| KRAS | 1 | 0 | 1 | 0.00 | 0.00 | 0.00 | 0.00 | 0.00 | 116.00 |
| LYN | 1 | 0 | 1 | 0.00 | 0.00 | 0.00 | 0.00 | 0.00 | 116.00 |
| LYZ | 1 | 0 | 1 | 0.00 | 0.00 | 0.00 | 0.00 | 0.00 | 251.00 |
| MAP2K1 | 1 | 0 | 1 | 0.00 | 0.00 | 0.00 | 0.00 | 0.00 | 251.00 |
| MAP2K2 | 1 | 0 | 1 | 0.00 | 0.00 | 0.00 | 0.00 | 0.00 | 251.00 |
| MAP2K6 | 1 | 0 | 1 | 0.00 | 0.00 | 0.00 | 0.00 | 0.00 | 251.00 |
| MAP3K14 | 1 | 0 | 1 | 0.00 | 0.00 | 0.00 | 0.00 | 0.00 | 251.00 |
| MAP3K5 | 1 | 0 | 1 | 0.00 | 0.00 | 0.00 | 0.00 | 0.00 | 251.00 |
| MAP4K3 | 1 | 0 | 1 | 0.00 | 0.00 | 0.00 | 0.00 | 0.00 | 251.00 |
| MAP4K4 | 1 | 0 | 1 | 0.00 | 0.00 | 0.00 | 0.00 | 0.00 | 251.00 |
| MAPK7 | 1 | 0 | 1 | 0.00 | 0.00 | 0.00 | 0.00 | 0.00 | 116.00 |
| MAPK9 | 1 | 0 | 1 | 0.00 | 0.00 | 0.00 | 0.00 | 0.00 | 251.00 |
| MAPKAPK3 | 1 | 0 | 1 | 0.00 | 0.00 | 0.00 | 0.00 | 0.00 | 251.00 |
| MAZ | 1 | 0 | 1 | 0.00 | 0.00 | 0.00 | 0.00 | 0.00 | 116.00 |
| MEF2 | 1 | 0 | 1 | 0.00 | 0.00 | 0.00 | 0.00 | 0.00 | 116.00 |
| MGAT4B | 1 | 0 | 1 | 0.00 | 0.00 | 0.00 | 0.00 | 0.00 | 116.00 |
| MGAT5 | 1 | 0 | 1 | 0.00 | 0.00 | 0.00 | 0.00 | 0.00 | 116.00 |
| MIF | 1 | 0 | 1 | 0.00 | 0.00 | 0.00 | 0.00 | 0.00 | 251.00 |
| MMP1 | 1 | 0 | 1 | 0.00 | 0.00 | 0.00 | 0.00 | 0.00 | 251.00 |
| MMP14 | 1 | 0 | 1 | 0.00 | 0.00 | 0.00 | 0.00 | 0.00 | 251.00 |
| MMP3 | 1 | 0 | 1 | 0.00 | 0.00 | 0.00 | 0.00 | 0.00 | 251.00 |
| MMP9 | 1 | 0 | 1 | 0.00 | 0.00 | 0.00 | 0.00 | 0.00 | 116.00 |
| MUC5AC | 1 | 0 | 1 | 0.00 | 0.00 | 0.00 | 0.00 | 0.00 | 251.00 |
| TLR5 | 0 | 1 | 1 | 0.00 | 0.25 | 0.00 | 6.00 | 3.97 | 16.00 |
| TLR6 | 0 | 1 | 1 | 0.00 | 0.25 | 0.00 | 6.00 | 3.97 | 16.00 |
| TLR7 | 0 | 1 | 1 | 0.00 | 0.25 | 0.00 | 6.00 | 3.97 | 16.00 |
| TLR8 | 0 | 1 | 1 | 0.00 | 0.25 | 0.00 | 6.00 | 3.97 | 16.00 |
| MYLK | 1 | 0 | 1 | 0.00 | 0.00 | 0.00 | 0.00 | 0.00 | 251.00 |
| NCAM1 | 1 | 0 | 1 | 0.00 | 0.00 | 0.00 | 0.00 | 0.00 | 251.00 |
| NFKB1 | 1 | 0 | 1 | 0.00 | 0.00 | 0.00 | 0.00 | 0.00 | 251.00 |
| NFKB2 | 1 | 0 | 1 | 0.00 | 0.00 | 0.00 | 0.00 | 0.00 | 251.00 |
| NOS1 | 1 | 0 | 1 | 0.00 | 0.00 | 0.00 | 0.00 | 0.00 | 251.00 |
| ORM1 | 1 | 0 | 1 | 0.00 | 0.00 | 0.00 | 0.00 | 0.00 | 116.00 |
| PAWR | 1 | 0 | 1 | 0.00 | 0.00 | 0.00 | 0.00 | 0.00 | 251.00 |
| PFN1 | 1 | 0 | 1 | 0.00 | 0.00 | 0.00 | 0.00 | 0.00 | 251.00 |
| PHB1 | 1 | 0 | 1 | 0.00 | 0.00 | 0.00 | 0.00 | 0.00 | 116.00 |
| PI3 | 1 | 0 | 1 | 0.00 | 0.00 | 0.00 | 0.00 | 0.00 | 3.00 |
| PLA2G2A | 1 | 0 | 1 | 0.00 | 0.00 | 0.00 | 0.00 | 0.00 | 251.00 |
| PLAU | 1 | 0 | 1 | 0.00 | 0.00 | 0.00 | 0.00 | 0.00 | 251.00 |
| PPP1R15A | 1 | 0 | 1 | 0.00 | 0.00 | 0.00 | 0.00 | 0.00 | 251.00 |
| PRKCI | 1 | 0 | 1 | 0.00 | 0.00 | 0.00 | 0.00 | 0.00 | 251.00 |
| PRKCZ | 1 | 0 | 1 | 0.00 | 0.00 | 0.00 | 0.00 | 0.00 | 251.00 |
| PRKD1 | 1 | 0 | 1 | 0.00 | 0.00 | 0.00 | 0.00 | 0.00 | 116.00 |
| Proteasome | 1 | 0 | 1 | 0.00 | 0.00 | 0.00 | 0.00 | 0.00 | 251.00 |
| PSMB9 | 1 | 0 | 1 | 0.00 | 0.00 | 0.00 | 0.00 | 0.00 | 251.00 |
| PSME1 | 1 | 0 | 1 | 0.00 | 0.00 | 0.00 | 0.00 | 0.00 | 251.00 |
| PTGES | 1 | 0 | 1 | 0.00 | 0.00 | 0.00 | 0.00 | 0.00 | 251.00 |
| PTHLH | 1 | 0 | 1 | 0.00 | 0.00 | 0.00 | 0.00 | 0.00 | 251.00 |
| PTX3 | 1 | 0 | 1 | 0.00 | 0.00 | 0.00 | 0.00 | 0.00 | 251.00 |
| RELA | 1 | 0 | 1 | 0.00 | 0.00 | 0.00 | 0.00 | 0.00 | 251.00 |
| RIPK1 | 1 | 0 | 1 | 0.00 | 0.00 | 0.00 | 0.00 | 0.00 | 251.00 |
| RIPK3 | 1 | 0 | 1 | 0.00 | 0.00 | 0.00 | 0.00 | 0.00 | 251.00 |
| SELE | 1 | 0 | 1 | 0.00 | 0.00 | 0.00 | 0.00 | 0.00 | 251.00 |
| SERPINE1 | 1 | 0 | 1 | 0.00 | 0.00 | 0.00 | 0.00 | 0.00 | 251.00 |
| SERPING1 | 1 | 0 | 1 | 0.00 | 0.00 | 0.00 | 0.00 | 0.00 | 251.00 |
| SERPINH1 | 1 | 0 | 1 | 0.00 | 0.00 | 0.00 | 0.00 | 0.00 | 251.00 |
| SLC6A4 | 1 | 0 | 1 | 0.00 | 0.00 | 0.00 | 0.00 | 0.00 | 251.00 |
| SLC6A6 | 1 | 0 | 1 | 0.00 | 0.00 | 0.00 | 0.00 | 0.00 | 251.00 |
| SMPD1 | 1 | 0 | 1 | 0.00 | 0.00 | 0.00 | 0.00 | 0.00 | 251.00 |
| SOCS1 | 1 | 0 | 1 | 0.00 | 0.00 | 0.00 | 0.00 | 0.00 | 38.00 |
| SP1 | 1 | 0 | 1 | 0.00 | 0.00 | 0.00 | 0.00 | 0.00 | 251.00 |
| SPP1 | 1 | 0 | 1 | 0.00 | 0.00 | 0.00 | 0.00 | 0.00 | 251.00 |
| ST3GAL6 | 1 | 0 | 1 | 0.00 | 0.00 | 0.00 | 0.00 | 0.00 | 251.00 |
| BMX | 0 | 1 | 1 | 0.00 | 0.31 | 0.00 | 5.00 | 3.26 | 38.00 |
| ETS1 | 0 | 1 | 1 | 0.00 | 0.31 | 0.00 | 5.00 | 3.26 | 38.00 |
| FGFR1 | 0 | 1 | 1 | 0.00 | 0.31 | 0.00 | 5.00 | 3.26 | 38.00 |
| FGFR3 | 0 | 1 | 1 | 0.00 | 0.31 | 0.00 | 5.00 | 3.26 | 38.00 |
| HDAC | 0 | 1 | 1 | 0.00 | 0.31 | 0.00 | 5.00 | 3.26 | 38.00 |
| HDAC1 | 0 | 1 | 1 | 0.00 | 0.31 | 0.00 | 5.00 | 3.26 | 38.00 |
| HDAC2 | 0 | 1 | 1 | 0.00 | 0.31 | 0.00 | 5.00 | 3.26 | 38.00 |
| HDAC3 | 0 | 1 | 1 | 0.00 | 0.31 | 0.00 | 5.00 | 3.26 | 38.00 |
| IFNGR1 | 0 | 1 | 1 | 0.00 | 0.31 | 0.00 | 5.00 | 3.26 | 38.00 |
| IFNL1 | 0 | 1 | 1 | 0.00 | 0.31 | 0.00 | 5.00 | 3.26 | 38.00 |
| IFNL2 | 0 | 1 | 1 | 0.00 | 0.31 | 0.00 | 5.00 | 3.26 | 38.00 |
| JAK2 | 0 | 1 | 1 | 0.00 | 0.31 | 0.00 | 5.00 | 3.26 | 38.00 |
| LPL | 0 | 1 | 1 | 0.00 | 0.31 | 0.00 | 5.00 | 3.26 | 38.00 |
| PDGFRB | 0 | 1 | 1 | 0.00 | 0.31 | 0.00 | 5.00 | 3.26 | 38.00 |
| PRKCD | 0 | 1 | 1 | 0.00 | 0.31 | 0.00 | 5.00 | 3.26 | 38.00 |
| PRKCE | 0 | 1 | 1 | 0.00 | 0.31 | 0.00 | 5.00 | 3.26 | 38.00 |
| PRMT1 | 0 | 1 | 1 | 0.00 | 0.31 | 0.00 | 5.00 | 3.26 | 38.00 |
| RET | 0 | 1 | 1 | 0.00 | 0.31 | 0.00 | 5.00 | 3.26 | 38.00 |
| TYK2 | 0 | 1 | 1 | 0.00 | 0.31 | 0.00 | 5.00 | 3.26 | 38.00 |
| STAT2 | 1 | 0 | 1 | 0.00 | 0.00 | 0.00 | 0.00 | 0.00 | 116.00 |
| TAP1 | 1 | 0 | 1 | 0.00 | 0.00 | 0.00 | 0.00 | 0.00 | 251.00 |
| TAP2 | 1 | 0 | 1 | 0.00 | 0.00 | 0.00 | 0.00 | 0.00 | 251.00 |
| THBS1 | 1 | 0 | 1 | 0.00 | 0.00 | 0.00 | 0.00 | 0.00 | 251.00 |
| TIMP1 | 1 | 0 | 1 | 0.00 | 0.00 | 0.00 | 0.00 | 0.00 | 251.00 |
| ADAM17 | 0 | 1 | 1 | 0.00 | 0.39 | 0.00 | 4.00 | 2.54 | 251.00 |
| AGTR1 | 0 | 1 | 1 | 0.00 | 0.39 | 0.00 | 4.00 | 2.54 | 251.00 |
| AKR1B1 | 0 | 1 | 1 | 0.00 | 0.39 | 0.00 | 4.00 | 2.54 | 251.00 |
| ALOX5 | 0 | 1 | 1 | 0.00 | 0.39 | 0.00 | 4.00 | 2.54 | 251.00 |
| Caspase | 0 | 1 | 1 | 0.00 | 0.39 | 0.00 | 4.00 | 2.54 | 251.00 |
| CENPJ | 0 | 1 | 1 | 0.00 | 0.39 | 0.00 | 4.00 | 2.54 | 251.00 |
| DCN | 0 | 1 | 1 | 0.00 | 0.39 | 0.00 | 4.00 | 2.54 | 251.00 |

| Gene | C1 | C2 | C3 | V1 | V2 | V3 | V4 | V5 | V6 |
|------|----|----|----|------|------|------|------|------|--------|
| DCTN1 | 0 | 1 | 1 | 0.00 | 0.39 | 0.00 | 4.00 | 2.54 | 251.00 |
| DHX9 | 0 | 1 | 1 | 0.00 | 0.39 | 0.00 | 4.00 | 2.54 | 251.00 |
| EGR4 | 0 | 1 | 1 | 0.00 | 0.39 | 0.00 | 4.00 | 2.54 | 251.00 |
| HMGB1 | 0 | 1 | 1 | 0.00 | 0.39 | 0.00 | 4.00 | 2.54 | 251.00 |
| HNRNPU | 0 | 1 | 1 | 0.00 | 0.39 | 0.00 | 4.00 | 2.54 | 251.00 |
| HSPA4 | 0 | 1 | 1 | 0.00 | 0.39 | 0.00 | 4.00 | 2.54 | 251.00 |
| HSPB1 | 0 | 1 | 1 | 0.00 | 0.39 | 0.00 | 4.00 | 2.54 | 251.00 |
| ICMT | 0 | 1 | 1 | 0.00 | 0.39 | 0.00 | 4.00 | 2.54 | 251.00 |
| IRF9 | 0 | 1 | 1 | 0.00 | 0.39 | 0.00 | 4.00 | 2.54 | 251.00 |
| MAP3K7 | 0 | 1 | 1 | 0.00 | 0.39 | 0.00 | 4.00 | 2.54 | 251.00 |
| MMP12 | 0 | 1 | 1 | 0.00 | 0.39 | 0.00 | 4.00 | 2.54 | 251.00 |
| NAMPT | 0 | 1 | 1 | 0.00 | 0.39 | 0.00 | 4.00 | 2.54 | 251.00 |
| NFAT | 0 | 1 | 1 | 0.00 | 0.39 | 0.00 | 4.00 | 2.54 | 251.00 |
| NOX4 | 0 | 1 | 1 | 0.00 | 0.39 | 0.00 | 4.00 | 2.54 | 251.00 |
| NRAS | 0 | 1 | 1 | 0.00 | 0.39 | 0.00 | 4.00 | 2.54 | 251.00 |
| OLR1 | 0 | 1 | 1 | 0.00 | 0.39 | 0.00 | 4.00 | 2.54 | 251.00 |
| PDE3B | 0 | 1 | 1 | 0.00 | 0.39 | 0.00 | 4.00 | 2.54 | 251.00 |
| PLA2G4A | 0 | 1 | 1 | 0.00 | 0.39 | 0.00 | 4.00 | 2.54 | 251.00 |
| RALBP1 | 0 | 1 | 1 | 0.00 | 0.39 | 0.00 | 4.00 | 2.54 | 251.00 |
| TARDBP | 0 | 1 | 1 | 0.00 | 0.39 | 0.00 | 4.00 | 2.54 | 251.00 |
| TONSL | 0 | 1 | 1 | 0.00 | 0.39 | 0.00 | 4.00 | 2.54 | 251.00 |
| TP53 | 1 | 0 | 1 | 0.00 | 0.00 | 0.00 | 0.00 | 0.00 | 251.00 |
| TRAF1 | 1 | 0 | 1 | 0.00 | 0.00 | 0.00 | 0.00 | 0.00 | 251.00 |
| TRAF3 | 1 | 0 | 1 | 0.00 | 0.00 | 0.00 | 0.00 | 0.00 | 251.00 |
| TRAF6 | 1 | 0 | 1 | 0.00 | 0.00 | 0.00 | 0.00 | 0.00 | 16.00 |
| TRPV1 | 1 | 0 | 1 | 0.00 | 0.00 | 0.00 | 0.00 | 0.00 | 3.00 |
| TYMP | 1 | 0 | 1 | 0.00 | 0.00 | 0.00 | 0.00 | 0.00 | 251.00 |
| VIM | 1 | 0 | 1 | 0.00 | 0.00 | 0.00 | 0.00 | 0.00 | 251.00 |
| ABCC2 | 1 | 0 | 1 | 0.00 | 0.00 | 0.00 | 0.00 | 0.00 | 251.00 |
| AHSG | 1 | 0 | 1 | 0.00 | 0.00 | 0.00 | 0.00 | 0.00 | 116.00 |
| ALB | 1 | 0 | 1 | 0.00 | 0.00 | 0.00 | 0.00 | 0.00 | 116.00 |
| APOE | 1 | 0 | 1 | 0.00 | 0.00 | 0.00 | 0.00 | 0.00 | 116.00 |
| C3AR1 | 1 | 0 | 1 | 0.00 | 0.00 | 0.00 | 0.00 | 0.00 | 251.00 |
| CASP9 | 1 | 0 | 1 | 0.00 | 0.00 | 0.00 | 0.00 | 0.00 | 116.00 |
| CCND1 | 1 | 0 | 1 | 0.00 | 0.00 | 0.00 | 0.00 | 0.00 | 7.00 |
| ZNF175 | 0 | 1 | 1 | 0.00 | 0.30 | 0.00 | 5.00 | 3.29 | 10.00 |
| CD33 | 1 | 0 | 1 | 0.00 | 0.00 | 0.00 | 0.00 | 0.00 | 116.00 |
| CP | 1 | 0 | 1 | 0.00 | 0.00 | 0.00 | 0.00 | 0.00 | 251.00 |
| CST3 | 1 | 0 | 1 | 0.00 | 0.00 | 0.00 | 0.00 | 0.00 | 251.00 |
| CTNNB1 | 1 | 0 | 1 | 0.00 | 0.00 | 0.00 | 0.00 | 0.00 | 251.00 |
| ERG | 1 | 0 | 1 | 0.00 | 0.00 | 0.00 | 0.00 | 0.00 | 251.00 |
| F12 | 1 | 0 | 1 | 0.00 | 0.00 | 0.00 | 0.00 | 0.00 | 116.00 |
| FOXO1 | 1 | 0 | 1 | 0.00 | 0.00 | 0.00 | 0.00 | 0.00 | 116.00 |
| FSCN1 | 1 | 0 | 1 | 0.00 | 0.00 | 0.00 | 0.00 | 0.00 | 251.00 |
| GADD45A | 1 | 0 | 1 | 0.00 | 0.00 | 0.00 | 0.00 | 0.00 | 116.00 |
| GLI3 | 1 | 0 | 1 | 0.00 | 0.00 | 0.00 | 0.00 | 0.00 | 5.00 |
| IGFBP3 | 1 | 0 | 1 | 0.00 | 0.00 | 0.00 | 0.00 | 0.00 | 251.00 |
| BMP6 | 0 | 1 | 1 | 0.00 | 0.28 | 0.00 | 6.00 | 3.52 | 116.00 |
| GH1 | 0 | 1 | 1 | 0.00 | 0.28 | 0.00 | 6.00 | 3.52 | 116.00 |
| MEOX2 | 0 | 1 | 1 | 0.00 | 0.28 | 0.00 | 6.00 | 3.52 | 116.00 |
| PPARG | 0 | 1 | 1 | 0.00 | 0.28 | 0.00 | 6.00 | 3.52 | 116.00 |
| SOD1 | 0 | 1 | 1 | 0.00 | 0.28 | 0.00 | 6.00 | 3.52 | 116.00 |
| INSR | 1 | 0 | 1 | 0.00 | 0.00 | 0.00 | 0.00 | 0.00 | 251.00 |
| JUP | 1 | 0 | 1 | 0.00 | 0.00 | 0.00 | 0.00 | 0.00 | 251.00 |
| NCOA1 | 1 | 0 | 1 | 0.00 | 0.00 | 0.00 | 0.00 | 0.00 | 116.00 |
| NFKBIB | 1 | 0 | 1 | 0.00 | 0.00 | 0.00 | 0.00 | 0.00 | 251.00 |
| NOS2 | 1 | 0 | 1 | 0.00 | 0.00 | 0.00 | 0.00 | 0.00 | 251.00 |
| OCLN | 1 | 0 | 1 | 0.00 | 0.00 | 0.00 | 0.00 | 0.00 | 251.00 |
| PDE4 | 1 | 0 | 1 | 0.00 | 0.00 | 0.00 | 0.00 | 0.00 | 5.00 |
| PPM1B | 1 | 0 | 1 | 0.00 | 0.00 | 0.00 | 0.00 | 0.00 | 251.00 |
| PROCR | 1 | 0 | 1 | 0.00 | 0.00 | 0.00 | 0.00 | 0.00 | 251.00 |
| PROS1 | 1 | 0 | 1 | 0.00 | 0.00 | 0.00 | 0.00 | 0.00 | 116.00 |
| RB1 | 1 | 0 | 1 | 0.00 | 0.00 | 0.00 | 0.00 | 0.00 | 251.00 |
| RETN | 1 | 0 | 1 | 0.00 | 0.00 | 0.00 | 0.00 | 0.00 | 251.00 |
| SCARB1 | 1 | 0 | 1 | 0.00 | 0.00 | 0.00 | 0.00 | 0.00 | 251.00 |
| SCD | 1 | 0 | 1 | 0.00 | 0.00 | 0.00 | 0.00 | 0.00 | 251.00 |
| SLC10A1 | 1 | 0 | 1 | 0.00 | 0.00 | 0.00 | 0.00 | 0.00 | 251.00 |
| SLC2A4 | 1 | 0 | 1 | 0.00 | 0.00 | 0.00 | 0.00 | 0.00 | 251.00 |
| SLCO1A2 | 1 | 0 | 1 | 0.00 | 0.00 | 0.00 | 0.00 | 0.00 | 251.00 |
| PIAS1 | 0 | 1 | 1 | 0.00 | 0.31 | 0.00 | 5.00 | 3.26 | 38.00 |
| TERT | 1 | 0 | 1 | 0.00 | 0.00 | 0.00 | 0.00 | 0.00 | 19.00 |
| TFAP2C | 1 | 0 | 1 | 0.00 | 0.00 | 0.00 | 0.00 | 0.00 | 251.00 |
| THBD | 1 | 0 | 1 | 0.00 | 0.00 | 0.00 | 0.00 | 0.00 | 251.00 |
| AMPK | 0 | 1 | 1 | 0.00 | 0.39 | 0.00 | 4.00 | 2.54 | 251.00 |
| EGF | 0 | 1 | 1 | 0.00 | 0.39 | 0.00 | 4.00 | 2.54 | 251.00 |
| IL10RA | 0 | 1 | 1 | 0.00 | 0.39 | 0.00 | 4.00 | 2.54 | 251.00 |
| P2RY6 | 0 | 1 | 1 | 0.00 | 0.39 | 0.00 | 4.00 | 2.54 | 251.00 |
| PEBP1 | 0 | 1 | 1 | 0.00 | 0.39 | 0.00 | 4.00 | 2.54 | 251.00 |
| PIAS3 | 0 | 1 | 1 | 0.00 | 0.39 | 0.00 | 4.00 | 2.54 | 251.00 |
| PNPLA3 | 0 | 1 | 1 | 0.00 | 0.39 | 0.00 | 4.00 | 2.54 | 251.00 |
| PRKCH | 0 | 1 | 1 | 0.00 | 0.39 | 0.00 | 4.00 | 2.54 | 251.00 |
| RHOB | 0 | 1 | 1 | 0.00 | 0.39 | 0.00 | 4.00 | 2.54 | 251.00 |
| RNF216 | 0 | 1 | 1 | 0.00 | 0.39 | 0.00 | 4.00 | 2.54 | 251.00 |
| SALL4 | 0 | 1 | 1 | 0.00 | 0.39 | 0.00 | 4.00 | 2.54 | 251.00 |
| SERPINA1 | 0 | 1 | 1 | 0.00 | 0.39 | 0.00 | 4.00 | 2.54 | 251.00 |
| SFTPA1 | 0 | 1 | 1 | 0.00 | 0.39 | 0.00 | 4.00 | 2.54 | 251.00 |
| SH3BGRL3 | 0 | 1 | 1 | 0.00 | 0.39 | 0.00 | 4.00 | 2.54 | 251.00 |
| SOD2 | 0 | 1 | 1 | 0.00 | 0.39 | 0.00 | 4.00 | 2.54 | 251.00 |
| TNFAIP3 | 0 | 1 | 1 | 0.00 | 0.39 | 0.00 | 4.00 | 2.54 | 251.00 |
| TXN | 0 | 1 | 1 | 0.00 | 0.39 | 0.00 | 4.00 | 2.54 | 251.00 |
| USP31 | 0 | 1 | 1 | 0.00 | 0.39 | 0.00 | 4.00 | 2.54 | 251.00 |
| VIPR2 | 0 | 1 | 1 | 0.00 | 0.39 | 0.00 | 4.00 | 2.54 | 251.00 |
| XIAP | 0 | 1 | 1 | 0.00 | 0.39 | 0.00 | 4.00 | 2.54 | 251.00 |
| ZFAND5 | 0 | 1 | 1 | 0.00 | 0.39 | 0.00 | 4.00 | 2.54 | 251.00 |
| ZFP36 | 0 | 1 | 1 | 0.00 | 0.39 | 0.00 | 4.00 | 2.54 | 251.00 |

**Table S3.** Replicate fine-tunign results for BERT, Llama 1B and Llama 3B parameter LLM

| BERT Model replicate trials | | | | | | | | Llama 3.2 1B Model replicate trials | | | | | | | |
|---|---|---|---|---|---|---|---|---|---|---|---|---|---|---|---|
| | Trial #1 | Trial #2 | Trial #3 | Trial #4 | Trial #5 | Average | Std Error | Trial #1 | Trial #2 | Trial #3 | Trial #4 | Trial #5 | Average | Std Error | p value wrt BERT |
| Total Positive Questions | 23 | 23 | 23 | 23 | 23 | 23.00 | 0.00 | 23 | 23 | 23 | 23 | 23 | 23.00 | 0.00 | |
| Total Negative Questions | 23 | 23 | 23 | 23 | 23 | 23.00 | 0.00 | 23 | 23 | 23 | 23 | 23 | 23.00 | 0.00 | |
| Total Correct Positive | 15 | 15 | 21 | 0 | 5 | 11.20 | 3.80 | 0 | 23 | 13 | 23 | 0 | 11.80 | 5.15 | |
| Total Correct Negative | 14 | 10 | 6 | 23 | 16 | 13.80 | 2.87 | 23 | 0 | 11 | 0 | 23 | 11.40 | 5.14 | |
| | | | | | | | | | | | | | | | |
| Overall Accuracy Positive (%) | 65.22 | 65.22 | 91.30 | 0.00 | 21.74 | 48.70 | 16.52 | 0.00 | 100.00 | 56.52 | 100.00 | 0.00 | 51.30 | 22.40 | 0.928 |
| Overall Accuracy Negative (% | 60.87 | 43.48 | 26.09 | 100.00 | 69.57 | 60.00 | 12.48 | 100.00 | 0.00 | 47.83 | 0.00 | 100.00 | 49.57 | 22.36 | 0.697 |
| Overall Accuracy (%) | 63.04 | 54.35 | 58.70 | 50.00 | 45.65 | 54.35 | 3.07 | 50.00 | 50.00 | 52.17 | 50.00 | 50.00 | 50.43 | 0.43 | 0.274 |
| | | | | | | | | | | | | | | | |
| Average Entropy Positive | 0.91 | 0.87 | 1.00 | 1.00 | 0.96 | 0.95 | 0.02 | 0.96 | 0.32 | 0.02 | 0.26 | 0.58 | 0.43 | 0.16 | 0.031 |
| Average Entropy Negative | 0.89 | 0.85 | 1.00 | 1.00 | 0.83 | 0.91 | 0.04 | 0.96 | 0.32 | 0.02 | 0.26 | 0.61 | 0.43 | 0.16 | 0.040 |

Table S4. Active learning recruitment of new examples

| Name | Indegree | Outdegree | EdgeCount | Trial #1 Correct high E + incorrect low E | | Trial #2 Correct high E + incorrect low E | | Trial #3 Correct high E + incorrect low E | | Trial #1 Incorrect low E only | | Trial #2 Incorrect low entropy | | Trial #3 Incorrect low entropy | |
|---|---|---|---|---|---|---|---|---|---|---|---|---|---|---|---|
| | | | | Frequency as Source | Frequency as target | Frequency as Source | Frequency as target | Frequency as Source | Frequency as target | Frequency as Source | Frequency as target | Frequency as Source | Frequency as target | Frequency as Source | Frequency as target |
| TNF | 96 | 177 | 273 | 41 | 20 | 44 | 21 | 32 | 33 | 45 | 23 | 36 | 26 | 34 | 24 |
| IL6 | 59 | 67 | 126 | 21 | 8 | 13 | 12 | 15 | 12 | 16 | 10 | 11 | 12 | 11 | 7 |
| STAT1 | 35 | 7 | 42 | 3 | 5 | 2 | 5 | 1 | 3 | 3 | 7 | 2 | 5 | 1 | 4 |
| CRP | 13 | 7 | 20 | 2 | 2 | 1 | 2 | 1 | 2 | 4 | 1 | 1 | 3 | 2 | 2 |
| MYD88 | 10 | 7 | 17 | 0 | 0 | 0 | 0 | 0 | 0 | 0 | 3 | 1 | 1 | 1 | 0 |
| CD14 | 8 | 3 | 11 | 0 | 3 | 0 | 4 | 0 | 4 | 0 | 3 | 1 | 2 | 1 | 1 |
| NFkappaB | 6 | 4 | 10 | 0 | 0 | 0 | 0 | 0 | 0 | 1 | 0 | 0 | 0 | 0 | 0 |
| IRF3 | 7 | 3 | 10 | 1 | 0 | 1 | 1 | 0 | 0 | 1 | 0 | 1 | 1 | 1 | 0 |
| IFI16 | 5 | 2 | 7 | 0 | 1 | 0 | 1 | 1 | 1 | 0 | 1 | 1 | 1 | 1 | 1 |
| IFNG | 2 | 4 | 6 | 1 | 0 | 2 | 0 | 0 | 0 | 1 | 0 | 1 | 1 | 0 | 0 |
| TLR9 | 4 | 1 | 5 | 0 | 0 | 0 | 0 | 0 | 0 | 1 | 0 | 1 | 1 | 1 | 1 |
| PTPA | 0 | 5 | 5 | 2 | 0 | 2 | 0 | 2 | 0 | 2 | 0 | 0 | 0 | 1 | 0 |
| AGT | 1 | 3 | 4 | 0 | 0 | 0 | 0 | 0 | 0 | 0 | 0 | 0 | 0 | 0 | 1 |
| IFNA1 | 2 | 2 | 4 | 1 | 0 | 1 | 1 | 1 | 0 | 1 | 1 | 0 | 1 | 0 | 2 |
| IL1B | 1 | 3 | 4 | 1 | 1 | 1 | 0 | 0 | 1 | 0 | 1 | 0 | 1 | 0 | 0 |
| EGFR | 2 | 2 | 4 | 0 | 1 | 0 | 1 | 0 | 0 | 0 | 1 | 0 | 0 | 1 | 0 |
| STAT3 | 2 | 2 | 4 | 1 | 0 | 1 | 0 | 0 | 1 | 2 | 0 | 1 | 0 | 2 | 0 |
| MAPK1 | 2 | 2 | 4 | 0 | 1 | 0 | 0 | 0 | 0 | 0 | 1 | 0 | 1 | 0 | 1 |
| MAPK3 | 2 | 2 | 4 | 0 | 1 | 0 | 1 | 1 | 1 | 0 | 1 | 0 | 1 | 1 | 0 |
| TLR2 | 2 | 2 | 4 | 0 | 0 | 0 | 0 | 0 | 0 | 0 | 1 | 0 | 0 | 0 | 0 |
| AP1 | 0 | 3 | 3 | 0 | 0 | 0 | 0 | 1 | 0 | 0 | 0 | 0 | 0 | 0 | 0 |
| CA2 | 3 | 0 | 3 | 0 | 0 | 0 | 0 | 0 | 1 | 0 | 0 | 0 | 0 | 0 | 0 |
| CCL2 | 2 | 1 | 3 | 0 | 0 | 0 | 0 | 0 | 0 | 0 | 0 | 0 | 0 | 0 | 1 |
| CD40LG | 1 | 2 | 3 | 1 | 0 | 1 | 0 | 0 | 0 | 1 | 0 | 0 | 0 | 1 | 0 |
| ERK | 1 | 2 | 3 | 0 | 1 | 0 | 1 | 1 | 1 | 0 | 1 | 1 | 0 | 1 | 0 |
| p38 | 0 | 3 | 3 | 0 | 0 | 0 | 0 | 0 | 0 | 0 | 1 | 0 | 0 | 0 | 0 |
| PTGS2 | 1 | 2 | 3 | 0 | 0 | 1 | 0 | 2 | 0 | 2 | 1 | 0 | 0 | 2 | 0 |
| CXCL12 | 1 | 2 | 3 | 0 | 0 | 1 | 0 | 1 | 0 | 0 | 0 | 0 | 0 | 0 | 0 |
| ESR1 | 3 | 0 | 3 | 0 | 2 | 0 | 3 | 0 | 0 | 0 | 3 | 0 | 0 | 0 | 1 |
| IL17A | 1 | 2 | 3 | 1 | 0 | 1 | 0 | 0 | 1 | 1 | 0 | 0 | 2 | 0 | 1 |
| IL18 | 1 | 2 | 3 | 1 | 1 | 0 | 1 | 0 | 0 | 1 | 1 | 0 | 1 | 0 | 1 |
| IL6ST | 1 | 2 | 3 | 0 | 0 | 1 | 1 | 0 | 1 | 1 | 0 | 0 | 1 | 0 | 0 |
| INS | 1 | 2 | 3 | 1 | 0 | 1 | 0 | 2 | 0 | 1 | 0 | 2 | 1 | 0 | 1 |
| NGF | 0 | 3 | 3 | 0 | 0 | 0 | 0 | 0 | 0 | 0 | 0 | 0 | 0 | 0 | 0 |
| TLR4 | 0 | 3 | 3 | 0 | 0 | 0 | 0 | 0 | 0 | 0 | 0 | 0 | 0 | 0 | 0 |
| TLR3 | 1 | 2 | 3 | 0 | 0 | 0 | 0 | 0 | 0 | 0 | 0 | 0 | 0 | 0 | 0 |
| IRF5 | 1 | 2 | 3 | 0 | 0 | 0 | 0 | 0 | 0 | 0 | 0 | 0 | 0 | 1 | 0 |
| TBK1 | 1 | 2 | 3 | 0 | 0 | 0 | 0 | 0 | 0 | 0 | 0 | 0 | 0 | 0 | 0 |
| POLI2F1 | 2 | 1 | 3 | 1 | 2 | 0 | 2 | 1 | 0 | 0 | 2 | 1 | 1 | 1 | 1 |
| ADIPOQ | 2 | 1 | 3 | 0 | 1 | 0 | 1 | 0 | 0 | 0 | 1 | 1 | 0 | 0 | 0 |
| ABCB1 | 2 | 0 | 2 | 0 | 0 | 0 | 0 | 0 | 0 | 0 | 0 | 0 | 0 | 0 | 0 |
| PKC | 0 | 2 | 2 | 0 | 0 | 0 | 0 | 1 | 0 | 0 | 0 | 0 | 0 | 0 | 0 |
| CASP8 | 2 | 0 | 2 | 0 | 1 | 0 | 1 | 0 | 0 | 0 | 1 | 0 | 0 | 0 | 0 |
| PI3K | 1 | 1 | 2 | 0 | 0 | 0 | 0 | 0 | 0 | 0 | 0 | 0 | 0 | 0 | 0 |
| PRKAC | 1 | 1 | 2 | 0 | 0 | 0 | 0 | 0 | 0 | 0 | 0 | 0 | 0 | 0 | 0 |
| CD1A | 2 | 0 | 2 | 0 | 0 | 0 | 2 | 0 | 0 | 0 | 0 | 0 | 0 | 0 | 0 |
| CD40 | 1 | 1 | 2 | 0 | 0 | 0 | 0 | 0 | 0 | 0 | 0 | 0 | 0 | 0 | 0 |
| CDKN1A | 2 | 0 | 2 | 0 | 1 | 0 | 0 | 0 | 0 | 0 | 1 | 0 | 1 | 0 | 1 |
| CEBPB | 2 | 0 | 2 | 0 | 0 | 0 | 0 | 1 | 0 | 0 | 0 | 0 | 0 | 0 | 1 |
| CHUK | 1 | 1 | 2 | 0 | 0 | 0 | 0 | 0 | 0 | 0 | 0 | 0 | 0 | 0 | 1 |
| CXCL8 | 1 | 1 | 2 | 1 | 0 | 1 | 0 | 0 | 0 | 1 | 0 | 1 | 0 | 0 | 0 |
| EP300 | 1 | 1 | 2 | 0 | 0 | 0 | 0 | 0 | 0 | 0 | 0 | 0 | 0 | 0 | 0 |
| CAMP | 0 | 2 | 2 | 0 | 0 | 0 | 0 | 0 | 0 | 0 | 0 | 1 | 0 | 0 | 0 |
| GSK3B | 1 | 1 | 2 | 0 | 0 | 0 | 0 | 1 | 0 | 1 | 0 | 1 | 0 | 0 | 0 |
| HCK | 1 | 1 | 2 | 0 | 0 | 0 | 0 | 0 | 0 | 0 | 0 | 0 | 0 | 0 | 0 |
| HSF1 | 2 | 0 | 2 | 0 | 0 | 0 | 0 | 0 | 0 | 0 | 0 | 0 | 0 | 0 | 1 |
| IFNB1 | 1 | 1 | 2 | 1 | 1 | 1 | 1 | 1 | 0 | 1 | 1 | 1 | 0 | 1 | 0 |
| IKBKB | 1 | 1 | 2 | 0 | 1 | 0 | 0 | 0 | 0 | 0 | 0 | 1 | 0 | 0 | 0 |
| IKBKG | 1 | 1 | 2 | 0 | 0 | 0 | 1 | 0 | 1 | 0 | 0 | 0 | 0 | 0 | 0 |
| ADRB2 | 0 | 2 | 2 | 0 | 0 | 0 | 0 | 0 | 0 | 0 | 0 | 1 | 0 | 1 | 0 |
| BSG | 0 | 2 | 2 | 0 | 0 | 0 | 0 | 1 | 0 | 0 | 0 | 0 | 0 | 0 | 0 |
| ERBB2 | 1 | 1 | 2 | 0 | 0 | 0 | 0 | 0 | 0 | 0 | 0 | 0 | 0 | 0 | 1 |
| IL13 | 0 | 2 | 2 | 1 | 0 | 0 | 0 | 0 | 0 | 1 | 0 | 0 | 0 | 0 | 0 |
| IL4 | 0 | 2 | 2 | 0 | 0 | 1 | 0 | 0 | 0 | 0 | 0 | 0 | 0 | 0 | 0 |
| OSM | 0 | 2 | 2 | 0 | 0 | 0 | 0 | 0 | 0 | 0 | 0 | 0 | 0 | 0 | 0 |
| PARP1 | 2 | 0 | 2 | 0 | 0 | 0 | 0 | 0 | 0 | 0 | 0 | 0 | 0 | 0 | 0 |
| PTK2 | 0 | 2 | 2 | 1 | 0 | 1 | 0 | 1 | 0 | 1 | 0 | 1 | 0 | 1 | 0 |
| PTPN11 | 0 | 2 | 2 | 1 | 0 | 2 | 0 | 1 | 0 | 2 | 0 | 1 | 0 | 1 | 0 |
| TGFB1 | 0 | 2 | 2 | 1 | 0 | 1 | 0 | 2 | 0 | 1 | 0 | 0 | 0 | 0 | 0 |
| TNFRSF1A | 1 | 1 | 2 | 0 | 0 | 0 | 0 | 0 | 0 | 0 | 0 | 1 | 0 | 0 | 0 |
| VEGF | 0 | 2 | 2 | 1 | 0 | 1 | 0 | 1 | 0 | 0 | 0 | 0 | 0 | 0 | 0 |
| XDH | 1 | 1 | 2 | 0 | 0 | 0 | 0 | 0 | 0 | 0 | 0 | 0 | 0 | 0 | 0 |
| IRAK1 | 1 | 1 | 2 | 0 | 0 | 0 | 0 | 0 | 0 | 0 | 0 | 0 | 1 | 0 | 0 |
| IRF7 | 1 | 1 | 2 | 0 | 0 | 0 | 0 | 0 | 0 | 0 | 0 | 0 | 0 | 0 | 0 |
| MAP3K1 | 1 | 1 | 2 | 1 | 0 | 0 | 0 | 0 | 1 | 1 | 1 | 0 | 0 | 0 | 0 |
| MAPK14 | 1 | 1 | 2 | 0 | 0 | 0 | 0 | 0 | 0 | 0 | 0 | 0 | 0 | 0 | 0 |
| TLR1 | 0 | 2 | 2 | 0 | 0 | 0 | 0 | 0 | 0 | 0 | 0 | 0 | 0 | 1 | 0 |
| POMC | 1 | 1 | 2 | 0 | 0 | 0 | 0 | 0 | 0 | 0 | 0 | 0 | 0 | 0 | 0 |
| PTEN | 1 | 1 | 2 | 1 | 0 | 1 | 0 | 0 | 0 | 0 | 0 | 0 | 0 | 1 | 0 |
| RIPK2 | 1 | 1 | 2 | 0 | 1 | 0 | 1 | 0 | 1 | 0 | 0 | 0 | 0 | 0 | 0 |
| RORA | 1 | 1 | 2 | 0 | 0 | 0 | 0 | 1 | 1 | 0 | 0 | 1 | 1 | 0 | 0 |
| SERPINA3 | 1 | 1 | 2 | 0 | 0 | 0 | 0 | 0 | 0 | 0 | 0 | 0 | 0 | 0 | 0 |
| IL10 | 0 | 2 | 2 | 1 | 0 | 1 | 0 | 1 | 0 | 1 | 0 | 1 | 0 | 1 | 0 |
| TGFA | 1 | 1 | 2 | 0 | 1 | 0 | 0 | 0 | 0 | 0 | 1 | 0 | 0 | 0 | 0 |
| TIMP3 | 1 | 1 | 2 | 0 | 0 | 0 | 1 | 0 | 0 | 0 | 0 | 0 | 0 | 0 | 0 |
| NFKBIA | 1 | 1 | 2 | 0 | 1 | 0 | 0 | 1 | 1 | 0 | 0 | 0 | 1 | 0 | 1 |
| NOS3 | 1 | 1 | 2 | 0 | 0 | 0 | 0 | 0 | 0 | 0 | 0 | 0 | 1 | 0 | 0 |
| TRAF2 | 1 | 1 | 2 | 1 | 1 | 1 | 1 | 1 | 0 | 1 | 1 | 0 | 0 | 0 | 1 |
| TNFRSF1B | 1 | 1 | 2 | 0 | 0 | 0 | 0 | 0 | 1 | 0 | 0 | 0 | 0 | 0 | 0 |
| VCAM1 | 2 | 0 | 2 | 0 | 1 | 0 | 0 | 0 | 1 | 0 | 1 | 0 | 0 | 0 | 1 |
| VEGFA | 2 | 0 | 2 | 0 | 1 | 0 | 1 | 0 | 1 | 0 | 1 | 0 | 0 | 0 | 0 |
| AKT | 2 | 0 | 2 | 0 | 0 | 0 | 0 | 0 | 0 | 0 | 0 | 0 | 0 | 0 | 0 |
| PPARA | 0 | 2 | 2 | 0 | 0 | 0 | 0 | 1 | 0 | 0 | 0 | 0 | 0 | 0 | 0 |
| SOCS3 | 0 | 2 | 2 | 1 | 0 | 1 | 0 | 0 | 0 | 1 | 0 | 1 | 0 | 0 | 0 |
| ADRB | 0 | 2 | 2 | 0 | 0 | 0 | 0 | 0 | 0 | 0 | 0 | 2 | 0 | 1 | 0 |
| A2M | 1 | 0 | 1 | 0 | 0 | 0 | 0 | 0 | 0 | 0 | 0 | 0 | 0 | 0 | 0 |
| ABCC1 | 1 | 0 | 1 | 0 | 0 | 0 | 0 | 0 | 1 | 0 | 0 | 0 | 1 | 0 | 1 |
| ABCG2 | 1 | 0 | 1 | 0 | 0 | 0 | 0 | 0 | 0 | 0 | 0 | 0 | 0 | 0 | 0 |
| ADAMTS5 | 1 | 0 | 1 | 0 | 0 | 0 | 0 | 0 | 0 | 0 | 0 | 0 | 1 | 0 | 1 |
| AIM2 | 1 | 0 | 1 | 0 | 0 | 0 | 0 | 0 | 1 | 0 | 0 | 0 | 0 | 0 | 0 |
| AKT1 | 1 | 0 | 1 | 0 | 0 | 0 | 0 | 0 | 0 | 0 | 0 | 0 | 0 | 0 | 0 |
| AKT2 | 1 | 0 | 1 | 0 | 0 | 0 | 0 | 0 | 1 | 0 | 0 | 0 | 0 | 0 | 0 |
| ALPL | 1 | 0 | 1 | 0 | 0 | 0 | 0 | 0 | 0 | 0 | 1 | 0 | 1 | 0 | 1 |
| ANXA1 | 1 | 0 | 1 | 0 | 0 | 0 | 0 | 0 | 0 | 0 | 0 | 0 | 0 | 0 | 0 |
| AREG | 1 | 0 | 1 | 0 | 0 | 0 | 1 | 0 | 0 | 0 | 0 | 0 | 0 | 0 | 0 |
| ATF2 | 1 | 0 | 1 | 0 | 0 | 0 | 0 | 0 | 0 | 0 | 0 | 0 | 0 | 0 | 0 |
| ATF3 | 1 | 0 | 1 | 0 | 1 | 0 | 0 | 0 | 0 | 0 | 0 | 0 | 0 | 0 | 1 |
| BMP2 | 1 | 0 | 1 | 0 | 0 | 0 | 0 | 0 | 0 | 0 | 0 | 0 | 0 | 0 | 0 |
| C1R | 1 | 0 | 1 | 0 | 0 | 0 | 0 | 0 | 0 | 0 | 0 | 0 | 1 | 0 | 0 |
| C3 | 1 | 0 | 1 | 0 | 0 | 0 | 0 | 0 | 0 | 0 | 0 | 0 | 0 | 0 | 0 |
| C5AR1 | 1 | 0 | 1 | 0 | 0 | 0 | 0 | 0 | 0 | 0 | 0 | 0 | 0 | 0 | 0 |
| C8A | 1 | 0 | 1 | 0 | 0 | 0 | 0 | 0 | 0 | 0 | 0 | 0 | 0 | 0 | 0 |
| C8G | 1 | 0 | 1 | 0 | 0 | 0 | 1 | 0 | 0 | 0 | 0 | 0 | 0 | 0 | 0 |
| NKX2-1 | 0 | 1 | 1 | 0 | 0 | 0 | 0 | 0 | 0 | 0 | 0 | 0 | 0 | 0 | 0 |
| CASP1 | 1 | 0 | 1 | 0 | 0 | 0 | 0 | 0 | 0 | 0 | 0 | 0 | 0 | 0 | 0 |
| CASP2 | 1 | 0 | 1 | 0 | 0 | 0 | 0 | 0 | 0 | 0 | 0 | 0 | 0 | 0 | 0 |
| CASP6 | 1 | 0 | 1 | 0 | 0 | 0 | 0 | 0 | 1 | 0 | 0 | 0 | 0 | 0 | 0 |
| CASP7 | 1 | 0 | 1 | 0 | 0 | 0 | 0 | 0 | 0 | 0 | 1 | 0 | 0 | 0 | 0 |
| CBL | 1 | 0 | 1 | 0 | 0 | 0 | 0 | 0 | 0 | 0 | 0 | 0 | 0 | 0 | 0 |
| CCL11 | 1 | 0 | 1 | 0 | 0 | 0 | 1 | 0 | 0 | 0 | 0 | 0 | 1 | 0 | 0 |
| CCL3 | 1 | 0 | 1 | 0 | 0 | 0 | 0 | 0 | 1 | 0 | 0 | 0 | 0 | 0 | 1 |
| CCR2 | 1 | 0 | 1 | 0 | 0 | 0 | 0 | 0 | 0 | 0 | 0 | 0 | 0 | 0 | 0 |
| CEBP | 0 | 1 | 1 | 0 | 0 | 0 | 0 | 0 | 0 | 0 | 0 | 0 | 0 | 0 | 0 |
| KITLG | 0 | 1 | 1 | 0 | 0 | 1 | 0 | 0 | 0 | 0 | 0 | 1 | 0 | 0 | 0 |
| CD163 | 1 | 0 | 1 | 0 | 0 | 0 | 0 | 0 | 0 | 0 | 0 | 0 | 0 | 0 | 1 |
| CD38 | 1 | 0 | 1 | 0 | 0 | 0 | 0 | 0 | 0 | 0 | 0 | 0 | 1 | 0 | 0 |
| CD55 | 1 | 0 | 1 | 0 | 0 | 0 | 0 | 0 | 0 | 0 | 1 | 0 | 0 | 0 | 0 |
| CD59 | 1 | 0 | 1 | 0 | 0 | 0 | 0 | 0 | 1 | 0 | 0 | 0 | 0 | 0 | 0 |
| CD80 | 1 | 0 | 1 | 0 | 0 | 0 | 0 | 0 | 0 | 0 | 0 | 0 | 0 | 0 | 0 |
| CD86 | 1 | 0 | 1 | 0 | 0 | 0 | 0 | 0 | 0 | 0 | 0 | 0 | 0 | 0 | 0 |
| CDKN1B | 1 | 0 | 1 | 0 | 0 | 0 | 0 | 0 | 0 | 0 | 0 | 0 | 0 | 0 | 0 |
| CFB | 1 | 0 | 1 | 0 | 0 | 0 | 0 | 0 | 0 | 0 | 0 | 0 | 0 | 0 | 0 |
| CFL2 | 1 | 0 | 1 | 0 | 0 | 0 | 0 | 0 | 0 | 0 | 0 | 0 | 0 | 0 | 1 |
| CK2 | 1 | 0 | 1 | 0 | 0 | 0 | 0 | 0 | 0 | 0 | 1 | 0 | 0 | 0 | 0 |
| CLEC3B | 1 | 0 | 1 | 0 | 0 | 0 | 0 | 0 | 0 | 0 | 0 | 0 | 0 | 0 | 0 |
| CLIC4 | 1 | 0 | 1 | 0 | 0 | 0 | 0 | 0 | 1 | 0 | 0 | 0 | 1 | 0 | 0 |
| CMA1 | 1 | 0 | 1 | 0 | 0 | 0 | 0 | 0 | 0 | 0 | 0 | 0 | 0 | 0 | 0 |
| CD32 | 0 | 1 | 1 | 0 | 0 | 0 | 0 | 0 | 0 | 0 | 0 | 0 | 0 | 0 | 0 |
| HNF1A | 0 | 1 | 1 | 0 | 0 | 0 | 0 | 0 | 0 | 0 | 0 | 0 | 0 | 0 | 0 |
| LEP | 0 | 1 | 1 | 0 | 0 | 0 | 0 | 0 | 0 | 0 | 0 | 0 | 0 | 0 | 0 |
| CSF1 | 1 | 0 | 1 | 0 | 0 | 0 | 0 | 0 | 0 | 0 | 0 | 0 | 0 | 0 | 0 |
| CSF2 | 1 | 0 | 1 | 0 | 0 | 0 | 0 | 0 | 0 | 0 | 0 | 0 | 0 | 0 | 0 |
| CSF3 | 1 | 0 | 1 | 0 | 0 | 0 | 0 | 0 | 0 | 0 | 0 | 0 | 1 | 0 | 0 |
| CTSB | 1 | 0 | 1 | 0 | 0 | 0 | 0 | 0 | 1 | 0 | 1 | 0 | 0 | 0 | 0 |
| CTSL | 1 | 0 | 1 | 0 | 0 | 0 | 0 | 0 | 0 | 0 | 0 | 0 | 0 | 0 | 0 |
| CTSS | 1 | 0 | 1 | 0 | 0 | 0 | 0 | 0 | 0 | 0 | 0 | 0 | 0 | 0 | 0 |
| CX3CL1 | 1 | 0 | 1 | 0 | 0 | 0 | 0 | 0 | 0 | 0 | 0 | 0 | 0 | 0 | 0 |
| CXCL10 | 1 | 0 | 1 | 0 | 0 | 0 | 0 | 0 | 0 | 0 | 0 | 0 | 0 | 0 | 1 |
| CXCL5 | 1 | 0 | 1 | 0 | 0 | 0 | 0 | 0 | 0 | 0 | 0 | 0 | 1 | 0 | 0 |
| CXCL9 | 1 | 0 | 1 | 0 | 0 | 0 | 1 | 0 | 0 | 0 | 0 | 1 | 0 | 0 | 0 |
| CXCR3 | 1 | 0 | 1 | 0 | 0 | 0 | 0 | 0 | 0 | 0 | 0 | 0 | 0 | 0 | 0 |
| DAPK1 | 1 | 0 | 1 | 0 | 0 | 0 | 0 | 0 | 0 | 0 | 0 | 0 | 0 | 0 | 0 |
| EHF | 1 | 0 | 1 | 0 | 0 | 0 | 0 | 0 | 0 | 0 | 0 | 0 | 0 | 0 | 0 |
| F2RL1 | 1 | 0 | 1 | 0 | 0 | 0 | 0 | 0 | 1 | 0 | 1 | 0 | 0 | 0 | 1 |
| F2RL3 | 1 | 0 | 1 | 0 | 0 | 0 | 1 | 0 | 0 | 0 | 0 | 0 | 0 | 0 | 0 |
| FAS | 1 | 0 | 1 | 0 | 0 | 0 | 0 | 0 | 0 | 0 | 0 | 0 | 0 | 0 | 0 |
| FCER2 | 1 | 0 | 1 | 0 | 0 | 0 | 0 | 0 | 0 | 0 | 0 | 0 | 0 | 0 | 0 |
| FCGR1A | 1 | 0 | 1 | 0 | 0 | 0 | 0 | 0 | 0 | 0 | 0 | 0 | 0 | 0 | 0 |
| FCGR3B | 1 | 0 | 1 | 0 | 0 | 0 | 0 | 0 | 1 | 0 | 0 | 0 | 0 | 0 | 0 |
| FN1 | 1 | 0 | 1 | 0 | 0 | 0 | 0 | 0 | 0 | 0 | 1 | 0 | 0 | 0 | 0 |
| FOXO4 | 1 | 0 | 1 | 0 | 0 | 0 | 0 | 0 | 0 | 0 | 0 | 0 | 0 | 0 | 0 |
| FUT4 | 1 | 0 | 1 | 0 | 0 | 0 | 0 | 0 | 0 | 0 | 0 | 0 | 0 | 0 | 0 |
| FYN | 1 | 0 | 1 | 0 | 0 | 0 | 0 | 0 | 0 | 0 | 0 | 0 | 0 | 0 | 1 |
| GBF1 | 1 | 0 | 1 | 0 | 0 | 0 | 0 | 0 | 0 | 0 | 0 | 0 | 0 | 0 | 0 |
| HBEGF | 1 | 0 | 1 | 0 | 0 | 0 | 0 | 0 | 1 | 0 | 1 | 0 | 0 | 0 | 1 |
| HIF1A | 1 | 0 | 1 | 0 | 0 | 0 | 0 | 0 | 0 | 0 | 0 | 0 | 0 | 0 | 1 |
| HLA-A | 1 | 0 | 1 | 0 | 0 | 0 | 0 | 0 | 0 | 0 | 0 | 0 | 0 | 0 | 0 |
| HLA-B | 1 | 0 | 1 | 0 | 0 | 0 | 0 | 0 | 0 | 0 | 0 | 0 | 0 | 0 | 0 |
| HLA-C | 1 | 0 | 1 | 0 | 0 | 0 | 0 | 0 | 0 | 0 | 0 | 0 | 0 | 0 | 0 |
| HLA-F | 1 | 0 | 1 | 0 | 1 | 0 | 0 | 0 | 0 | 0 | 0 | 0 | 0 | 0 | 0 |
| HP | 1 | 0 | 1 | 0 | 0 | 0 | 0 | 0 | 0 | 0 | 0 | 0 | 0 | 0 | 0 |
| HSD11B1 | 1 | 0 | 1 | 0 | 0 | 0 | 0 | 0 | 0 | 0 | 0 | 0 | 0 | 0 | 1 |
| HSP90B1 | 1 | 0 | 1 | 0 | 0 | 0 | 1 | 0 | 0 | 0 | 0 | 0 | 0 | 0 | 0 |
| HOXD3 | 0 | 1 | 1 | 0 | 0 | 0 | 0 | 0 | 0 | 0 | 0 | 0 | 0 | 0 | 0 |
| LIF | 0 | 1 | 1 | 0 | 0 | 0 | 0 | 0 | 0 | 0 | 0 | 0 | 0 | 0 | 0 |
| RAF1 | 0 | 1 | 1 | 0 | 0 | 0 | 0 | 0 | 0 | 0 | 1 | 0 | 1 | 0 | 0 |
| IKK_family | 1 | 0 | 1 | 0 | 0 | 0 | 0 | 0 | 0 | 0 | 0 | 0 | 0 | 0 | 0 |
| IL11 | 1 | 0 | 1 | 0 | 0 | 0 | 0 | 0 | 0 | 0 | 0 | 0 | 0 | 0 | 0 |
| IL12A | 1 | 0 | 1 | 0 | 0 | 0 | 0 | 0 | 0 | 0 | 0 | 0 | 0 | 0 | 0 |
| IL12B | 1 | 0 | 1 | 0 | 0 | 0 | 0 | 0 | 0 | 0 | 0 | 0 | 0 | 0 | 0 |
| IL4R | 1 | 0 | 1 | 0 | 0 | 0 | 0 | 0 | 0 | 0 | 0 | 0 | 0 | 0 | 0 |
| AHR | 0 | 1 | 1 | 0 | 0 | 0 | 0 | 0 | 0 | 0 | 0 | 0 | 0 | 0 | 0 |
| AR | 0 | 1 | 1 | 0 | 0 | 0 | 0 | 0 | 0 | 1 | 0 | 0 | 0 | 0 | 0 |
| F10 | 0 | 1 | 1 | 0 | 0 | 0 | 0 | 0 | 0 | 0 | 0 | 0 | 0 | 0 | 0 |
| F2 | 0 | 1 | 1 | 0 | 1 | 0 | 0 | 0 | 0 | 0 | 0 | 0 | 0 | 0 | 0 |
| GAB1 | 0 | 1 | 1 | 0 | 0 | 0 | 0 | 0 | 0 | 0 | 0 | 0 | 0 | 0 | 0 |
| IL17F | 0 | 1 | 1 | 0 | 0 | 0 | 0 | 0 | 0 | 0 | 0 | 0 | 0 | 0 | 0 |
| IL1R1 | 0 | 1 | 1 | 0 | 0 | 0 | 0 | 0 | 0 | 0 | 0 | 1 | 0 | 0 | 0 |
| IL6R | 0 | 1 | 1 | 0 | 0 | 0 | 0 | 0 | 0 | 0 | 0 | 0 | 0 | 0 | 0 |
| ITGA5 | 0 | 1 | 1 | 0 | 0 | 0 | 0 | 0 | 0 | 0 | 0 | 0 | 0 | 0 | 0 |
| KLF7 | 0 | 1 | 1 | 0 | 0 | 0 | 0 | 0 | 0 | 0 | 0 | 0 | 0 | 0 | 0 |
| LEPR | 0 | 1 | 1 | 0 | 0 | 0 | 0 | 0 | 0 | 0 | 0 | 0 | 0 | 0 | 0 |
| LMX1B | 0 | 1 | 1 | 0 | 0 | 0 | 0 | 0 | 0 | 0 | 0 | 0 | 0 | 0 | 0 |
| PIK3C3 | 0 | 1 | 1 | 0 | 0 | 0 | 0 | 0 | 0 | 0 | 0 | 0 | 0 | 0 | 0 |
| PNPT1 | 0 | 1 | 1 | 0 | 0 | 0 | 0 | 0 | 0 | 0 | 0 | 0 | 0 | 0 | 0 |

| | | | | | | | | | | | | | | | | | | | |
|---|---|---|---|---|---|---|---|---|---|---|---|---|---|---|---|---|---|---|---|
| ZFAND5 | 0 | 1 | 1 | | 0 | 0 | | 0 | 0 | | 0 | 0 | | 0 | 0 | | 1 | 0 | | 0 | 0 |
| ZFP36 | 0 | 1 | 1 | | 0 | 0 | | 0 | 0 | | 0 | 0 | | 0 | 0 | | 0 | 0 | | 0 | 0 |

| | Unique Sources | Unique Targets | | Unique Sources | Unique Targets | | Unique Sources | Unique Targets | | Unique Sources | Unique Targets | | Unique Sources | Unique Targets | | Unique Sources | Unique Targets | | | p-values Unique Sources | Unique Targets |
|---|---|---|---|---|---|---|---|---|---|---|---|---|---|---|---|---|---|---|---|---|---|
| | 27 | 30 | | 26 | 26 | | 32 | 21 | | 26 | 28 | | 32 | 20 | | 29 | 20 | Edge Count > 1 | 0.81 | 0.47 |
| | 16 | 44 | | 21 | 40 | | 26 | 40 | | 23 | 47 | | 19 | 37 | | 15 | 38 | Edge Count =1 | 0.62 | 0.86 |
| | 43 | 74 | | 47 | 66 | | 58 | 61 | | 49 | 75 | | 51 | 57 | | 44 | 58 | Total Unique | 0.81 | 0.63 |

**Table S5.** Sources and targets with high and low edge count recruited uniquely by each policy

| Name | EdgeCount | Sources unique to High/Low Entropy | Targets unique to High/Low Entropy | Sources unique to Low Entropy | Targets unique to Low Entropy |
|---|---|---|---|---|---|
| TNF | 273 | 0 | 0 | 0 | 0 |
| IL6 | 126 | 0 | 0 | 0 | 0 |
| STAT1 | 42 | 0 | 0 | 0 | 0 |
| CRP | 20 | 0 | 0 | 0 | 0 |
| MYD88 | 17 | 0 | 0 | 1 | 0 |
| CD14 | 11 | 0 | 0 | 0 | 0 |
| NFkappaB | 10 | 0 | 0 | 0 | 0 |
| IRF3 | 10 | 0 | 0 | 0 | 0 |
| IFI16 | 7 | 0 | 0 | 0 | 0 |
| IFNG | 6 | 0 | 0 | 0 | 1 |
| TLR9 | 5 | 1 | 0 | 0 | 0 |
| PTPA | 5 | 0 | 0 | 0 | 0 |
| AGT | 4 | 1 | 0 | 0 | 1 |
| IFNA1 | 4 | 0 | 0 | 0 | 0 |
| IL1B | 4 | 0 | 0 | 0 | 0 |
| EGFR | 4 | 0 | 0 | 1 | 0 |
| STAT3 | 4 | 0 | 1 | 0 | 0 |
| MAPK1 | 4 | 0 | 0 | 1 | 0 |
| MAPK3 | 4 | 0 | 0 | 0 | 0 |
| TLR2 | 4 | 1 | 0 | 0 | 0 |
| AP1 | 3 | 1 | 0 | 0 | 0 |
| CA2 | 3 | 0 | 1 | 0 | 0 |
| CCL2 | 3 | 0 | 0 | 0 | 1 |
| CD40LG | 3 | 0 | 0 | 0 | 0 |
| ERK | 3 | 0 | 0 | 0 | 0 |
| p38 | 3 | 0 | 0 | 0 | 0 |
| PTGS2 | 3 | 0 | 0 | 0 | 1 |
| CXCL12 | 3 | 1 | 0 | 0 | 0 |
| ESR1 | 3 | 0 | 0 | 0 | 0 |
| IL17A | 3 | 0 | 0 | 0 | 0 |
| IL18 | 3 | 0 | 0 | 0 | 0 |
| IL6ST | 3 | 1 | 0 | 0 | 0 |
| INS | 3 | 0 | 0 | 0 | 0 |
| NGF | 3 | 0 | 0 | 0 | 0 |
| TLR4 | 3 | 0 | 0 | 0 | 0 |
| TLR3 | 3 | 0 | 0 | 0 | 0 |
| IRF5 | 3 | 0 | 0 | 1 | 0 |
| TBK1 | 3 | 0 | 0 | 0 | 0 |
| POU2F1 | 3 | 0 | 0 | 0 | 0 |
| ADIPOQ | 3 | 0 | 0 | 1 | 0 |
| ABCB1 | 2 | 0 | 0 | 0 | 0 |
| PKC | 2 | 1 | 0 | 0 | 0 |
| CASP8 | 2 | 0 | 0 | 0 | 0 |
| PI3K | 2 | 0 | 0 | 0 | 0 |
| PRKAC | 2 | 1 | 0 | 0 | 0 |
| CD1A | 2 | 0 | 1 | 0 | 0 |
| CD40 | 2 | 0 | 0 | 0 | 0 |
| CDKN1A | 2 | 0 | 0 | 0 | 0 |
| CEBPB | 2 | 0 | 0 | 0 | 1 |
| CHUK | 2 | 1 | 0 | 0 | 1 |
| CXCL8 | 2 | 0 | 0 | 0 | 0 |
| EP300 | 2 | 0 | 0 | 0 | 1 |
| CAMP | 2 | 0 | 0 | 0 | 0 |
| GSK3B | 2 | 0 | 0 | 0 | 0 |
| HCK | 2 | 0 | 0 | 0 | 0 |
| HSF1 | 2 | 0 | 1 | 0 | 0 |
| IFNB1 | 2 | 0 | 0 | 0 | 0 |
| IKBKB | 2 | 0 | 0 | 0 | 0 |
| IKBKG | 2 | 0 | 0 | 0 | 0 |
| ADRB2 | 2 | 0 | 0 | 1 | 0 |
| BSG | 2 | 0 | 0 | 0 | 0 |
| ERBB2 | 2 | 0 | 0 | 0 | 0 |
| IL13 | 2 | 0 | 0 | 0 | 0 |
| IL4 | 2 | 0 | 0 | 0 | 0 |
| OSM | 2 | 1 | 0 | 0 | 0 |
| PARP1 | 2 | 1 | 0 | 0 | 0 |
| PTK2 | 2 | 0 | 0 | 0 | 0 |
| PTPN11 | 2 | 0 | 0 | 0 | 0 |
| TGFB1 | 2 | 0 | 0 | 0 | 0 |
| TNFRSF1A | 2 | 0 | 0 | 1 | 0 |
| VEGF | 2 | 1 | 0 | 0 | 0 |
| XDH | 2 | 0 | 0 | 0 | 0 |
| IRAK1 | 2 | 0 | 0 | 0 | 1 |
| IRF7 | 2 | 0 | 0 | 0 | 0 |
| MAP3K1 | 2 | 0 | 0 | 0 | 0 |
| MAPK14 | 2 | 0 | 0 | 0 | 0 |
| TLR1 | 2 | 0 | 0 | 1 | 0 |
| POMC | 2 | 0 | 0 | 0 | 0 |
| PTEN | 2 | 0 | 0 | 0 | 0 |
| RIPK2 | 2 | 1 | 1 | 0 | 0 |
| RORA | 2 | 0 | 0 | 0 | 0 |
| SERPINA3 | 2 | 0 | 0 | 0 | 0 |
| IL10 | 2 | 0 | 0 | 0 | 0 |
| TGFA | 2 | 0 | 0 | 1 | 0 |
| TIMP3 | 2 | 0 | 1 | 0 | 0 |
| NFKBIA | 2 | 1 | 0 | 0 | 0 |
| NOS3 | 2 | 0 | 0 | 0 | 0 |
| TRAF2 | 2 | 0 | 0 | 0 | 0 |
| TNFRSF1B | 2 | 0 | 0 | 0 | 0 |
| VCAM1 | 2 | 0 | 0 | 0 | 0 |
| VEGFA | 2 | 0 | 1 | 0 | 0 |
| AKT | 2 | 0 | 0 | 0 | 0 |
| PPARA | 2 | 0 | 0 | 1 | 0 |
| SOCS3 | 2 | 0 | 0 | 0 | 0 |
| ADRB | 2 | 0 | 0 | 1 | 0 |
| A2M | 1 | 0 | 0 | 0 | 0 |
| ABCC1 | 1 | 0 | 0 | 0 | 0 |
| ABCG2 | 1 | 0 | 0 | 0 | 1 |
| ADAMTS5 | 1 | 0 | 0 | 0 | 1 |
| AIM2 | 1 | 0 | 1 | 0 | 0 |
| AKT1 | 1 | 0 | 0 | 0 | 0 |
| AKT2 | 1 | 0 | 0 | 0 | 0 |
| ALPL | 1 | 0 | 0 | 0 | 1 |
| ANXA1 | 1 | 0 | 0 | 0 | 0 |
| AREG | 1 | 0 | 0 | 0 | 0 |

Edge Count >1

|  |  |  |  |  |  |
|---|---|---|---|---|---|
| ATF2 | 1 | 0 | 0 | 0 | 0 |
| ATF3 | 1 | 0 | 0 | 0 | 0 |
| BMP2 | 1 | 0 | 0 | 0 | 0 |
| C1R | 1 | 0 | 0 | 0 | 1 |
| C3 | 1 | 0 | 0 | 0 | 0 |
| C5AR1 | 1 | 0 | 0 | 0 | 0 |
| C8A | 1 | 0 | 0 | 0 | 0 |
| C8G | 1 | 0 | 1 | 0 | 0 |
| NKX2-1 | 1 | 0 | 0 | 0 | 0 |
| CASP1 | 1 | 0 | 0 | 0 | 0 |
| CASP2 | 1 | 0 | 0 | 0 | 0 |
| CASP6 | 1 | 0 | 1 | 0 | 0 |
| CASP7 | 1 | 0 | 0 | 0 | 0 |
| CBL | 1 | 0 | 0 | 0 | 1 |
| CCL11 | 1 | 0 | 1 | 0 | 0 |
| CCL3 | 1 | 0 | 0 | 0 | 0 |
| CCR2 | 1 | 0 | 0 | 0 | 0 |
| CEBP | 1 | 1 | 0 | 0 | 0 |
| KITLG | 1 | 0 | 0 | 0 | 0 |
| CD163 | 1 | 0 | 0 | 0 | 1 |
| CD38 | 1 | 0 | 0 | 0 | 1 |
| CD55 | 1 | 0 | 0 | 0 | 1 |
| CD59 | 1 | 0 | 0 | 0 | 0 |
| CD80 | 1 | 0 | 0 | 0 | 0 |
| CD86 | 1 | 0 | 0 | 0 | 0 |
| CDKN1B | 1 | 0 | 0 | 0 | 0 |
| CFB | 1 | 0 | 0 | 0 | 0 |
| CFL2 | 1 | 0 | 0 | 0 | 1 |
| CK2 | 1 | 0 | 0 | 0 | 1 |
| CLEC3B | 1 | 0 | 0 | 0 | 0 |
| CLIC4 | 1 | 0 | 0 | 0 | 0 |
| CMA1 | 1 | 0 | 0 | 0 | 0 |
| CD32 | 1 | 0 | 0 | 0 | 0 |
| HNF1A | 1 | 0 | 0 | 0 | 0 |
| LEP | 1 | 0 | 0 | 0 | 0 |
| CSF1 | 1 | 0 | 0 | 0 | 0 |
| CSF2 | 1 | 0 | 0 | 0 | 0 |
| CSF3 | 1 | 0 | 0 | 0 | 1 |
| CTSB | 1 | 0 | 0 | 0 | 0 |
| CTSL | 1 | 0 | 0 | 0 | 0 |
| CTSS | 1 | 0 | 0 | 0 | 0 |
| CX3CL1 | 1 | 0 | 0 | 0 | 0 |
| CXCL10 | 1 | 0 | 0 | 0 | 1 |
| CXCL5 | 1 | 0 | 0 | 0 | 1 |
| CXCL9 | 1 | 0 | 0 | 0 | 0 |
| CXCR3 | 1 | 0 | 1 | 0 | 0 |
| DAPK1 | 1 | 0 | 0 | 0 | 0 |
| EHF | 1 | 0 | 0 | 0 | 0 |
| F2RL1 | 1 | 0 | 0 | 0 | 0 |
| F2RL3 | 1 | 0 | 0 | 0 | 0 |
| FAS | 1 | 0 | 0 | 0 | 0 |
| FCER2 | 1 | 0 | 0 | 0 | 0 |
| FCGR1A | 1 | 0 | 0 | 0 | 0 |
| FCGR3B | 1 | 0 | 0 | 0 | 0 |
| FN1 | 1 | 0 | 0 | 0 | 0 |
| FOXO4 | 1 | 0 | 0 | 0 | 1 |
| FUT4 | 1 | 0 | 0 | 0 | 0 |
| FYN | 1 | 0 | 0 | 0 | 1 |
| GBF1 | 1 | 0 | 0 | 0 | 0 |
| HBEGF | 1 | 0 | 0 | 0 | 0 |
| HIF1A | 1 | 0 | 0 | 0 | 0 |
| HLA-A | 1 | 0 | 1 | 0 | 0 |
| HLA-B | 1 | 0 | 0 | 0 | 0 |
| HLA-C | 1 | 0 | 0 | 0 | 0 |
| HLA-F | 1 | 0 | 1 | 0 | 0 |
| HP | 1 | 0 | 1 | 0 | 0 |
| HSD11B1 | 1 | 0 | 0 | 0 | 1 |
| HSP90B1 | 1 | 0 | 0 | 0 | 0 |
| HOXD3 | 1 | 0 | 0 | 0 | 0 |
| LIF | 1 | 1 | 0 | 0 | 0 |
| RAF1 | 1 | 0 | 0 | 1 | 0 |
| IKK_family | 1 | 0 | 0 | 0 | 0 |
| IL11 | 1 | 0 | 1 | 0 | 0 |
| IL12A | 1 | 0 | 0 | 0 | 0 |
| IL12B | 1 | 0 | 0 | 0 | 1 |
| IL4R | 1 | 0 | 0 | 0 | 0 |
| AHR | 1 | 0 | 0 | 0 | 0 |
| AR | 1 | 0 | 0 | 0 | 0 |
| F10 | 1 | 0 | 0 | 0 | 0 |
| F2 | 1 | 0 | 0 | 0 | 0 |
| GAB1 | 1 | 0 | 0 | 0 | 0 |
| IL17F | 1 | 1 | 0 | 0 | 0 |
| IL1R1 | 1 | 0 | 0 | 1 | 0 |
| IL6R | 1 | 0 | 0 | 0 | 0 |
| ITGA5 | 1 | 0 | 0 | 0 | 0 |
| KLF7 | 1 | 0 | 0 | 0 | 0 |
| LEPR | 1 | 0 | 0 | 0 | 0 |
| LMX1B | 1 | 0 | 0 | 0 | 0 |
| PIK3C3 | 1 | 0 | 0 | 0 | 0 |
| PNPT1 | 1 | 0 | 0 | 0 | 0 |
| PTGER4 | 1 | 0 | 0 | 0 | 0 |
| TIRAP | 1 | 0 | 0 | 0 | 0 |
| IRAK4 | 1 | 0 | 0 | 0 | 1 |
| BEGAIN | 1 | 0 | 0 | 1 | 0 |
| IKBKE | 1 | 0 | 0 | 0 | 0 |
| MAVS | 1 | 0 | 0 | 0 | 0 |
| TICAM1 | 1 | 0 | 0 | 0 | 0 |
| ITGA6 | 1 | 0 | 0 | 0 | 0 |
| ITGAV | 1 | 0 | 0 | 0 | 0 |
| JAK | 1 | 0 | 0 | 0 | 0 |
| JUN | 1 | 0 | 0 | 0 | 1 |
| KRAS | 1 | 0 | 0 | 0 | 0 |
| LYN | 1 | 0 | 1 | 0 | 0 |
| LYZ | 1 | 0 | 0 | 0 | 0 |
| MAP2K1 | 1 | 0 | 0 | 0 | 1 |
| MAP2K2 | 1 | 0 | 0 | 0 | 0 |
| MAP2K6 | 1 | 0 | 0 | 0 | 0 |
| MAP3K14 | 1 | 0 | 0 | 0 | 0 |

| Gene | C1 | C2 | C3 | C4 | C5 |
|---|---|---|---|---|---|
| MAP3K5 | 1 | 0 | 0 | 0 | 0 |
| MAP4K3 | 1 | 0 | 1 | 0 | 0 |
| MAP4K4 | 1 | 0 | 0 | 0 | 0 |
| MAPK7 | 1 | 0 | 0 | 0 | 0 |
| MAPK9 | 1 | 0 | 0 | 0 | 1 |
| MAPKAPK3 | 1 | 0 | 0 | 0 | 0 |
| MAZ | 1 | 0 | 0 | 0 | 0 |
| MEF2 | 1 | 0 | 1 | 0 | 0 |
| MGAT4B | 1 | 0 | 1 | 0 | 0 |
| MGAT5 | 1 | 0 | 0 | 0 | 1 |
| MIF | 1 | 0 | 0 | 0 | 0 |
| MMP1 | 1 | 0 | 0 | 0 | 0 |
| MMP14 | 1 | 0 | 0 | 0 | 0 |
| MMP3 | 1 | 0 | 0 | 0 | 0 |
| MMP9 | 1 | 0 | 0 | 0 | 0 |
| MUC5AC | 1 | 0 | 0 | 0 | 0 |
| TLR5 | 1 | 0 | 0 | 0 | 0 |
| TLR6 | 1 | 0 | 0 | 0 | 0 |
| TLR7 | 1 | 0 | 0 | 1 | 0 |
| TLR8 | 1 | 0 | 0 | 0 | 0 |
| MYLK | 1 | 0 | 0 | 0 | 0 |
| NCAM1 | 1 | 0 | 0 | 0 | 0 |
| NFKB1 | 1 | 0 | 0 | 0 | 0 |
| NFKB2 | 1 | 0 | 0 | 0 | 0 |
| NOS1 | 1 | 0 | 0 | 0 | 0 |
| ORM1 | 1 | 0 | 1 | 0 | 0 |
| PAWR | 1 | 0 | 0 | 0 | 0 |
| PFN1 | 1 | 0 | 1 | 0 | 0 |
| PHB1 | 1 | 0 | 1 | 0 | 0 |
| PI3 | 1 | 0 | 0 | 0 | 0 |
| PLA2G2A | 1 | 0 | 0 | 0 | 1 |
| PLAU | 1 | 0 | 0 | 0 | 0 |
| PPP1R15A | 1 | 0 | 0 | 0 | 0 |
| PRKCI | 1 | 0 | 0 | 0 | 0 |
| PRKCZ | 1 | 0 | 0 | 0 | 0 |
| PRKD1 | 1 | 0 | 0 | 0 | 0 |
| Proteasome | 1 | 0 | 0 | 0 | 0 |
| PSMB9 | 1 | 0 | 0 | 0 | 0 |
| PSME1 | 1 | 0 | 0 | 0 | 0 |
| PTGES | 1 | 0 | 0 | 0 | 0 |
| PTHLH | 1 | 0 | 0 | 0 | 0 |
| PTX3 | 1 | 0 | 0 | 0 | 0 |
| RELA | 1 | 0 | 0 | 0 | 1 |
| RIPK1 | 1 | 0 | 0 | 0 | 0 |
| RIPK3 | 1 | 0 | 0 | 0 | 0 |
| SELE | 1 | 0 | 0 | 0 | 0 |
| SERPINE1 | 1 | 0 | 0 | 0 | 0 |
| SERPING1 | 1 | 0 | 0 | 0 | 0 |
| SERPINH1 | 1 | 0 | 0 | 0 | 1 |
| SLC6A4 | 1 | 0 | 0 | 0 | 0 |
| SLC6A6 | 1 | 0 | 0 | 0 | 0 |
| SMPD1 | 1 | 0 | 0 | 0 | 0 |
| SOCS1 | 1 | 0 | 0 | 0 | 0 |
| SP1 | 1 | 0 | 1 | 0 | 0 |
| SPP1 | 1 | 0 | 0 | 0 | 0 |
| ST3GAL6 | 1 | 0 | 0 | 0 | 0 |
| BMX | 1 | 0 | 0 | 0 | 0 |
| ETS1 | 1 | 0 | 0 | 0 | 0 |
| FGFR1 | 1 | 0 | 0 | 1 | 0 |
| FGFR3 | 1 | 1 | 0 | 0 | 0 |
| HDAC | 1 | 0 | 0 | 0 | 0 |
| HDAC1 | 1 | 0 | 0 | 0 | 0 |
| HDAC2 | 1 | 0 | 0 | 0 | 0 |
| HDAC3 | 1 | 0 | 0 | 0 | 0 |
| IFNGR1 | 1 | 0 | 0 | 0 | 0 |
| IFNL1 | 1 | 1 | 0 | 0 | 0 |
| IFNL2 | 1 | 1 | 0 | 0 | 0 |
| JAK2 | 1 | 0 | 0 | 1 | 0 |
| LPL | 1 | 0 | 0 | 0 | 0 |
| PDGFRB | 1 | 0 | 0 | 0 | 0 |
| PRKCD | 1 | 0 | 0 | 0 | 0 |
| PRKCE | 1 | 0 | 0 | 0 | 0 |
| PRMT1 | 1 | 0 | 0 | 1 | 0 |
| RET | 1 | 0 | 0 | 1 | 0 |
| TYK2 | 1 | 0 | 0 | 0 | 0 |
| STAT2 | 1 | 0 | 0 | 0 | 0 |
| TAP1 | 1 | 0 | 0 | 0 | 0 |
| TAP2 | 1 | 0 | 0 | 0 | 0 |
| THBS1 | 1 | 0 | 0 | 0 | 0 |
| TIMP1 | 1 | 0 | 0 | 0 | 1 |
| ADAM17 | 1 | 0 | 0 | 0 | 0 |
| AGTR1 | 1 | 0 | 0 | 0 | 0 |
| AKR1B1 | 1 | 0 | 0 | 1 | 0 |
| ALOX5 | 1 | 0 | 0 | 0 | 0 |
| Caspase | 1 | 0 | 0 | 0 | 0 |
| CENPJ | 1 | 0 | 0 | 0 | 0 |
| DCN | 1 | 0 | 0 | 0 | 0 |
| DCTN1 | 1 | 0 | 0 | 0 | 0 |
| DHX9 | 1 | 0 | 0 | 0 | 0 |
| EGR4 | 1 | 0 | 0 | 0 | 0 |
| HMGB1 | 1 | 0 | 0 | 1 | 0 |
| HNRNPU | 1 | 0 | 0 | 0 | 0 |
| HSPA4 | 1 | 1 | 0 | 0 | 0 |
| HSPB1 | 1 | 0 | 0 | 1 | 0 |
| ICMT | 1 | 0 | 0 | 1 | 0 |
| IRF9 | 1 | 1 | 0 | 0 | 0 |
| MAP3K7 | 1 | 0 | 0 | 0 | 0 |
| MMP12 | 1 | 0 | 0 | 0 | 0 |
| NAMPT | 1 | 0 | 0 | 0 | 0 |
| NFAT | 1 | 1 | 0 | 0 | 0 |
| NOX4 | 1 | 0 | 0 | 1 | 0 |
| NRAS | 1 | 0 | 0 | 1 | 0 |
| OLR1 | 1 | 0 | 0 | 0 | 0 |
| PDE3B | 1 | 1 | 0 | 0 | 0 |
| PLA2G4A | 1 | 1 | 0 | 0 | 0 |
| RALBP1 | 1 | 0 | 0 | 0 | 0 |
| TARDBP | 1 | 0 | 0 | 1 | 0 |
| TONSL | 1 | 0 | 0 | 0 | 0 |

|  |  |  |  |  |
|---|---|---|---|---|
| TP53 | 1 | 0 | 1 | 0 | 0 |
| TRAF1 | 1 | 0 | 0 | 0 | 0 |
| TRAF3 | 1 | 0 | 0 | 0 | 1 |
| TRAF6 | 1 | 0 | 0 | 0 | 0 |
| TRPV1 | 1 | 0 | 0 | 0 | 0 |
| TYMP | 1 | 0 | 0 | 0 | 1 |
| VIM | 1 | 0 | 1 | 0 | 0 |
| ABCC2 | 1 | 0 | 0 | 0 | 0 |
| AHSG | 1 | 0 | 0 | 0 | 0 |
| ALB | 1 | 0 | 1 | 0 | 0 |
| APOE | 1 | 0 | 0 | 0 | 0 |
| C3AR1 | 1 | 0 | 0 | 0 | 0 |
| CASP9 | 1 | 0 | 0 | 0 | 0 |
| CCND1 | 1 | 0 | 0 | 0 | 0 |
| ZNF175 | 1 | 0 | 0 | 0 | 0 |
| CD33 | 1 | 0 | 0 | 0 | 0 |
| CP | 1 | 0 | 0 | 0 | 0 |
| CST3 | 1 | 0 | 0 | 0 | 0 |
| CTNNB1 | 1 | 0 | 0 | 0 | 0 |
| ERG | 1 | 0 | 0 | 0 | 1 |
| F12 | 1 | 0 | 0 | 0 | 0 |
| FOXO1 | 1 | 0 | 0 | 0 | 0 |
| FSCN1 | 1 | 0 | 0 | 0 | 0 |
| GADD45A | 1 | 0 | 0 | 0 | 0 |
| GLI3 | 1 | 0 | 0 | 0 | 0 |
| IGFBP3 | 1 | 0 | 0 | 0 | 0 |
| BMP6 | 1 | 0 | 0 | 0 | 0 |
| GH1 | 1 | 0 | 0 | 0 | 0 |
| MEOX2 | 1 | 1 | 0 | 0 | 0 |
| PPARG | 1 | 0 | 0 | 0 | 0 |
| SOD1 | 1 | 0 | 0 | 0 | 0 |
| INSR | 1 | 0 | 0 | 0 | 0 |
| JUP | 1 | 0 | 0 | 0 | 0 |
| NCOA1 | 1 | 0 | 0 | 0 | 0 |
| NFKBIB | 1 | 0 | 1 | 0 | 0 |
| NOS2 | 1 | 0 | 0 | 0 | 1 |
| OCLN | 1 | 0 | 0 | 0 | 0 |
| PDE4 | 1 | 0 | 1 | 0 | 0 |
| PPM1B | 1 | 0 | 0 | 0 | 0 |
| PROCR | 1 | 0 | 0 | 0 | 0 |
| PROS1 | 1 | 0 | 0 | 0 | 0 |
| RB1 | 1 | 0 | 0 | 0 | 0 |
| RETN | 1 | 0 | 1 | 0 | 0 |
| SCARB1 | 1 | 0 | 0 | 0 | 0 |
| SCD | 1 | 0 | 0 | 0 | 0 |
| SLC10A1 | 1 | 0 | 0 | 0 | 0 |
| SLC2A4 | 1 | 0 | 0 | 0 | 0 |
| SLCO1A2 | 1 | 0 | 0 | 0 | 0 |
| PIAS1 | 1 | 0 | 0 | 0 | 0 |
| TERT | 1 | 0 | 0 | 0 | 0 |
| TFAP2C | 1 | 0 | 0 | 0 | 0 |
| THBD | 1 | 0 | 1 | 0 | 0 |
| AMPK | 1 | 0 | 0 | 0 | 0 |
| EGF | 1 | 0 | 0 | 0 | 0 |
| IL10RA | 1 | 1 | 0 | 0 | 0 |
| P2RY6 | 1 | 1 | 0 | 0 | 0 |
| PEBP1 | 1 | 0 | 0 | 0 | 0 |
| PIAS3 | 1 | 0 | 0 | 0 | 0 |
| PNPLA3 | 1 | 1 | 0 | 0 | 0 |
| PRKCH | 1 | 0 | 0 | 0 | 0 |
| RHOB | 1 | 0 | 0 | 1 | 0 |
| RNF216 | 1 | 0 | 0 | 0 | 0 |
| SALL4 | 1 | 0 | 0 | 0 | 0 |
| SERPINA1 | 1 | 0 | 0 | 0 | 0 |
| SFTPA1 | 1 | 0 | 0 | 0 | 0 |
| SH3BGRL3 | 1 | 0 | 0 | 0 | 0 |
| SOD2 | 1 | 0 | 0 | 0 | 0 |
| TNFAIP3 | 1 | 0 | 0 | 1 | 0 |
| TXN | 1 | 0 | 0 | 0 | 0 |
| USP31 | 1 | 0 | 0 | 0 | 0 |
| VIPR2 | 1 | 0 | 0 | 1 | 0 |
| XIAP | 1 | 1 | 0 | 0 | 0 |
| ZFAND5 | 1 | 0 | 0 | 1 | 0 |
| ZFP36 | 1 | 0 | 0 | 0 | 0 |

| | Sources Unique to High/Low Entropy | Targets Unique to High/Low Entropy | Sources Unique to Low Entropy | Targets Unique to Low Entropy | |
|---|---|---|---|---|---|
| | 14 | 7 | 10 | 8 | Edge Count >1 |
| | 16 | 24 | 19 | 30 | Edge Count = 1 |

**Table S6.** Sources and targets with high and low neighborhood connectivity recruited uniquely by each policy

| Name | EdgeCount | Neighborhood Connectivity | Sources unique to High/Low Entropy | Targets unique to High/Low Entropy | Sources unique to Low Entropy | Targets unique to Low Entropy |
|---|---|---|---|---|---|---|
| CHUK | 2 | 251.00 | 1 | 0 | 0 | 1 |
| IKBKB | 2 | 251.00 | 0 | 0 | 0 | 0 |
| IKBKG | 2 | 251.00 | 0 | 0 | 0 | 0 |
| MAP3K1 | 2 | 251.00 | 0 | 0 | 0 | 0 |
| PTEN | 2 | 251.00 | 0 | 0 | 0 | 0 |
| RIPK2 | 2 | 251.00 | 1 | 1 | 0 | 0 |
| TIMP3 | 2 | 251.00 | 0 | 1 | 0 | 0 |
| NOS3 | 2 | 251.00 | 0 | 0 | 0 | 0 |
| TRAF2 | 2 | 251.00 | 0 | 0 | 0 | 0 |
| TNFRSF1B | 2 | 251.00 | 0 | 0 | 0 | 0 |
| AKT2 | 1 | 251.00 | 0 | 0 | 0 | 0 |
| ALPL | 1 | 251.00 | 0 | 0 | 0 | 1 |
| AREG | 1 | 251.00 | 0 | 0 | 0 | 0 |
| ATF2 | 1 | 251.00 | 0 | 0 | 0 | 0 |
| ATF3 | 1 | 251.00 | 0 | 0 | 0 | 0 |
| BMP2 | 1 | 251.00 | 0 | 0 | 0 | 0 |
| C1R | 1 | 251.00 | 0 | 0 | 0 | 1 |
| C3 | 1 | 251.00 | 0 | 0 | 0 | 0 |
| C5AR1 | 1 | 251.00 | 0 | 0 | 0 | 0 |
| CASP1 | 1 | 251.00 | 0 | 0 | 0 | 0 |
| CASP2 | 1 | 251.00 | 0 | 0 | 0 | 0 |
| CASP6 | 1 | 251.00 | 0 | 1 | 0 | 0 |
| CASP7 | 1 | 251.00 | 0 | 0 | 0 | 0 |
| CCL11 | 1 | 251.00 | 0 | 1 | 0 | 0 |
| CCL3 | 1 | 251.00 | 0 | 0 | 0 | 0 |
| CD38 | 1 | 251.00 | 0 | 0 | 0 | 1 |
| CD55 | 1 | 251.00 | 0 | 0 | 0 | 1 |
| CD59 | 1 | 251.00 | 0 | 0 | 0 | 0 |
| CD80 | 1 | 251.00 | 0 | 0 | 0 | 0 |
| CD86 | 1 | 251.00 | 0 | 0 | 0 | 0 |
| CFB | 1 | 251.00 | 0 | 0 | 0 | 0 |
| CFL2 | 1 | 251.00 | 0 | 0 | 0 | 1 |
| CK2 | 1 | 251.00 | 0 | 0 | 0 | 1 |
| CLEC3B | 1 | 251.00 | 0 | 0 | 0 | 0 |
| CLIC4 | 1 | 251.00 | 0 | 0 | 0 | 0 |
| CSF1 | 1 | 251.00 | 0 | 0 | 0 | 0 |
| CSF2 | 1 | 251.00 | 0 | 0 | 0 | 0 |
| CSF3 | 1 | 251.00 | 0 | 0 | 0 | 1 |
| CTSB | 1 | 251.00 | 0 | 0 | 0 | 0 |
| CX3CL1 | 1 | 251.00 | 0 | 0 | 0 | 0 |
| CXCL10 | 1 | 251.00 | 0 | 0 | 0 | 1 |
| CXCL5 | 1 | 251.00 | 0 | 0 | 0 | 1 |
| CXCL9 | 1 | 251.00 | 0 | 0 | 0 | 0 |
| CXCR3 | 1 | 251.00 | 0 | 1 | 0 | 0 |
| DAPK1 | 1 | 251.00 | 0 | 0 | 0 | 0 |
| EHF | 1 | 251.00 | 0 | 0 | 0 | 0 |
| F2RL1 | 1 | 251.00 | 0 | 0 | 0 | 0 |
| F2RL3 | 1 | 251.00 | 0 | 0 | 0 | 0 |
| FAS | 1 | 251.00 | 0 | 0 | 0 | 0 |
| FCER2 | 1 | 251.00 | 0 | 0 | 0 | 0 |
| FCGR3B | 1 | 251.00 | 0 | 0 | 0 | 0 |
| FOXO4 | 1 | 251.00 | 0 | 0 | 0 | 1 |
| FUT4 | 1 | 251.00 | 0 | 0 | 0 | 0 |
| HIF1A | 1 | 251.00 | 0 | 0 | 0 | 0 |
| HLA-A | 1 | 251.00 | 0 | 1 | 0 | 0 |
| HLA-B | 1 | 251.00 | 0 | 0 | 0 | 0 |
| HLA-C | 1 | 251.00 | 0 | 0 | 0 | 0 |
| HLA-F | 1 | 251.00 | 0 | 1 | 0 | 0 |
| HSD11B1 | 1 | 251.00 | 0 | 0 | 0 | 1 |
| IL11 | 1 | 251.00 | 0 | 1 | 0 | 0 |
| IL12A | 1 | 251.00 | 0 | 0 | 0 | 0 |
| IL12B | 1 | 251.00 | 0 | 0 | 0 | 1 |
| IL4R | 1 | 251.00 | 0 | 0 | 0 | 0 |
| ITGA6 | 1 | 251.00 | 0 | 0 | 0 | 0 |
| ITGAV | 1 | 251.00 | 0 | 0 | 0 | 0 |
| JAK | 1 | 251.00 | 0 | 0 | 0 | 0 |
| JUN | 1 | 251.00 | 0 | 0 | 0 | 1 |
| LYZ | 1 | 251.00 | 0 | 0 | 0 | 0 |
| MAP2K1 | 1 | 251.00 | 0 | 0 | 0 | 1 |
| MAP2K2 | 1 | 251.00 | 0 | 0 | 0 | 0 |
| MAP2K6 | 1 | 251.00 | 0 | 0 | 0 | 0 |
| MAP3K14 | 1 | 251.00 | 0 | 0 | 0 | 0 |
| MAP3K5 | 1 | 251.00 | 0 | 0 | 0 | 0 |
| MAP4K3 | 1 | 251.00 | 0 | 1 | 0 | 0 |
| MAP4K4 | 1 | 251.00 | 0 | 0 | 0 | 0 |
| MAPK9 | 1 | 251.00 | 0 | 0 | 0 | 1 |
| MAPKAPK3 | 1 | 251.00 | 0 | 0 | 0 | 0 |
| MIF | 1 | 251.00 | 0 | 0 | 0 | 0 |
| MMP1 | 1 | 251.00 | 0 | 0 | 0 | 0 |
| MMP14 | 1 | 251.00 | 0 | 0 | 0 | 0 |
| MMP3 | 1 | 251.00 | 0 | 0 | 0 | 0 |
| MUC5AC | 1 | 251.00 | 0 | 0 | 0 | 0 |
| MYLK | 1 | 251.00 | 0 | 0 | 0 | 0 |
| NCAM1 | 1 | 251.00 | 0 | 0 | 0 | 0 |
| NFKB1 | 1 | 251.00 | 0 | 0 | 0 | 0 |
| NFKB2 | 1 | 251.00 | 0 | 0 | 0 | 0 |
| NOS1 | 1 | 251.00 | 0 | 0 | 0 | 0 |
| PAWR | 1 | 251.00 | 0 | 0 | 0 | 0 |
| PFN1 | 1 | 251.00 | 0 | 1 | 0 | 0 |
| PLA2G2A | 1 | 251.00 | 0 | 0 | 0 | 1 |
| PLAU | 1 | 251.00 | 0 | 0 | 0 | 0 |
| PPP1R15A | 1 | 251.00 | 0 | 0 | 0 | 0 |
| PRKCI | 1 | 251.00 | 0 | 0 | 0 | 0 |
| PRKCZ | 1 | 251.00 | 0 | 0 | 0 | 0 |
| Proteasome | 1 | 251.00 | 0 | 0 | 0 | 0 |
| PSMB9 | 1 | 251.00 | 0 | 0 | 0 | 0 |
| PSME1 | 1 | 251.00 | 0 | 0 | 0 | 0 |
| PTGES | 1 | 251.00 | 0 | 0 | 0 | 0 |
| PTHLH | 1 | 251.00 | 0 | 0 | 0 | 0 |
| PTX3 | 1 | 251.00 | 0 | 0 | 0 | 0 |
| RELA | 1 | 251.00 | 0 | 0 | 0 | 1 |
| RIPK1 | 1 | 251.00 | 0 | 0 | 0 | 0 |
| RIPK3 | 1 | 251.00 | 0 | 0 | 0 | 0 |
| SELE | 1 | 251.00 | 0 | 0 | 0 | 0 |
| SERPINE1 | 1 | 251.00 | 0 | 0 | 0 | 0 |
| SERPING1 | 1 | 251.00 | 0 | 0 | 0 | 0 |
| SERPINH1 | 1 | 251.00 | 0 | 0 | 0 | 1 |
| SLC6A4 | 1 | 251.00 | 0 | 0 | 0 | 0 |
| SLC6A6 | 1 | 251.00 | 0 | 0 | 0 | 0 |
| SMPD1 | 1 | 251.00 | 0 | 0 | 0 | 0 |
| SP1 | 1 | 251.00 | 0 | 1 | 0 | 0 |
| SPP1 | 1 | 251.00 | 0 | 0 | 0 | 0 |
| ST3GAL6 | 1 | 251.00 | 0 | 0 | 0 | 0 |
| TAP1 | 1 | 251.00 | 0 | 0 | 0 | 0 |
| TAP2 | 1 | 251.00 | 0 | 0 | 0 | 0 |
| THBS1 | 1 | 251.00 | 0 | 0 | 0 | 0 |
| TIMP1 | 1 | 251.00 | 0 | 0 | 0 | 1 |
| ADAM17 | 1 | 251.00 | 0 | 0 | 0 | 0 |
| AGTR1 | 1 | 251.00 | 0 | 0 | 0 | 0 |
| AKR1B1 | 1 | 251.00 | 0 | 0 | 1 | 0 |
| ALOX5 | 1 | 251.00 | 0 | 0 | 0 | 0 |
| Caspase | 1 | 251.00 | 0 | 0 | 0 | 0 |
| CENPJ | 1 | 251.00 | 0 | 0 | 0 | 0 |
| DCN | 1 | 251.00 | 0 | 0 | 0 | 0 |
| DCTN1 | 1 | 251.00 | 0 | 0 | 0 | 0 |
| DHX9 | 1 | 251.00 | 0 | 0 | 0 | 0 |
| EGR4 | 1 | 251.00 | 0 | 0 | 0 | 0 |
| HMGB1 | 1 | 251.00 | 0 | 0 | 1 | 0 |
| HNRNPU | 1 | 251.00 | 0 | 0 | 0 | 0 |
| HSPA4 | 1 | 251.00 | 1 | 0 | 0 | 0 |
| HSPB1 | 1 | 251.00 | 0 | 0 | 1 | 0 |

*Neighborhood Connectivity = max or 251*

|  |  |  |  |  |  |  |
|---|---|---|---|---|---|---|
| ICMT | 1 | 251.00 | 0 | 0 | 1 | 0 |
| IRF9 | 1 | 251.00 | 1 | 0 | 0 | 0 |
| MAP3K7 | 1 | 251.00 | 0 | 0 | 0 | 0 |
| MMP12 | 1 | 251.00 | 0 | 0 | 0 | 0 |
| NAMPT | 1 | 251.00 | 0 | 0 | 0 | 0 |
| NFAT | 1 | 251.00 | 1 | 0 | 0 | 0 |
| NOX4 | 1 | 251.00 | 0 | 0 | 1 | 0 |
| NRAS | 1 | 251.00 | 0 | 0 | 1 | 0 |
| OLR1 | 1 | 251.00 | 0 | 0 | 0 | 0 |
| PDE3B | 1 | 251.00 | 1 | 0 | 0 | 0 |
| PLA2G4A | 1 | 251.00 | 1 | 0 | 0 | 0 |
| RALBP1 | 1 | 251.00 | 0 | 0 | 0 | 0 |
| TARDBP | 1 | 251.00 | 0 | 0 | 1 | 0 |
| TONSL | 1 | 251.00 | 0 | 0 | 0 | 0 |
| TP53 | 1 | 251.00 | 0 | 1 | 0 | 0 |
| TRAF1 | 1 | 251.00 | 0 | 0 | 0 | 0 |
| TRAF3 | 1 | 251.00 | 0 | 0 | 0 | 1 |
| TYMP | 1 | 251.00 | 0 | 0 | 0 | 1 |
| VIM | 1 | 251.00 | 0 | 1 | 0 | 0 |
| ABCC2 | 1 | 251.00 | 0 | 0 | 0 | 0 |
| C3AR1 | 1 | 251.00 | 0 | 0 | 0 | 0 |
| CP | 1 | 251.00 | 0 | 0 | 0 | 0 |
| CST3 | 1 | 251.00 | 0 | 0 | 0 | 0 |
| CTNNB1 | 1 | 251.00 | 0 | 0 | 0 | 0 |
| ERG | 1 | 251.00 | 0 | 0 | 0 | 1 |
| FSCN1 | 1 | 251.00 | 0 | 0 | 0 | 0 |
| IGFBP3 | 1 | 251.00 | 0 | 0 | 0 | 0 |
| INSR | 1 | 251.00 | 0 | 0 | 0 | 0 |
| JUP | 1 | 251.00 | 0 | 0 | 0 | 0 |
| NFKBIB | 1 | 251.00 | 0 | 1 | 0 | 0 |
| NOS2 | 1 | 251.00 | 0 | 0 | 0 | 1 |
| OCLN | 1 | 251.00 | 0 | 0 | 0 | 0 |
| PPM1B | 1 | 251.00 | 0 | 0 | 0 | 0 |
| PROCR | 1 | 251.00 | 0 | 0 | 0 | 0 |
| RB1 | 1 | 251.00 | 0 | 0 | 0 | 0 |
| RETN | 1 | 251.00 | 0 | 1 | 0 | 0 |
| SCARB1 | 1 | 251.00 | 0 | 0 | 0 | 0 |
| SCD | 1 | 251.00 | 0 | 0 | 0 | 0 |
| SLC10A1 | 1 | 251.00 | 0 | 0 | 0 | 0 |
| SLC2A4 | 1 | 251.00 | 0 | 0 | 0 | 0 |
| SLCO1A2 | 1 | 251.00 | 0 | 0 | 0 | 0 |
| TFAP2C | 1 | 251.00 | 0 | 0 | 0 | 0 |
| THBD | 1 | 251.00 | 0 | 1 | 0 | 0 |
| AMPK | 1 | 251.00 | 0 | 0 | 0 | 0 |
| EGF | 1 | 251.00 | 0 | 0 | 0 | 0 |
| IL10RA | 1 | 251.00 | 1 | 0 | 0 | 0 |
| P2RY6 | 1 | 251.00 | 1 | 0 | 0 | 0 |
| PEBP1 | 1 | 251.00 | 0 | 0 | 0 | 0 |
| PIAS3 | 1 | 251.00 | 0 | 0 | 0 | 0 |
| PNPLA3 | 1 | 251.00 | 1 | 0 | 0 | 0 |
| PRKCH | 1 | 251.00 | 0 | 0 | 0 | 0 |
| RHOB | 1 | 251.00 | 0 | 0 | 1 | 0 |
| RNF216 | 1 | 251.00 | 0 | 0 | 0 | 0 |
| SALL4 | 1 | 251.00 | 0 | 0 | 0 | 0 |
| SERPINA1 | 1 | 251.00 | 0 | 0 | 0 | 0 |
| SFTPA1 | 1 | 251.00 | 0 | 0 | 0 | 0 |
| SH3BGRL3 | 1 | 251.00 | 0 | 0 | 0 | 0 |
| SOD2 | 1 | 251.00 | 0 | 0 | 0 | 0 |
| TNFAIP3 | 1 | 251.00 | 0 | 0 | 1 | 0 |
| TXN | 1 | 251.00 | 0 | 0 | 0 | 0 |
| USP31 | 1 | 251.00 | 0 | 0 | 0 | 0 |
| VIPR2 | 1 | 251.00 | 0 | 0 | 1 | 0 |
| XIAP | 1 | 251.00 | 1 | 0 | 1 | 0 |
| ZFAND5 | 1 | 251.00 | 0 | 0 | 1 | 0 |
| ZFP36 | 1 | 251.00 | 0 | 0 | 0 | 0 |
| CD40LG | 3 | 183.50 | 0 | 0 | 0 | 0 |
| IL17A | 3 | 183.50 | 0 | 0 | 0 | 0 |
| IL18 | 3 | 183.50 | 0 | 0 | 0 | 0 |
| INS | 3 | 183.50 | 0 | 0 | 0 | 0 |
| POU2F1 | 3 | 183.50 | 0 | 0 | 0 | 0 |
| ADIPOQ | 3 | 183.50 | 0 | 0 | 1 | 0 |
| ABCB1 | 2 | 183.50 | 0 | 0 | 0 | 0 |
| CD1A | 2 | 183.50 | 0 | 1 | 0 | 0 |
| CD40 | 2 | 183.50 | 0 | 0 | 0 | 0 |
| CEBPB | 2 | 183.50 | 0 | 0 | 0 | 1 |
| HSF1 | 2 | 183.50 | 0 | 1 | 0 | 0 |
| ADRB2 | 2 | 183.50 | 0 | 0 | 1 | 0 |
| BSG | 2 | 183.50 | 0 | 0 | 0 | 0 |
| ERBB2 | 2 | 183.50 | 0 | 0 | 0 | 0 |
| IL13 | 2 | 183.50 | 0 | 0 | 0 | 0 |
| PARP1 | 2 | 183.50 | 1 | 0 | 0 | 0 |
| PTK2 | 2 | 183.50 | 0 | 0 | 0 | 0 |
| TNFRSF1A | 2 | 183.50 | 0 | 0 | 1 | 0 |
| VEGF | 2 | 183.50 | 1 | 0 | 0 | 0 |
| XDH | 2 | 183.50 | 0 | 0 | 0 | 0 |
| MAPK14 | 2 | 183.50 | 0 | 0 | 0 | 0 |
| RORA | 2 | 183.50 | 0 | 0 | 0 | 0 |
| SERPINA3 | 2 | 183.50 | 0 | 0 | 0 | 0 |
| VEGFA | 2 | 183.50 | 0 | 1 | 0 | 0 |
| CDKN1A | 2 | 144.50 | 0 | 0 | 0 | 0 |
| IL10 | 2 | 144.50 | 0 | 0 | 0 | 0 |
| ADRB | 2 | 144.50 | 0 | 0 | 1 | 0 |
| EGFR | 4 | 135.00 | 0 | 0 | 1 | 0 |
| MAPK3 | 4 | 135.00 | 0 | 0 | 0 | 0 |
| PTGS2 | 3 | 135.00 | 0 | 0 | 0 | 1 |
| ESR1 | 3 | 135.00 | 0 | 0 | 0 | 0 |
| NFKBIA | 2 | 135.00 | 1 | 0 | 0 | 0 |
| VCAM1 | 2 | 135.00 | 0 | 0 | 0 | 0 |
| CASP8 | 2 | 133.50 | 0 | 0 | 0 | 0 |
| TLR1 | 2 | 133.50 | 0 | 0 | 1 | 0 |
| EP300 | 2 | 130.50 | 0 | 0 | 0 | 1 |
| IFNB1 | 2 | 130.50 | 0 | 0 | 0 | 0 |
| IL1B | 4 | 128.67 | 0 | 0 | 0 | 0 |
| CCL2 | 3 | 128.67 | 0 | 0 | 0 | 1 |
| TGFA | 2 | 128.00 | 0 | 0 | 0 | 0 |
| GSK3B | 2 | 126.50 | 0 | 0 | 0 | 0 |
| CXCL12 | 3 | 125.67 | 1 | 0 | 0 | 0 |
| TLR4 | 3 | 125.67 | 0 | 0 | 0 | 0 |
| STAT3 | 4 | 124.67 | 0 | 1 | 0 | 0 |
| AP1 | 3 | 123.33 | 1 | 0 | 0 | 0 |
| AGT | 4 | 122.67 | 1 | 0 | 0 | 1 |
| NGF | 3 | 122.67 | 0 | 0 | 0 | 0 |
| HCK | 2 | 116.00 | 0 | 0 | 0 | 0 |
| POMC | 2 | 116.00 | 0 | 0 | 0 | 0 |
| A2M | 1 | 116.00 | 0 | 0 | 0 | 0 |
| ABCC1 | 1 | 116.00 | 0 | 0 | 0 | 0 |
| ABCG2 | 1 | 116.00 | 0 | 0 | 0 | 1 |
| ADAMTS5 | 1 | 116.00 | 0 | 0 | 0 | 1 |
| AKT1 | 1 | 116.00 | 0 | 0 | 0 | 0 |
| ANXA1 | 1 | 116.00 | 0 | 0 | 0 | 0 |
| C8A | 1 | 116.00 | 0 | 0 | 0 | 0 |
| C8G | 1 | 116.00 | 0 | 1 | 0 | 0 |
| CBL | 1 | 116.00 | 0 | 0 | 0 | 1 |
| CD163 | 1 | 116.00 | 0 | 0 | 0 | 1 |
| CDKN1B | 1 | 116.00 | 0 | 0 | 0 | 0 |
| CMA1 | 1 | 116.00 | 0 | 0 | 0 | 0 |
| CTSL | 1 | 116.00 | 0 | 0 | 0 | 0 |
| CTSS | 1 | 116.00 | 0 | 0 | 0 | 0 |
| FN1 | 1 | 116.00 | 0 | 0 | 0 | 0 |
| FYN | 1 | 116.00 | 0 | 0 | 0 | 1 |
| GBF1 | 1 | 116.00 | 0 | 0 | 0 | 0 |
| HBEGF | 1 | 116.00 | 0 | 0 | 0 | 0 |
| HP | 1 | 116.00 | 0 | 1 | 0 | 0 |
| HSP90B1 | 1 | 116.00 | 0 | 0 | 0 | 0 |

| Gene | Count | Value | Sources Unique to High/Low Entropy | Targets Unique to High/Low Entropy | Sources Unique to Low Entropy | Targets Unique to Low Entropy |
|---|---|---|---|---|---|---|
| AHR | 1 | 116.00 | 0 | 0 | 0 | 0 |
| AR | 1 | 116.00 | 0 | 0 | 0 | 0 |
| F10 | 1 | 116.00 | 0 | 0 | 0 | 0 |
| F2 | 1 | 116.00 | 0 | 0 | 0 | 0 |
| GAB1 | 1 | 116.00 | 0 | 0 | 0 | 0 |
| IL17F | 1 | 116.00 | 1 | 0 | 1 | 0 |
| IL1R1 | 1 | 116.00 | 0 | 0 | 0 | 0 |
| IL6R | 1 | 116.00 | 0 | 0 | 0 | 0 |
| ITGA5 | 1 | 116.00 | 0 | 0 | 0 | 0 |
| KLF7 | 1 | 116.00 | 0 | 0 | 0 | 0 |
| LEPR | 1 | 116.00 | 0 | 0 | 0 | 0 |
| LMX1B | 1 | 116.00 | 0 | 0 | 0 | 0 |
| PIK3C3 | 1 | 116.00 | 0 | 0 | 0 | 0 |
| PNPT1 | 1 | 116.00 | 0 | 0 | 0 | 0 |
| PTGER4 | 1 | 116.00 | 0 | 0 | 0 | 0 |
| TIRAP | 1 | 116.00 | 0 | 0 | 0 | 0 |
| KRAS | 1 | 116.00 | 0 | 0 | 0 | 0 |
| LYN | 1 | 116.00 | 0 | 1 | 0 | 0 |
| MAPK7 | 1 | 116.00 | 0 | 0 | 0 | 0 |
| MAZ | 1 | 116.00 | 0 | 0 | 0 | 0 |
| MEF2 | 1 | 116.00 | 0 | 1 | 0 | 0 |
| MGAT4B | 1 | 116.00 | 0 | 1 | 0 | 0 |
| MGAT5 | 1 | 116.00 | 0 | 0 | 0 | 1 |
| MMP9 | 1 | 116.00 | 0 | 0 | 0 | 0 |
| ORM1 | 1 | 116.00 | 0 | 1 | 0 | 0 |
| PHB1 | 1 | 116.00 | 0 | 1 | 0 | 0 |
| PRKD1 | 1 | 116.00 | 0 | 0 | 0 | 0 |
| STAT2 | 1 | 116.00 | 0 | 0 | 0 | 0 |
| AHSG | 1 | 116.00 | 0 | 0 | 0 | 0 |
| ALB | 1 | 116.00 | 0 | 1 | 0 | 0 |
| APOE | 1 | 116.00 | 0 | 0 | 0 | 0 |
| CASP9 | 1 | 116.00 | 0 | 0 | 0 | 0 |
| CD33 | 1 | 116.00 | 0 | 0 | 0 | 0 |
| F12 | 1 | 116.00 | 0 | 0 | 0 | 0 |
| FOXO1 | 1 | 116.00 | 0 | 0 | 0 | 0 |
| GADD45A | 1 | 116.00 | 0 | 0 | 0 | 0 |
| BMP6 | 1 | 116.00 | 0 | 0 | 0 | 0 |
| GH1 | 1 | 116.00 | 0 | 0 | 0 | 0 |
| MEOX2 | 1 | 116.00 | 1 | 0 | 0 | 0 |
| PPARG | 1 | 116.00 | 0 | 0 | 0 | 0 |
| SOD1 | 1 | 116.00 | 0 | 0 | 0 | 0 |
| NCOA1 | 1 | 116.00 | 0 | 0 | 0 | 0 |
| PROS1 | 1 | 116.00 | 0 | 0 | 0 | 0 |
| MAPK1 | 4 | 102.50 | 0 | 0 | 1 | 0 |
| TLR2 | 4 | 92.33 | 1 | 0 | 0 | 0 |
| IFNG | 6 | 85.60 | 0 | 0 | 0 | 1 |
| IL6ST | 3 | 77.00 | 1 | 0 | 0 | 0 |
| OSM | 2 | 77.00 | 1 | 0 | 0 | 0 |
| PTPN11 | 2 | 77.00 | 0 | 0 | 0 | 0 |
| SOCS3 | 2 | 77.00 | 0 | 0 | 0 | 0 |
| IL4 | 2 | 67.50 | 0 | 0 | 0 | 0 |
| PPARA | 2 | 67.50 | 0 | 0 | 1 | 0 |
| PI3K | 2 | 63.00 | 0 | 0 | 0 | 0 |
| TGFB1 | 2 | 63.00 | 0 | 0 | 0 | 0 |
| AKT | 2 | 60.50 | 0 | 0 | 0 | 0 |
| PKC | 2 | 59.50 | 1 | 0 | 0 | 0 |
| NFkappaB | 10 | 57.13 | 0 | 0 | 0 | 0 |
| ERK | 3 | 46.67 | 0 | 0 | 0 | 0 |
| p38 | 3 | 45.33 | 0 | 0 | 0 | 0 |
| SOCS1 | 1 | 38.00 | 0 | 0 | 0 | 0 |
| BMX | 1 | 38.00 | 0 | 0 | 0 | 0 |
| ETS1 | 1 | 38.00 | 0 | 0 | 0 | 0 |
| FGFR1 | 1 | 38.00 | 0 | 0 | 1 | 0 |
| FGFR3 | 1 | 38.00 | 1 | 0 | 0 | 0 |
| HDAC | 1 | 38.00 | 0 | 0 | 0 | 0 |
| HDAC1 | 1 | 38.00 | 0 | 0 | 0 | 0 |
| HDAC2 | 1 | 38.00 | 0 | 0 | 0 | 0 |
| HDAC3 | 1 | 38.00 | 0 | 0 | 0 | 0 |
| IFNGR1 | 1 | 38.00 | 0 | 0 | 0 | 0 |
| IFNL1 | 1 | 38.00 | 1 | 0 | 0 | 0 |
| IFNL2 | 1 | 38.00 | 1 | 0 | 0 | 0 |
| JAK2 | 1 | 38.00 | 0 | 0 | 1 | 0 |
| LPL | 1 | 38.00 | 0 | 0 | 0 | 0 |
| PDGFRB | 1 | 38.00 | 0 | 0 | 0 | 0 |
| PRKCD | 1 | 38.00 | 0 | 0 | 0 | 0 |
| PRKCE | 1 | 38.00 | 0 | 0 | 0 | 0 |
| PRMT1 | 1 | 38.00 | 0 | 0 | 1 | 0 |
| RET | 1 | 38.00 | 0 | 0 | 1 | 0 |
| TYK2 | 1 | 38.00 | 0 | 0 | 0 | 0 |
| PIAS1 | 1 | 38.00 | 0 | 0 | 0 | 0 |
| IFI16 | 7 | 37.57 | 0 | 0 | 0 | 0 |
| CD14 | 11 | 27.40 | 0 | 0 | 0 | 0 |
| IRF5 | 3 | 27.00 | 0 | 0 | 1 | 0 |
| IRAK1 | 2 | 27.00 | 0 | 0 | 0 | 1 |
| CRP | 20 | 21.32 | 0 | 0 | 0 | 0 |
| IFNA1 | 4 | 19.33 | 0 | 0 | 0 | 0 |
| CCR2 | 1 | 19.00 | 0 | 0 | 0 | 0 |
| CD32 | 1 | 19.00 | 0 | 0 | 0 | 0 |
| HNF1A | 1 | 19.00 | 0 | 0 | 0 | 0 |
| LEP | 1 | 19.00 | 0 | 0 | 0 | 0 |
| FCGR1A | 1 | 19.00 | 0 | 0 | 0 | 0 |
| TERT | 1 | 19.00 | 0 | 0 | 0 | 0 |
| IKK_family | 1 | 16.00 | 0 | 0 | 0 | 0 |
| IRAK4 | 1 | 16.00 | 0 | 0 | 0 | 1 |
| TLR5 | 1 | 16.00 | 0 | 0 | 0 | 0 |
| TLR6 | 1 | 16.00 | 0 | 0 | 0 | 0 |
| TLR7 | 1 | 16.00 | 0 | 0 | 1 | 0 |
| TLR8 | 1 | 16.00 | 0 | 0 | 0 | 0 |
| TRAF6 | 1 | 16.00 | 0 | 0 | 0 | 0 |
| CXCL8 | 2 | 12.00 | 0 | 0 | 0 | 0 |
| IRF7 | 2 | 10.50 | 0 | 0 | 0 | 0 |
| CEBP | 1 | 10.00 | 1 | 0 | 0 | 0 |
| KITLG | 1 | 10.00 | 0 | 0 | 0 | 0 |
| BEGAIN | 1 | 10.00 | 0 | 0 | 1 | 0 |
| IKBKE | 1 | 10.00 | 0 | 0 | 0 | 0 |
| MAVS | 1 | 10.00 | 0 | 0 | 0 | 0 |
| TICAM1 | 1 | 10.00 | 0 | 0 | 0 | 0 |
| ZNF175 | 1 | 10.00 | 0 | 0 | 0 | 0 |
| STAT1 | 42 | 8.39 | 0 | 0 | 0 | 0 |
| HOXD3 | 1 | 7.00 | 0 | 0 | 0 | 0 |
| LIF | 1 | 7.00 | 1 | 0 | 0 | 0 |
| RAF1 | 1 | 7.00 | 0 | 0 | 1 | 0 |
| CCND1 | 1 | 7.00 | 0 | 0 | 0 | 0 |
| TLR9 | 5 | 6.00 | 1 | 0 | 0 | 0 |
| TLR3 | 3 | 6.00 | 0 | 0 | 0 | 0 |
| PRKAC | 2 | 6.00 | 1 | 0 | 0 | 0 |
| TBK1 | 3 | 5.00 | 0 | 0 | 0 | 0 |
| GLI3 | 1 | 5.00 | 0 | 0 | 0 | 0 |
| PDE4 | 1 | 5.00 | 0 | 1 | 0 | 0 |
| AIM2 | 1 | 3.00 | 0 | 1 | 0 | 0 |
| NKX2-1 | 1 | 3.00 | 0 | 0 | 0 | 0 |
| PI3 | 1 | 3.00 | 0 | 0 | 0 | 0 |
| TRPV1 | 1 | 3.00 | 0 | 0 | 0 | 0 |
| MYD88 | 17 | 2.38 | 0 | 0 | 1 | 0 |
| PTPA | 5 | 2.20 | 0 | 0 | 0 | 0 |
| CA2 | 3 | 2.00 | 0 | 1 | 0 | 0 |
| CAMP | 2 | 2.00 | 0 | 0 | 0 | 0 |
| IRF3 | 10 | 1.90 | 0 | 0 | 0 | 0 |
| IL6 | 126 | 1.79 | 0 | 0 | 0 | 0 |
| TNF | 273 | 1.57 | 0 | 0 | 0 | 0 |
|  |  |  | Sources Unique to High/Low Entropy | Targets Unique to High/Low Entropy | Sources Unique to Low Entropy | Targets Unique to Low Entropy |  |
|  |  |  | 11 | 16 | 11 | 24 | Neighborhood Connectivity = max |
|  |  |  | 19 | 15 | 18 | 14 | Neighborhood Connectivity < max |